\documentclass[11pt]{article}
\usepackage{amsmath}
\usepackage[colorlinks=true, allcolors=blue]{hyperref}

\makeatletter
\newcommand{\removelatexerror}{\let\@latex@error\@gobble}
\makeatother
\usepackage{amsmath}
\usepackage{amsthm}
\usepackage{amsfonts}
\usepackage{amssymb}
\usepackage{graphicx}
\usepackage{dsfont}
\usepackage{mathtools}
\usepackage{cleveref}
\usepackage{pgfplots}
\usepackage{bbm}
\usepackage[noend]{algpseudocode}
\usepackage{stmaryrd}
\newcommand{\cceil}[1]{\llceil #1 \rrceil}
\usepackage{algorithm}
\usepackage{complexity}
\usepackage{enumitem}
\usepackage{tikz}
\usetikzlibrary{positioning}
\usepackage{thm-restate}
\usepackage{float}
\usetikzlibrary{calc}
\usepackage{thm-restate}
\usetikzlibrary{shapes,decorations.markings}
\usepackage{caption}
\usepackage{subcaption}
\usepackage{comment}
\newcommand{\rank}{\mathrm{rank}}

\newtheorem{theorem}{Theorem}
\newtheorem{observation}[theorem]{Observation}
\newtheorem{definition}[theorem]{Definition}
\newtheorem{claim}[theorem]{Claim}
\newtheorem{remark}[theorem]{Remark}
\newtheorem{them}[theorem]{Theorem}
\newtheorem{lem}[theorem]{Lemma}
\newtheorem{cor}[theorem]{Corollary}
\newtheorem{fact}[theorem]{Fact}

\renewcommand{\E}{\mathbb{E}}
\newcommand{\U}{\mathcal{U}}
\usepackage[english]{babel}
\usepackage[T1]{fontenc}
\usepackage{caption}

\newcommand{\repeatcaption}[2]{%
  \renewcommand{\thefigure}{\ref{#1}}%
  \captionsetup{list=no}%
  \caption{#2 (repeated from page \pageref{#1})}%
  \addtocounter{figure}{-1}%
}
\usepackage[margin=1in]{geometry}

\algblockdefx{Withprob}{EndWithprob}[1]{\textbf{with probability} #1 \textbf{do}}{}

\makeatletter
\ifthenelse{\equal{\ALG@noend}{t}}%
  {\algtext*{EndWithprob}}
  {}%
\makeatother

\title{Near-Optimal Relative Error Streaming Quantile Estimation \\ via Elastic Compactors}
\author{Elena Gribelyuk\footnote{Princeton University, email: \href{mailto:eg5539@princeton.edu}{\url{eg5539@princeton.edu}}. Supported by NSF CAREER award CCF-2339942.} \and Pachara Sawettamalya\footnote{Princeton University. Email: \href{mailto:pachara@princeton.edu}{\url{pachara@princeton.edu}}. Supported by NSF CAREER award CCF-2339942.} \and Hongxun Wu\footnote{UC Berkeley. Email: \href{mailto:wuhx@berkeley.edu}{\url{wuhx@berkeley.edu}}. Supported by Avishay Tal's Sloan Research Fellowship, NSF CAREER Award CCF-2145474, and Jelani Nelson's NSF award CCF-2427808.} \and Huacheng Yu\footnote{Princeton University. Email:\href{mailto:yuhch123@gmail.com}{\url{yuhch123@gmail.com}}. Supported by NSF CAREER award CCF-2339942.}}
\date{}
\begin{document}
\maketitle
\begin{abstract} 
    Computing the approximate quantiles or ranks of a stream is a fundamental task in data monitoring. Given a stream of elements $x_1, x_2, \dots, x_n$ and a query $x$, a relative-error quantile estimation algorithm can estimate the rank of $x$ with respect to the stream, up to a multiplicative $\pm \epsilon \cdot \rank(x)$ error. Notably, this requires the sketch to obtain more precise estimates for the ranks of elements on the tails of the distribution, as compared to the additive $\pm \epsilon n$ error regime. This is particularly favorable for some practical applications, such as anomaly detection. 

    Previously, the best known algorithms for relative error achieved space $\tilde O(\epsilon^{-1} \log^{1.5}(\epsilon n))$ (Cormode, Karnin, Liberty, Thaler, Vesel{\`y}, 2021) and $\tilde O(\epsilon^{-2} \log(\epsilon n))$ (Zhang, Lin, Xu, Korn, Wang, 2006). In this work, we present a nearly-optimal streaming algorithm for the relative-error quantile estimation problem using $\tilde O(\epsilon^{-1} \log(\epsilon n))$ space, which almost matches the trivial $\Omega(\epsilon^{-1} \log (\epsilon n))$ space lower bound. 

    To surpass the $\Omega(\epsilon^{-1} \log^{1.5}(\epsilon n))$ barrier of the previous approach, our algorithm crucially relies on a new data structure, called an \emph{elastic compactor}, which can be dynamically resized over the course of the stream. Interestingly, we design a space allocation scheme which adaptively allocates space to each compactor based on the ``hardness'' of the input stream. This approach allows us to avoid using the maximal space \emph{simultaneously} for every compactor and facilitates the improvement in the total space complexity. 
    
    Along the way, we also propose and study a new problem called the Top Quantiles Problem, which only requires the sketch to provide estimates for the ranks of elements in a fixed-length tail of the distribution. This problem serves as an important subproblem in our algorithm, though it is also an interesting problem of its own right.
\end{abstract}

\thispagestyle{empty}
\newpage
\tableofcontents
\pagenumbering{roman}
\newpage
\pagenumbering{arabic}

\section{Introduction}
Learning the distribution of data that are represented as a stream is an important task in streaming data analysis.
A concrete problem that captures this task is the \emph{streaming quantile estimation} problem. %

Given a stream of elements $\pi=(x_1,x_2,\ldots,x_n)$, the quantile estimation problem asks us to process the stream, while maintaining a \emph{small memory} that stores a few input elements, such that at the end of the stream, for any given query $y$, the algorithm must output an approximation of the \emph{rank} of $y$ in $\pi$ with high probability, i.e., an approximation of $\rank_{\pi}(y):=\left|\left\{i\in[n]: x_i<y\right\}\right|$.\footnote{In this work, we consider algorithms in \emph{comparison-based model}, wherein stream elements are drawn from a universe equipped with a total-ordering. At any time, the algorithm may only performed comparisons between any two elements stored in memory, and does not depend on the true value of each element.}

The problem has been extensively studied \cite{greenwald2001space, karnin2016optimal, agrawal1995, ostrovsky2015, manku1998, manku2004, ioannidis1999} when we allow \emph{additive error}, i.e., the algorithm outputs an estimate $\widehat{\rank}_\pi(y)=\rank_{\pi}(y)\pm\epsilon n$ with high probability.
Optimal bounds are known in this setting: Karnin, Lang, and Liberty proposed an algorithm with $O((1/\epsilon) \log \log (1/\delta)
)$ space, which matches the best space one can hope for even for \emph{offline} algorithms when the failure probability $\delta$ is a constant \cite{karnin2016optimal}.\footnote{An offline algorithm sees the elements all at once and computes a small sketch that can answer rank queries approximately.}

On the other hand, oftentimes, the application needs to accurately learn the tail distribution of the data stream. For instance, this need arises when monitoring network latencies: the distribution of response times is often very long-tailed, and understanding the occasional, yet problematic, high response times is a key purpose of the task~\cite{cormode2005effective}.
For such applications, algorithms that guarantee \emph{relative errors} give high accuracy on the tail distribution, and are thus more aligned with this stricter requirement.
That is, the algorithm must return $\widehat{\rank}_\pi(y)=(1\pm \epsilon)\rank_{\pi}(y)$.\footnote{The definition as-is gives higher accuracy for queries with small ranks. By running the algorithm with a reversed total-ordering of stream elements, we can obtain high accuracy at the tail of the distribution.} The relative-error quantile estimation task also arises when approximately counting the inversions in a stream~\cite{gupta2003counting}.

The optimal bound for offline algorithms with relative error is $\Theta(\epsilon^{-1}\log(\epsilon n))$, by simply storing elements with ranks $\{1,2,\ldots,\epsilon^{-1}\}$ and $\{\epsilon^{-1}(1+\epsilon), \epsilon^{-1}(1+\epsilon)^2,\ldots\}$. Best-known streaming algorithms are Multi-Layer Randomization (``MR'' algorithm) by Zhang, Lin, Xu, Korn and Wang~\cite{zhang2006space} with  $O(\epsilon^{-2} \log(\epsilon^2 n))$ space, and a recent breakthrough~\cite{cormode2023relative} by Cormode, Karnin, Liberty, Thaler and Vesel\`{y} with $O(\epsilon^{-1}\log^{1.5}(\epsilon n))$ space (we will refer to it as the CKLTV algorithm below).

The MR algorithm~\cite{zhang2006space} maintains logarithmically many sketches of size $O(\epsilon^{-2})$.
Each sketch is responsible for queries with rank in $[\epsilon^{-2} 2^i, \epsilon^{-2} 2^{i+1})$ for some $i$. More recently,
Cormode et al.~\cite{cormode2023relative} introduced ``relative compactors''. Roughly speaking, a relative compactor takes a stream of elements as input and outputs a shorter stream such that the rank of any query in the \emph{input} stream can be approximated with \emph{small relative error} based on its rank in the \emph{output} stream (see \Cref{sec:full-details-elastic} for a more detailed overview).
Then, the algorithm of \cite{cormode2023relative} ``connects'' logarithmically many relative compactors, i.e., the output stream of the previous relative compactor is fed (online, in the streaming sense) to the next relative compactor as its input stream. 
However, both of the aforementioned algorithms (\cite{cormode2023relative}, \cite{zhang2006space}) have the optimal dependence on one of the parameters $\epsilon$ and $n$, while are suboptimal by a polynomial factor on the other. 

A natural question is whether we can improve the sketch size in the MR algorithm, or improve the relative compactor space in CKLTV, so that the offline optimal space for this problem can also be achieved in streaming.
It turns out that the answer to this question is \emph{yes and no}.
Neither of the two subroutines can be improved in general, due to a lower bound that we prove in \Cref{sec:lb}.
On the other hand, the ``bad'' streams that lead to such lower bounds inherently cannot be combined -- the algorithm maintains logarithmically many sketches of relative compactors, there are bad streams that would force one of them to consume large space, but not all of them simultaneously.

Based on this observation, we propose a new streaming algorithm for quantile estimation with relative errors using nearly optimal space $\tilde{O}(\epsilon^{-1}\log \epsilon n)$\footnote{The $\tilde{O}$ hides the dominated $\log(1/\epsilon)$, $\log \log n$ and $\log \log (1/\delta)$ terms. For a precise space bound, see \Cref{thm:main}.}, nearly matching the trivial $\Omega(\epsilon^{-1} \log (\epsilon n))$ offline lower bound\footnote{The $\Omega(\epsilon^{-1} \log (\epsilon n))$ lower bound can be shown by inserting $\epsilon^{-1} \log(\epsilon n)$ many distinct elements $x_1 < x_2 < \dots < x_{\epsilon^{-1} \cdot \log(\epsilon n)}$ where for any $1 \leq i \leq \log(\epsilon n)$, the elements $x_{\epsilon^{-1} (i - 1) + 1}, \dots, x_{\epsilon^{-1}  i}$ are inserted $2^i$ times each. Any algorithm, even offline one that can see the entire stream, must keep all elements $x_1, x_2, \dots, x_{\epsilon^{-1} \cdot \log(\epsilon n)}$ in memory to satisfy the error guarantee.}.
We use the framework of MR, and maintain logarithmically many sketches such that each sketch is responsible for answering queries with rank in $[\epsilon^{-1} 2^i, \epsilon^{-1}2^{i+1})$.
Now, each sketch is implemented using a collection of new data structures, which we call \emph{elastic compactors}.
Elastic compactors are inspired by relative compactors, but they have one crucial additional feature: they are resizable. Since our sketch for each of the (logarithmically-many) scales will be built using these elastic compactors, we will actually be able to resize the entire sketch for each scale as needed. Depending on how ``difficult'' the input stream for each sketch is, our algorithm dynamically allocates space to the sketches, and resizes them to the current space on-the-fly as the stream is observed.
Whenever a piece of ``bad input stream'' targeting a specific sketch occurs (i.e. the one that we construct in \Cref{sec:lb}), the algorithm automatically allocates more space to that sketch temporarily, while still guaranteeing that the total space of all sketches is always bounded by $\tilde{O}(\epsilon^{-1}\log \epsilon n)$ with high probability.

\begin{them}\label{thm:main} %
 Let $0 < \delta \leq 0.5$ and $0 < \epsilon \leq 1$. There is a randomized, comparison-based, one-pass streaming algorithm that, when processing a stream $\pi$ consisting of $n$ elements, produces a sketch satisfying the following: for any query $x \in \mathcal{U}$, the sketch returns an estimate $\widehat \rank_{\pi}(x)$ for $\rank_{\pi}(x)$ such that with probability $1 - \delta$,
$$|\widehat \rank_{\pi}(x) - \rank_{\pi}(x) | \leq \epsilon \cdot \rank_{\pi}(x),$$    
where the probability is over the internal randomness of the streaming algorithm. Moreover, the total space used by the sketch is 
$$O(\epsilon^{-1} \log (\epsilon n) \cdot \log (1/\epsilon 
) \cdot  (\log \log n + \log (1/\epsilon)) \cdot (\log \log 1/\delta)^3).$$
    
\end{them}

There are several consequences of our construction, which to recall, consists of logarithmically-many resizable sketches for each scale $[\epsilon^{-1}2^i, \epsilon^{-1} 2^{i+1})$. First, since the sketch can be easily resized, our algorithm actually does not need to know the stream length $n$ in advance, and the same algorithm works as the stream length increases. Another important feature in practice is \textit{mergeability}, i.e. it is useful to be able to summarize two  substreams $\pi_1$ and $\pi_2$ separately into sketches $\mathcal{M}_1, \mathcal{M}_2$, and then create a \textit{merged} sketch $\mathcal{M}$ which applies to the combined stream $\pi = \pi_1 \bigsqcup \pi_2$ with similar error and space guarantees. Currently, it is not clear whether our relative-error quantiles sketch is fully-mergeable (See \Cref{sec:open-prob} for more discussion).

\paragraph{All-quantiles estimation.} As a straight-forward corollary of \Cref{thm:main}, we obtain a sketch that satisfies the \emph{all-quantiles guarantee}, meaning that for all queries $x \in \mathcal{U}$ \textit{simultaneously}, the sketch provides an accurate estimate with high probability. The proof proceeds by a standard union bound over an $\epsilon$-net, and is nearly identical to argument given in Appendix B of \cite{cormode2023relative}.

\begin{cor}
Let $0 < \delta \leq 0.5$ and $0 < \epsilon \leq 1$. There is a randomized, comparison-based, one pass streaming algorithm that, when processing a stream $\pi$ consisting of $n$ elements, produces a sketch satisfying the all-quantiles guarantee: 
for all queries $x \in \mathcal{U}$ \textit{simultaneously}, the sketch returns an estimate $\widehat \rank_{\pi}(x)$ such that with probability $1 - \delta$,
$$|\widehat \rank_{\pi}(x) - \rank_{\pi}(x) | \geq \epsilon \cdot \rank_{\pi}(x),$$
where the probability is over the internal randomness of the streaming algorithm. The total space used by the sketch is 

$$O\left(\epsilon^{-1} \log(\epsilon n) \cdot \log (1/\epsilon 
) \cdot (\log \log n + \log(1/\epsilon)) \cdot \left(\log \log \left(\frac{\log(\epsilon n)}{\delta \epsilon}\right)\right)^3\right)$$

\end{cor}

\subsection{Further Related Works}

\paragraph{Deterministic Sketches.} In the deterministic additive-error setting, Greenwald and Khanna constructed the GK sketch that stores $O(\epsilon^{-1} \log(\epsilon n))$ elements  \cite{greenwald2001space}; more recently, \cite{cormode2020tight} showed that the GK sketch is optimal and \cite{simpleGK} gave a simplification of the GK sketch which still achieves optimal space. In the relative-error case, Zhang et al.~\cite{zhang2007efficient} proposed a deterministic algorithm that uses $O(\epsilon^{-1}\log^3 (\epsilon n))$ space. This algorithm maintains logarithmically many sketches based on the \emph{chronological order} of the elements, and keeps merging sketches with similar sizes. Currently, the best known lower bound is $\Omega \left(\frac{\log^2 (\epsilon n)}{\epsilon} \right)$ \cite{cormode2020tight}.

\paragraph{Sketches with Known Universe.}
Additionally, some works focus on the case when the universe $\mathcal{U}$ is known in advance to the streaming algorithm \cite{cormode2006space, shrivastava2004medians}. In the additive error regime, the classical \emph{$q$-digest} algorithm gave an optimal deterministic quantile summary using $O(\epsilon^{-1} \log |\U|)$ words of memory. A recent work of \cite{gupta2024optimal} improved this bound to $O(\epsilon^{-1})$ words, achieving an optimal space if the stream length $n \leq \textrm{poly}(|\mathcal{U}|)$. For the relative error setting, ~\cite{cormode2006space} designed a deterministic bq-summary algorithm using $O\left(\frac{\log (\epsilon n) \log |\U|}{\epsilon}\right)$ words of memory, while the offline lower bound is only $\Omega \left(\frac{\log \epsilon n}{\epsilon}\right)$ words.

\paragraph{Resizable Sketches.}
During the past ten years, there has been a flurry of works on ``resizable sketches,'' wherein the goal is to design sketches that provide a fixed guarantee on the accuracy while allowing the space allocation to be dynamically adjusted throughout the runtime of the algorithm. This is especially important in practice, where the sketch size may start out being very small, but may need to grow sublinearly until it reaches some fixed maximum size. In particular, resizable sketches have been designed for filters \cite{pagh2013, infinifilter} \footnote{In a talk on ``Resizable Sketches'' at the Simons Institute Workshop on ``Sketching and Algorithm Design'' in October 2023, it was mentioned that there are also expandable sketches for the $k$-minimum values problem and the well-known Misra-Gries sketch for deterministic heavy-hitter detection \cite{simonsworkshop}.}. Recently in \cite{simonsworkshop}, it was posed as an open question to design resizable sketches for the quantile estimation problem. In fact, we actually answer this question since our final sketch for the relative-error quantile estimation problem is actually resizable. Thus, we show that our sketch can achieve near-optimal space $\tilde O(\epsilon^{-1} \log(\epsilon n))$ while also having this practically-favorable ``resizability'' feature.

\section{Technical Overview}\label{sec:tech_overview}
\subsection{Relative Compactors}
Before describing our algorithm, it is helpful to have a quick overview of the (relative) compactors, which form the main building block for the recent relative-error quantile estimation algorithm of~\cite{cormode2023relative}.
A compactor $C$ takes an input stream $\pi$, keeps a small number of elements in memory, and outputs a substream $\pi'$ of length at most $|\pi|/2$, while ``preserving'' the rank of any query $x$ approximately: \begin{equation}\label{eqn:estimate_rank}
    \rank_{\pi}(x)\approx 2\cdot \rank_{\pi'}(x)+\rank_C(x).
\end{equation}
That is, except for the few elements kept in the memory at the end of the stream, the compactor reduces the length of the stream by half. If we view the elements in the output as having \emph{weight two}, then the rank of any $x$ is preserved approximately.
For \emph{relative compactors}, we want the approximation in this building block to eventually lead to an overall ``small relative error,'' i.e., the difference of the two sides of~\eqref{eqn:estimate_rank} needs to be smaller for $x$ with a smaller rank.

To implement such a reduction in the stream length, a relative compactor maintains a \emph{sorted array}.
The array is divided into $b$ blocks of size $k$, hence, of total size $s = k\cdot b$.
The compactor keeps inserting elements from the stream $\pi$ to the array.
When the array is full, the algorithm performs a \emph{compaction}, which empties a few blocks. 
To perform a compaction, 
\begin{itemize}
    \item the algorithm first decides the number of \emph{blocks} $\ell$ to compact;
    \item then for the \emph{largest $\ell$ blocks} (i.e., the largest $\ell\cdot k$ elements), it outputs either \emph{all even-indexed elements} or \emph{all odd-indexed elements} in those blocks, each with probability $1/2$;
    \item finally, it removes all elements in those blocks. 
\end{itemize}

\begin{figure}
    \centering
    
\includegraphics{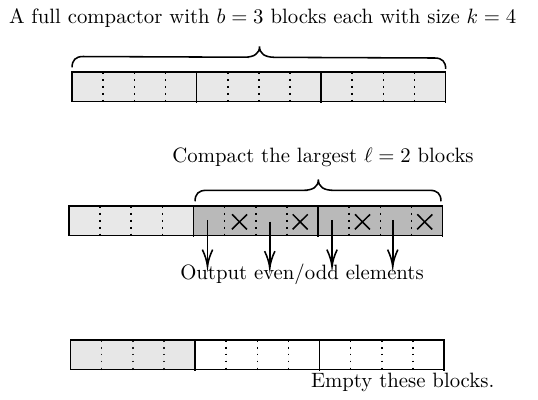}

    \caption{Compacting the largest $\ell$ blocks in a relative compactor. }
    \label{fig:relative-compactor}
\end{figure}

See \Cref{fig:relative-compactor} for an example. These compactions are crucial in order to make room for future insertions, yet they are the only source of errors in rank estimation.
Consider any fixed query $x$ and the error incurred to~\eqref{eqn:estimate_rank} during a compaction.
If all $\ell$ blocks only contain elements \emph{greater than} $x$, then clearly, this operation does not incur any error.
Otherwise, since the compaction outputs every other element, it makes an error of at most $1$ (in its absolute value).
In fact, this only happens when an \emph{odd} number of elements smaller than $x$ is involved in the compaction, and the error is $1$ or $-1$ with probability $1/2$. At the very least, when this happens, the compaction has to involve at least one element that is smaller than $x$. We adopt the terminology of~\cite{cormode2023relative} and call the elements smaller than $x$, the \emph{important elements}, and we call the compactions which involve important elements the \emph{important compactions.} Note all these definitions are with respect to a fixed query $x$.

The crucial piece of the algorithm is to set the number of compaction blocks $\ell$ each time, so that for any fixed query $x$, the total error is small, or equivalently, there are only few {important compactions}.
One way to do this is to \emph{sample $\ell$ from a geometric distribution} with $p=1/2$ independently each time.\footnote{This means we have $\Pr[\ell=t]=2^{-t}$ for integer $t\geq 1$.}
It is not hard to see that if we set the number of blocks $b\gg \log n$, then for each compaction, \emph{conditioned on} it involving at least one element smaller than $x$, there is a constant probability that the compaction involves (at least) one entire block of elements smaller than $x$, due to the geometric distribution.\footnote{When there is less than one entire block of important elements, the statement is technically not true. However, in this case, all these elements will be in the smallest block, and when $b\gg\log n$, there is a negligible probability that a compaction involves this block, i.e., we may assume the condition never happens. \label{footnote:determinstic}}
Intuitively, this implies that if a compaction incurs error, then it involves $\Omega(k)$ elements smaller than $x$ on average.
Since there are only $\rank_{\pi}(x)$ elements smaller than $x$, this intuition leads to an $O(\rank_{\pi}(x)/k)$ upper bound on the number of important compactions on average.\footnote{The work of~\cite{cormode2023relative} has another deterministic strategy for choosing $\ell$ with the same bound on the number of important compactions. The algorithm fixes a sequence of integers, called the \emph{compaction schedule}, and chooses deterministically the next integer from the sequence as $\ell$ each time. We in fact build on this deterministic version. See Section~\ref{sec:overview_RC} for an overview of the deterministic version.}
Since each such compaction incurs a $\pm 1$ error independently, one can conclude that the difference on the two sides of~\eqref{eqn:estimate_rank} is typically
\begin{equation}\label{eqn:rank_error}
    O(\sqrt{\rank_{\pi}(x)/k}).
\end{equation}
Furthermore, for $x$ with a very small rank, there is \emph{no} important compaction with high probability, in which case, the error is zero.

Their final streaming algorithm simply maintains a chain of $O(\log n)$ relative compactors.
The first compactor reads the input stream directly; the output of the $i$-th compactor is passed to the $(i+1)$-st compactor; the input to the last compactor is sufficiently short so that it never needs to perform compactions, hence, it has no output.

By~\eqref{eqn:estimate_rank}, the rank of a query $x$ is approximately reduced by half between the input and output of a relative compactor. The error analysis divides the chain of compactors into two parts based on the rank of $x$ in the input of the compactor:
\begin{enumerate}
    \item[(a)] for the first few compactors, when the query still has large rank, the \emph{relative error} is tiny due to the square root in~\eqref{eqn:rank_error};
    \item[(b)] after these compactors, the rank gets small enough so that there are no more important compactions -- the error is zero.
\end{enumerate}
It turns out that by setting $b=2\log n$ and $k=\Theta\left(\frac{1}{\epsilon\cdot \sqrt{\log n}}\right)$, 
the space is $O(\epsilon^{-1}\log^{1.5} n)$, and the accumulated \emph{relative error} from all compactors is at most $O(\epsilon)$.

To see this, consider a query of rank $r$, the relative error from the first compactor is $O(\sqrt{r/k}/r)=O(\sqrt{1/rk})$ by~\eqref{eqn:rank_error}, the relative error from $j$-th compactor is $O(\sqrt{2^j/rk})$ due to the reduction in $r$.
The errors form a geometric series, and the sum is dominated by the last term.
Note because $b \gg \log n$, when the rank becomes smaller than $\frac{b}{2}\cdot k$, there is no important compactions with respect to this query with high probability.
Hence, the last term in the geometric series has $r/2^j=\Theta(bk)$, the sum gives the total relative error $O(\sqrt{1/bk^2})=O(\epsilon)$.

\subsection{Our approach}
Our algorithm maintains $O(\log n)$ sketches, such that the $i$-th sketch is maintained for elements with rank roughly $\epsilon^{-1}\cdot 2^i$ with respect to the \emph{input stream so far}.
In particular, we will ensure that the sketches are in order: All elements in sketch $i$ are smaller than those in sketch $i+1$.
For each element from the input stream, we insert it into the sketch so that the ordering is preserved, i.e., we insert it into the largest sketch where the smallest element is smaller than the new element.
We will also ensure that the sketches always store approximately the right number of elements: When a total number of $\Omega(\epsilon^{-1}\cdot 2^i)$ elements have been inserted into sketch $i$, i.e., sketch $i$ ``overflows,'' we will move the largest half of its elements to sketch $i+1$, so that sketch $i$ only contains elements of current rank roughly $\epsilon^{-1}\cdot 2^i$ and all sketches are still in order.\footnote{Note that the \emph{sketches} will necessarily have to only store a subset of the inserted elements to save space, so some elements may have already been deleted when the time comes for them to be moved.
We will make this step more explicit later in the overview.}

Now let us focus on one such sketch, and see what properties we will need from it.
First, we can insert an element into the sketch, delete the largest elements, and can make approximate rank queries at the end.
Since all sketches are in order, sketch $i$ is only relevant for queries with rank $\Theta(\epsilon^{-1}\cdot 2^i)$.
Hence, we can afford an additive error of $O(2^i)$ from sketch $i$.
Note that while it looks similar to additive error quantile estimation, there is in fact a key difference -- we may insert up to $n\gg \epsilon^{-1}\cdot 2^i$ elements to a single sketch.
Even though we only need to focus on queries of rank $O(\epsilon^{-1}\cdot 2^i)$, there may actually be much more elements inserted in total.
This key difference makes the sketches with $O(1/\epsilon)$ space~\cite{karnin2016optimal} with additive error inapplicable to this setting. 

Consider the $i$-th sketch, observe that we can always first sample each insertion to it with probability $\Theta(\epsilon^{-1} \cdot 2^{-i})$.
This is because for a query of (final) rank $\Theta(\epsilon^{-1}\cdot 2^i)$, its rank after the sampling would be roughly $\Theta(1/\epsilon^2)\pm O(1/\epsilon)$, i.e., the relative error incurred due to sampling is $O(\epsilon)$.
By maintaining the \emph{smallest} $O(1/\epsilon^2)$ elements that survive the sampling, one could answer all such queries with small required error.
This is the main idea of~\cite{zhang2006space} with space $O(\epsilon^{-2}\cdot \log n)$, but it has suboptimal dependence on $\epsilon$.

To further improve the space, we could maintain a chain of $O(\log 1/\epsilon)$ relative compactors for all \emph{sampled elements} with $k=\Theta\left(\frac{1}{\epsilon\cdot \sqrt{\log n}}\right)$ and $b=O(\log n)$.
That is, we feed all sampled elements to the first compactor, and connect the output of compactor $j$ to the input of compactor $j+1$.
By the same error analysis as in the last subsection, for a query of rank roughly $r=\Theta(1/\epsilon^2)$ (after sampling), the total relative error incurred by the compactors is
\[
    \sqrt{1/rk}+\sqrt{2/rk}+\sqrt{2^2/rk}+\cdots+\sqrt{1/bk^2}=O(\epsilon).
\]
Note that when the sketch reaches its capacity, we can simply move the largest elements from all compactors to the next sketch.
This allows us to delete the largest elements.
However, since we set the parameters of each compactor same as before (this is necessary, since otherwise the relative error just from the last compactor would already be more than $\epsilon$), and there will be more than one relative compactor for each of the $\log n$ sketches, this does not give an improvement as is.

Although this data structure still uses $\tilde{\Omega}(\epsilon^{-1}\log^{1.5} n)$ space, this reformulation of the algorithm in \cite{cormode2023relative} facilitates the space improvement in our work. 
In particular, we observe the following.
\begin{itemize}
    \item Let $n_i$ be the number of insertions to the $i$-th sketch. Inside the relative compactors of the $i$-th sketch, we need really $b = \Theta(\log n_i)$ instead of $ \Theta(\log n)$ blocks. In order to force the $i$-th sketch to actually use \emph{all} $\Theta(\epsilon^{-1} \cdot \sqrt{\log n})$ space, we must insert $n_i = n^{\Theta(1)}$ elements into it.

    \item Suppose we indeed frequently insert many elements to sketch $i$, then this makes the task of later sketches easier. This is because for every $O(\epsilon^{-1}\cdot 2^i)$ insertions to sketch $i$, we will have to move the largest half to the next sketch $i+1$.\footnote{As sketch $i$ only has a bounded memory, we cannot remember exactly the largest half of elements inserted and move them to the next sketch $i + 1$. However, we can move our current sketch of the largest half from sketch $i$ to sketch $i + 1$. In fact, it turns out to be easy to integrate this sketch into sketch $i+1$, as it only has error $2^i$, smaller than that of sketch $i + 1$.} Note that these elements are smaller than any element in sketch $i + 1$. Also, when this happens $O(1)$ times, sketch $i+1$ must have reached its capacity, and will have to move the largest half of its elements to sketch $i+2$. 
    
    In particular, this means that all elements that were previous stored in sketch $i+1$ will be ``pushed away'' to later sketches after sketch $i$ overflows a constant number of times. As original elements in sketch $i + 1$ are pushed away, we can now \emph{effectively start the count $n_{i + 1}$ from zero again} and use a smaller space for sketch $i + 1$.

\end{itemize}

Combining these two observations, we argue that not \emph{all sketches need as much as $\Theta(\epsilon^{-1}\cdot\sqrt{\log n})$ space}.
If there is a large number and very frequent insertions to sketch $i$, then for a large number of sketches after $i$, they would only need $b=O(1)$ and $O(1/\epsilon)$ space. 
As a proof-of-concept, consider an input stream such that between any two adjacent insertions to sketch $i$, there are roughly $\beta_{i+1}$ insertions to sketch $i+1$.
Then by the above argument, before all current elements in sketch $i+1$ are pushed away to $i+2$ due to the overflows from sketch $i$, there will be only roughly $\beta_{i+1}\cdot \epsilon^{-1}\cdot 2^{i+1}$ new insertions to sketch $i+1$.
It turns out that we can maintain a smaller sketch $i+1$ by setting the number of blocks $b$ to $\Theta(\log \beta_{i+1})$, and 
a sketch of size $O(\epsilon^{-1}\cdot\sqrt{\log\beta_{i+1}})$ would be sufficient.
Since the stream length is $n$, implying that $\prod_i \beta_i\approx n$, the total sketch size is $\sum_i O(\epsilon^{-1}\cdot\sqrt{\log\beta_{i}})\leq\sum_i O(\epsilon^{-1}\cdot{\log\beta_{i}})\leq O(\epsilon^{-1}\cdot \log n)$.

To carry out the full details for general streams, we face the following concrete challenges.
\begin{itemize}
    \item The stream may insert new elements to a sketch at different rates at different times. It may even temporarily stop inserting into a sketch, and resume at a later time.
    This means that the sketch sizes need to actively and dynamically change over time based on how frequent the insertions to the sketches currently are. 
    \item We have no way to predict the insertion frequencies to different sketches, yet we need to allocate the \emph{fixed amount of total space} to different sketches so that each of them has small error as required.
\end{itemize}
In the following two subsections, we give an overview of how we tackle the challenges.

\subsection{Elastic Compactors} It turns out that the relative compactors can naturally be made resizable: when we need to downsize it to make space for other sketches, we simply do a compaction to empty the largest blocks, and then resize the array.
The main question then is under what conditions we can have the same error bound as before. 

To be more specific, we introduce elastic compactors.
In addition to what a relative compactor can do, the user may request an elastic compactor to resize to a given size, possibly after every element from the input stream.
Elastic compactors are constructed based on relative compactors.
We fix the block size $k$, and let the array resize by only adjusting the number of blocks $b$ (this turns out to be sufficient for our application).
Let us assume after every insertion, the user may specify a new number of blocks $b_j$ that the array has to resize to.
\begin{itemize}
    \item To resize, if the array has at most $kb_j$ elements, then the new size is large enough to contain all elements, we simply set the number of blocks to $b_j$.
    \item Otherwise, the algorithm samples the number of blocks to do compaction from a geometric distribution \emph{conditioned on} having at most $b_j$ nonempty blocks, i.e., it compacts the largest $b_{j-1}-b_j$ blocks with probability $1/2$, the largest $b_{j-1}-b_j+1$ blocks with probability $1/4$, etc.
\end{itemize}
Note that when $b_j$ remains the same for all $j$, this is the same as relative compactors.
The error analysis is also similar.
Consider any fixed query $x$, and consider a compaction. 
Conditioned on the compaction involving at least one element smaller than $x$, there is a constant probability that it involves at least one entire block of $k$ elements.
The only exception is when the compaction decides to empty the entire array.
We discussed in the review of relative compactors that when $b\gg \log n$, this does not happen except with negligible probability.
A simple calculation shows that as long as $\sum_j 2^{-b_j}\ll 1$, we have the same guarantee.
Hence, the same bound on the number of important compactions holds, and so does the error bound.

In other words, throughout our algorithm, suppose we require the compactor to resize to size $s_1, s_2, \dots, s_\ell$, then as long as
\begin{equation}\label{eqn:space_bound}
    \sum_{j=1}^\ell 2^{-s_j/k}\ll 1,
\end{equation}
the error bound of~\eqref{eqn:rank_error} holds.
Such elastic compactors can be used to construct the $O(\log n)$ sketches we discussed above.
It turns out that by replacing the $O(\log 1/\epsilon)$ relative compactors in each sketch by elastic compactors and setting $k=O(1/\epsilon)$, the sketches can also be resized (with a factor of $O(\log 1/\epsilon)$ more space than an elastic compactor) while having the same error guarantee as before.\footnote{A single elastic compactor may use more space than before by setting $k=O(1/\epsilon)$. In the worst case, it may use space as much as $O(\epsilon^{-1}\log n)$, but we will show that the total space is bounded.}

\subsection{Allocating space}
As we discussed earlier, the stream may insert elements to each sketch at different frequencies at different times.
Therefore, we may have to allocate a different amount of space to each sketch at different times.
To demonstrate the space allocation strategy, let us first consider the following special case.

\paragraph{A tree instance.}
Suppose the input stream inserts elements in the following way.
\begin{itemize}
    \item It inserts a batch of $n^{0.1}$ elements to the $1$st sketch almost consecutively, except that it may ``pause'' at any $100$ time points in the middle.
    \item During each ``pause,'' the stream inserts a batch of $n^{0.1}$ elements to the $2$nd sketch almost consecutively, and it may again ``pause'' $100$ times during each batch.
    \item During each ``pause'' in inserting elements to sketch $i$, the stream inserts $n^{0.1}$ elements to sketch $i+1$ and later sketches recursively. We recurse for $i = 1, 2, \dots, 0.1 \log n$. At the end of the recursion when $i = 0.1 \log n $, we just inserts $n^{0.1}$ elements to sketch $i+1$ but not later sketches.
\end{itemize}
Let us also assume that \emph{there are sufficiently many insertions to sketch $i$ between adjacent pauses of sketch $i$}.
In this case, the stream has a tree structure: the root corresponds to the insertion (sub)stream to sketch $1$, each child of the root corresponds to an insertion substream to sketch $2$ during each pause in the root, etc.
When there are sufficiently many insertions between adjacent pauses, different subtrees do not interfere with each other.

For such streams, we must drastically resize its sketches during each pause.
This is because during each pause in inserting to sketch $i$, the stream only inserts elements to larger sketches.
Hence, when the pause ends, the elements that sketch $i$ are supposed to maintain remain unchanged, and the insertions to sketch $i$ will continue from there.
Effectively, we are inserting $n^{0.1}$ elements to it, which forces the sketch to use $\Omega(\epsilon^{-1}\cdot\sqrt{\log n})$ space \emph{if we do not resize it} during the pause.
It leads to a total space of $\Omega(\epsilon^{-1}\cdot\log^{1.5} n)$ at the bottom of the recursion, when we have $0.1 \log n$ levels of pauses happening at the same time.

Our solution to this special case is to downsize the sketches during its pauses according to its length and the length of its current child.
That is, in the above tree view of the stream, let us consider the length of the stream corresponding to each node.
For a node $u$ with length $l_u$ and its child $v$ with length $l_v$, we will resize the sketch maintained for $u$ to $\tilde{O}(k\log (l_u/l_v))$ during the pause at $v$ (for now, let us assume that we know these lengths in advance).
This guarantees that the requested space sequence satisfies~\eqref{eqn:space_bound}, since $\sum_{v: \textrm{child of } u} l_v<l_u$.
Hence, the sketches provide the same error guarantee as before.
Importantly, the total space of all sketches is always bounded by $O(k\log n)=O(\epsilon^{-1} \log n)$, because the spaces allocated to the sketches form a telescoping sum and the length of the root is $n$.

\paragraph{General streams.}
Our final algorithm, which works for general streams, uses a similar idea.
In the above tree example, adjacent children of a node are essentially independent because we assumed that sufficiently many \emph{small} elements are inserted in between.
We first make an analogue of it for general streams.
Consider sketch $i$ and a time $t_1$.
Suppose sketch $i-1$ overflows a constant number of times between $t_1$ and some later time $t_2$, then all elements in sketch $i$ at $t_1$ will have been pushed away to later sketches.
This effectively \emph{resets} sketch $i$, making the sketch at time $t_2$ ``independent'' of the sketch at time $t_1$.
Hence, (for the analysis) we will divide the timeline of sketch $i$ into intervals of various lengths based on its ``resets,'' i.e., we start a new interval when sketch $i$ resets.
These intervals will correspond to (the substreams of) the nodes in the tree instance, but they do not form an exact tree structure as before.
However, we observe that each interval of sketch $i$ only intersects $O(1)$ intervals of sketch $i-1$ (this is because when sketch $i-1$ resets $O(1)$ times due to the overflow of sketch $i-2$, it must also have overflown $O(1)$ times itself, causing sketch $i$ to reset).
Hence, we can view each interval of sketch $i$ as having $O(1)$ ``parents,'' i.e., the intervals of sketch $i-1$ it intersects with.
We inductively assign a weight to each interval, generalizing the length, such that the weight of an interval $I$ is the sum of all weights of intervals that have $I$ as one parent, and the weight of an interval of the last sketch is one.
Now suppose we know the weights of all intervals in advance.
Then when sketch $i$ is currently in an interval $u$ with weight $w_u$ and sketch $i+1$ is in interval $v$ with weight $w_v$ (note that then, $u$ is a parent of $v$), we allocate space $O(k\log (w_u/w_v))$ to sketch $i$.
A similar analysis gives the same error bound.
By using the fact that the weight of any interval is at most $n^{O(1)}$, since each interval has $O(1)$ parents, the total space is again $O(k\log n)$ by a telescoping sum.

To calculate the weights, we need full information about the stream, including how the intervals intersect in the future.
We show that this is not necessary, the same algorithm works (up to factors of $\log\log n$ and $\log1/\epsilon$) even if we simply use the interval intersection information so far to calculate the weights. 
See Section~\ref{sec:dynamic-space} for more details. 
\section{Preliminaries}

All logarithms in this paper are base $2$. Without loss of generality, we will also assume that $1 / \epsilon, n, $ and $1/\delta$ are all powers of $2$.  

\paragraph{The Comparison-based Model.}  In this work, we focus on the \textit{comparison-based model}: this means that at any time $t$, the memory of the streaming algorithm is a tuple $(\mathcal{M}_t, I_t)$ where $\mathcal{M}_t$ is a subset of the stream elements and at any time, the algorithm can only compare two elements in $\mathcal{M}_t$. $I_t$ contains auxiliary information which is stored by the algorithm (e.g. information about previous comparisons, etc). All lower bounds in this paper are lower bounding $|\mathcal{M}_t|$. They continue to hold even when $|I_t|$ is unbounded. 

\paragraph{Probability.} We will use the following probability fact in our analysis. 

    \begin{fact} \label{fact:stadnard-deviation}
        For any two (not necessarily independent) random variables $X$ and $Y$, when we add them, their standard deviation at most adds, that is, 
        $$\E[(X+ Y)^2]^{1/2} \leq \E[X^2]^{1/2} + \E[Y^2]^{1/2}.$$
    \end{fact}
    \begin{proof}
        \begin{align*}\E[(X+Y)^2] &= \E[X^2] + \E[Y^2] + 2\E[XY] \\ 
        &\leq \E[X^2] + \E[Y^2] + 2\E[X^2]^{1/2}\E[Y^2]^{1/2} \tag{by Cauchy-Schwarz} \\
        &= \left(\E[X^2]^{1/2} + \E[Y^2]^{1/2}\right)^2.
        \end{align*}
    \end{proof}

We will also need properties of subgaussian random variables. A mean-zero variable $X$ is $\sigma^2$-subgaussian if $\E[e^{\lambda X}] \leq e^{\frac{\lambda^2 \sigma^2}{2}}$ for all $\lambda \in \mathbb{R}$. If a random variable $X$ is not mean-zero, we say it is $\sigma^2$-subgaussian if the mean zero variable $(X - \E[X])$ is $\sigma^2$-subgaussian .

\begin{fact} \label{fact:sum-subgaussian}
    If $X$ is $\sigma^2_1$-subgaussian and random variable $Y$ is an independent $\sigma_2^2$-subgaussian. Then $X + Y$ is $\sigma_1^2 + \sigma_2^2$-subgaussian. 
\end{fact}

\section{Description of the Algorithm} \label{sec:algo-description}
Before we present our full algorithm, we first provide a high-level overview of our strategy.
\paragraph{High-level Overview.}
Recall the definition of an $\epsilon$-relative-error quantile sketch: for any query $x \in \U$, after processing a stream $\pi$ the sketch must return an estimate $\widehat{\rank}_{\pi}(x)$ such that $$|\widehat{\rank}_{\pi}(x) - \rank_{\pi}(x)| \leq \epsilon \cdot \rank_{\pi}(x)$$ holds with probability at least $1-\delta$. Importantly, observe that (up to a factor of $2$) the relative error guarantee is equivalent to the following: for all queries $x$ such that $\rank(x) \in [\epsilon^{-1} \cdot 2^{i - 1}, \epsilon^{-1} \cdot 2^i]$, the answer of the sketch can tolerate an \textit{absolute} error of at most $2^i$.  

\begin{figure}[H]
    \centering

\begin{center}

\tikzset{every picture/.style={line width=0.75pt}} %

\begin{tikzpicture}[x=0.75pt,y=0.75pt,yscale=-1,xscale=1]

\draw    (87.86,190.29) -- (460.86,189.29) ;
\draw [shift={(463.86,189.29)}, rotate = 179.85] [fill={rgb, 255:red, 0; green, 0; blue, 0 }  ][line width=0.08]  [draw opacity=0] (8.93,-4.29) -- (0,0) -- (8.93,4.29) -- cycle    ;
\draw  [dash pattern={on 4.5pt off 4.5pt}]  (129.86,109.29) -- (129.86,209.29) ;
\draw  [dash pattern={on 4.5pt off 4.5pt}]  (188.86,109.29) -- (188.86,210.29) ;
\draw  [dash pattern={on 4.5pt off 4.5pt}]  (280.86,109.29) -- (280.86,210.29) ;
\draw  [dash pattern={on 4.5pt off 4.5pt}]  (430.86,112.29) -- (430.86,213.29) ;

\draw (88,139) node [anchor=north west][inner sep=0.75pt]   [align=left] {$\displaystyle \cdots $};
\draw (141,129) node [anchor=north west][inner sep=0.75pt]   [align=left] {$\displaystyle H_{i-1}$};
\draw (228,130) node [anchor=north west][inner sep=0.75pt]   [align=left] {$\displaystyle H_{i} \ $};
\draw (444,137) node [anchor=north west][inner sep=0.75pt]   [align=left] {$\displaystyle \cdots $};
\draw (346,130) node [anchor=north west][inner sep=0.75pt]   [align=left] {$\displaystyle H_{i+1} \ $};
\draw (453,197) node [anchor=north west][inner sep=0.75pt]   [align=left] {rank};
\draw (95,223) node [anchor=north west][inner sep=0.75pt]  [font=\small] [align=left] {$\displaystyle \epsilon ^{-1} \cdot 2^{\ i-2\ }$};
\draw (169,222) node [anchor=north west][inner sep=0.75pt]  [font=\small] [align=left] {$\displaystyle \epsilon ^{-1} \cdot 2^{\ i-1\ }$};
\draw (260,221) node [anchor=north west][inner sep=0.75pt]  [font=\small] [align=left] {$\displaystyle \epsilon ^{-1} \cdot 2^{\ i\ }$};
\draw (407,223) node [anchor=north west][inner sep=0.75pt]  [font=\small] [align=left] {$\displaystyle \epsilon ^{-1} \cdot 2^{\ i+1\ }$};
\draw (51,85) node [anchor=north west][inner sep=0.75pt]   [align=left] {error $ $};
\draw (136,84) node [anchor=north west][inner sep=0.75pt]  [font=\small] [align=left] {$\displaystyle \leq 2^{\ i-1\ }$};
\draw (221,83) node [anchor=north west][inner sep=0.75pt]  [font=\small] [align=left] {$\displaystyle \leq 2^{\ i\ }$};
\draw (341,83) node [anchor=north west][inner sep=0.75pt]  [font=\small] [align=left] {$\displaystyle \leq 2^{\ i+1\ }$};

\end{tikzpicture}
\end{center}

    \caption{Our basic strategy.}
    \label{fig:basic}
\end{figure}
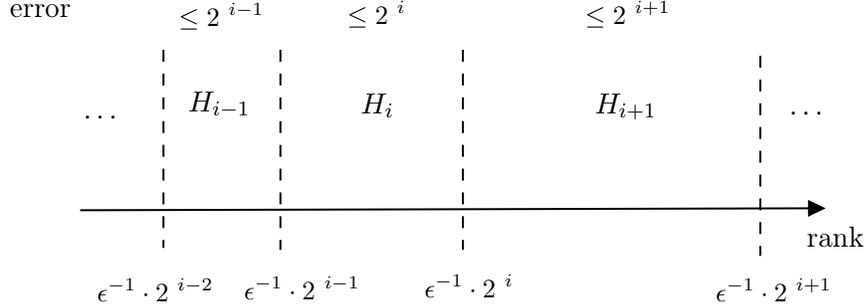

The general strategy of our algorithm naturally follows from the above observation: we decompose the
relative-error quantile estimation problem into  $\lceil \log_2(\epsilon n) \rceil$-many absolute-error quantile problems. In particular, we partition the rank-space into $\lceil \log_2(\epsilon n) \rceil$ \textit{ranges}, where range $i$ contains all inserted elements with rank (roughly) between $[\epsilon^{-1} \cdot 2^{i - 1}, \epsilon^{-1} \cdot 2^i]$ with respect to all elements inserted so far. Then, we maintain a separate sub-sketch $H_i$ for each range $i$ and maintain the guarantee that each sub-sketch $H_i$ will have absolute error of at most $2^i$ (See \Cref{fig:basic}). For each new element $x_t$ that appears in the stream, we find the range $i$ such that $x_t$ is smaller than all elements in $H_{i+1}$ and larger than or equal to all elements in $H_{i-1}$, and we insert $x_t$ into that sub-sketch $H_i$. At any point in the stream, if $H_i$ contains many more than $O(\epsilon^{-1} \cdot 2^i)$ elements, we remove the largest elements from $H_i$ and insert them into the next sub-sketch $H_{i + 1}$. 

Observe that in this construction, each sub-sketch $H_i$ is different from an (insertion-only) additive-error quantile sketch, as we are periodically deleting the largest elements in $H_i$ and moving them into $H_{i+1}$. We call such a sub-sketch $H_i$ a \emph{top-quantiles sketch}.\footnote{In fact, we formalize this as a separate problem of independent interest (See \Cref{sec:top-quantile}).}\footnote{Note that the name ``Top Quantiles'' refers to the task of estimating the \textit{bottom} ranks. This is generally in contrast with the rest of our work, in which we consider $\textrm{Quantile}(x) = \rank(x)/n$.}   Moreover, to achieve a total space of $\widetilde{O}(\epsilon^{-1} \log(\epsilon n))$, on average we can only afford to allocate $\widetilde{O}(1/\epsilon)$ space per sub-sketch $H_i$. However, as we show in \Cref{lem:lb-top-quantile}, there is a $\Omega\left(\epsilon^{-1} \cdot \sqrt{\log (\epsilon n)}\right)$ space lower bound for any such top-quantiles sketch $H_i$. This becomes the main technical barrier of our work. 

To get around this barrier, we make the following key observation: this lower bound says that for each sub-sketch, there \emph{exists} an input stream that forces the sketch to use maximum space \emph{at some time point} during the execution of the algorithm. However, the lower bound in \Cref{lem:lb-top-quantile} does not imply the existence of a single hard input stream  that simultaneously forces \emph{every sub-sketch} $H_i$ for rank interval $[\epsilon^{-1} \cdot 2^{i-1}, \epsilon^{-1} \cdot 2^i]$ to achieve its maximum space usage. As a result, there is still hope to design a near-optimal $\tilde O(\epsilon^{-1} \log(\epsilon n))$ algorithm.

Our data structure $H_i$ is based on the relative compactor from the previous relative error quantile sketch of ~\cite{cormode2023relative}. There are two main deviations of our algorithm from previous work: (1) as discussed earlier, we maintain a separate data structure (or ``sub-sketch'') for stream elements that are roughly in rank range $[\epsilon^{-1} \cdot 2^{i-1}, \epsilon^{-1} \cdot 2^i]$ for each $i = 1,..., \lceil \log(\epsilon n)\rceil$, and (2) our sketch is made up of \textit{elastic} compactors which are \textit{dynamically resizable}. In particular, our sub-sketches can change their space usage dynamically  based on ``demand'': for example, if the number of insertions into range $i$ is very high, we may choose to strategically allocate more space to $H_i$ while allocating less space to data structures for other ranges. Using this idea, the main task becomes to define a scheme to dynamically allocate space to the sub-sketches and constantly resize them. Surprisingly, this approach results in a reduced total space of $\widetilde{O}\left(\epsilon^{-1} \cdot  \log (\epsilon n)\right)$ without contradicting the aforementioned lower bound.  
The rest of this section is organized as follows. 
\begin{itemize}
    \item In \Cref{sec:compactor}, we will introduce the most basic component of our resizable sketch, which constitutes our \emph{elastic compactor}. However, we defer the detailed implementation of this data structure to \Cref{sec:full-details-elastic}. 
    \item Then, in \Cref{sec:top-quantile}, we will explain how to construct each sub-sketch $H_i$ from these elastic compactors.
    \item Finally, in \Cref{sec:relative-error}, we describe the strategy for dynamically allocating space to different sub-sketches $H_i$, which finishes the high-level overview of our algorithm. 
\end{itemize}

Throughout this section, we only provide bounds on the expected squared error, and some proofs are deferred to \Cref{sec:expectation}. For the high probability bounds, we refer the reader to \Cref{sec:high-prob}. 

\subsection{Elastic Compactors} \label{sec:compactor}
The most basic building block of our algorithm is what we call an \emph{elastic compactor}. Specifically, this is a compactor whose size can be dynamically adjusted. For our case, we will eventually choose (and dynamically adjust) the size of the elastic compactor in an online manner, based on the input stream. See \Cref{table:toy-invariants} for the definition. 

\begin{table}
    \centering
    \begin{tabular}{|p{12cm}|}
    \hline \\
    \multicolumn{1}{|c|}{\textbf{Elastic Compactor}} \\~\\
    \textbf{Parameters.} Let $C$ be an $(s,k)$-elastic compactor: $C$ may store at most $s$ elements in memory, and let $k$ be a parameter related to the total error allowed for $C$. \\~\\
    \textbf{Input:}  A stream $\pi$ of elements in $\U$. \\
    \textbf{Output:} As a result of each \textit{compaction}, $C$ can output elements to a \textit{compacted} stream $\pi'$. In the end, $\pi'$ contains at most $|\pi| / 2$ elements.\\~\\
    \textbf{Operations:} 
    \begin{itemize}
        \item[-] $\Call{Resize}{C, s'}$: Expand or compress the space of the compactor $C$ to $s'$ and set $s \gets s'$.
        
        Whenever a resize operation compresses the space to a smaller size (i.e. $s' < s$), we will perform compaction, and some elements stored in $C$ will be outputted to the output stream $\pi'$.

        \item[-] $\Call{Insert}{C, x_1, x_2, \dots, x_m}$: Add $x_1, x_2, \dots, x_m$ to the input stream $\pi$ of compactor $C$, where $m \leq s$. We will implement it using $\Call{Resize}{C, s'}$:
        \begin{itemize}
            \item First, we call $\Call{Resize}{C, 2s}$, insert $x_1, x_2, \dots, x_m$ into the new open space of size $s$ in memory.
            \item Then, we call $\Call{Resize}{C, s}$ to compress the space back to the original size. 
        \end{itemize}
        Note that some elements may be outputted to the output stream $\pi'$ during this last compression.

        \item[-] $\Call{Rank}{}_C(x)$: Return the rank of $x$ among the elements $\pi$ that are still in the memory of compactor $C$. 
        
        Let $\widehat{{\rank_{\pi}}}(x)$ denote the estimated rank of $x$ with respect to stream $\pi$. Then, we define $\widehat{\rank_{\pi}}(x) = \Call{Rank}{}_C(x) + 2 \cdot \rank_{\pi'}(x)$.

    \end{itemize}\\
    \hline
    \end{tabular}
    \caption{Elastic Compactor.}
    \label{table:toy-invariants}
\end{table}

\paragraph{Error Guarantee.} For any query $x \in \U$, we define the error associated to $x$ by
\begin{equation}\label{equ:Error-Compactor}
\Delta(x) \coloneqq \Big(\Call{Rank}{}_C(x) + 2 \cdot \rank_{\pi'}(x) \Big) - \rank_\pi(x)  
\end{equation}

We enforce the following guarantee for elastic compactors.

\begin{restatable}{lem}{compactor} \label{lem:compactor}
Let $\pi$ be the input stream and $s_1, s_2, \dots, s_\ell$ be the sequence of space parameters after each resize/insert operation. There is a randomized elastic compactor $C$ such that, if we condition on \begin{equation} \label{equ:compactor-cond}\sum_{t=1}^\ell 2^{- s_t / k} \leq 0.5,\end{equation} for any query $x \in \U$, $C$ achieves expected squared error
$$\E\left[\Delta(x)^2\right] \leq \frac{\rank_{\pi}(x)}{k},$$
where the expectation is taken over the randomness of the compactor.
\end{restatable}

Intuitively, \Cref{equ:compactor-cond} ensures that the space used by the compactor cannot be too small. Note that we modify the implementation of the relative compactor of~\cite{cormode2023relative} in our definition of the elastic compactor. We defer the proof of \Cref{lem:compactor} to \Cref{sec:full-details-elastic}.

\paragraph{Additional operations.} Finally, we define two more operations for our elastic compactor, which we call "reset" and "remove" operations. These will be required in \Cref{sec:full-details-elastic}.
\begin{itemize}
    \item $\Call{Reset}{C}$: This operation resets the sum in \Cref{equ:compactor-cond} that we have accumulated so far to zero, while maintaining all the elements that are stored in $C$. Essentially, by resetting the sum, we allow the compactor to handle more operations, but at the cost of introducing more error into the rank estimate.
    \item $\Call{Remove}{}_{\max}(C)$: This operation removes the largest element in the compactor $C$ and outputs it to another stream $\pi_{\mathrm{remove}}$. When this operation is present, we define the error attributed to $x \in \mathcal{U}$ to be
    \begin{equation} \label{equ:compactor-with-removal-error}
        \Delta(x) \coloneqq \Big(\Call{Rank}{}_C(x) + 2 \cdot \rank_{\pi'}(x) + \rank_{\pi_{\mathrm{remove}}}(x) \Big) - \rank_\pi(x).
    \end{equation}
\end{itemize}

Suppose we fix a query $x \in \U$. Then, we call a reset \emph{important} if at the time of reset, there is at least one element in the compactor $C$ that is smaller than $x$. Formally, we have the following guarantee. 

\begin{restatable}{lem}{compactorreset} \label{lem:compactor-reset}
Suppose between any two adjacent resets, the sequence of space constraints in the resize operations satisfies \Cref{equ:compactor-cond}. Let $S$ be an upper bound on the number of elements in the compactor at any time. Then for any query $x \in \U$, if there are $t$ important resets, the randomized compactor achieves expected squared error $$\E\left[\Delta(x)^2\right] \leq \frac{\rank_\pi(x) + t \cdot S}{k}.$$
Here $\Delta(x)$ is defined as in \Cref{equ:compactor-with-removal-error}. %
\end{restatable}

The proof of \Cref{lem:compactor-reset} is also deferred to \Cref{sec:full-details-elastic}.\footnote{In fact, \Cref{lem:compactor} is a direct application of \Cref{lem:compactor-reset} when $t=0$.}
 
\subsection{The Top Quantiles Sketch}\label{sec:top-quantile}

As mentioned before, our basic strategy is to maintain sub-sketches $H_i$'s that each handles elements of rank roughly around $R_i \coloneqq \epsilon^{-1} \cdot 2^i$ and has absolute error less than $2^i$. The main challenge of this task can be formalized as the following natural problem. 

\begin{table}[h]
    \centering
    \begin{tabular}{|p{12cm}|}
    \hline ~\\
    \multicolumn{1}{|c|}{\textbf{The Top-$R$ Quantiles Problem.}} \\~\\
    \textbf{Input.} A stream $\pi$ of length $|\pi| \gg R$.\\
    \textbf{Output.} For any query $x \in \U$ (given at the end of the stream) with the promise that $\rank_\pi(x) \leq R$, the sketch has to output an estimate  $\widehat{\rank}_\pi(x)$ such that,
    $$\left|\rank_\pi(x) - \widehat{\rank}_\pi(x)\right| \leq \epsilon \cdot R.$$
    \\
    \hline
    \end{tabular}
    \caption{The Top Quantiles Problem.}
    \label{table:top-quantiles}
\end{table}

Note when input stream length $|\pi| \leq R$, this problem can be solved by a standard additive-error quantile sketch, because the additive-error sketch will incur error at most $\epsilon \cdot |\pi| \leq \epsilon \cdot R$. However, this becomes a different problem when $|\pi| \gg R$. In fact, this is evidenced by the fact that the additive-error quantile sketch takes only $O(\epsilon^{-1} \log \log \delta)$ space (KLL sketch~\cite{karnin2016optimal}), while for this problem, as mentioned before, we have the following lower bound saying $\Omega(\epsilon^{-1} \sqrt{\log n})$ space is necessary (for a specific $\epsilon$) to achieve failure probability $\delta < 0.1$.

\begin{restatable}{lem}{lbtopquantile} \label{lem:lb-top-quantile}
Let $n = |\pi|$ be the stream length. When $\epsilon = 1 / \sqrt{\log n}$, any comparison-based algorithm that solves the top quantiles problem for $R = \log n$ with $\delta < 0.1$ failure probability requires at least $\Omega(\log n)$ space.   
\end{restatable}
The proof of \Cref{lem:lb-top-quantile} is deferred to \Cref{sec:lb}. Note that there are some natural motivations for studying this problem; e.g. when monitoring system latency, we may only care about an abnormal tail of the data and wish to construct sketches that incur the smaller error $\epsilon \cdot R$ instead of $\epsilon \cdot |\pi|$.

\paragraph*{Outline of our approach.} First, we design a resizable sketch $H$ which solves the top-$R$ quantiles problem. Later on, we will augment this sketch $H$ with additional features in order to obtain our final sub-sketches $H_i$'s, which will be used to solve the overall relative error quantile estimation problem. 

\paragraph{Compactor Hierarchy.} Similar to the KLL sketch~\cite{karnin2016optimal}, our sketch $H$ is a hierarchy that consists of a sampler, a chain of elastic compactors $C_1,..., C_{\log(1/\epsilon)-1}$, and finally a buffer $B$ which stores $O(1/\epsilon)$ elements, as shown in \Cref{fig:hierarchy} below.

\begin{figure}[H]
    \centering
\tikzset{every picture/.style={line width=0.75pt}} %

\begin{tikzpicture}[x=0.75pt,y=0.75pt,yscale=-1,xscale=1]

\draw   (70.58,67.68) -- (110.31,84.39) -- (109.69,122.26) -- (69.43,137.67) -- cycle ;
\draw    (120,104) -- (156.86,104) ;
\draw [shift={(158.86,104)}, rotate = 180] [color={rgb, 255:red, 0; green, 0; blue, 0 }  ][line width=0.75]    (10.93,-3.29) .. controls (6.95,-1.4) and (3.31,-0.3) .. (0,0) .. controls (3.31,0.3) and (6.95,1.4) .. (10.93,3.29)   ;
\draw   (171,67.29) -- (182.86,67.29) -- (182.86,147) -- (171,147) -- cycle ;
\draw    (171,83) -- (182.86,83) ;
\draw    (171,99) -- (182.86,99) ;
\draw    (172,115) -- (183.86,115) ;
\draw    (171,131) -- (182.86,131) ;
\draw   (226,66.29) -- (237.86,66.29) -- (237.86,146) -- (226,146) -- cycle ;
\draw    (226,82) -- (237.86,82) ;
\draw    (226,98) -- (237.86,98) ;
\draw    (227,114) -- (238.86,114) ;
\draw    (226,130) -- (237.86,130) ;
\draw    (192,104) -- (214.86,104) ;
\draw [shift={(216.86,104)}, rotate = 180] [color={rgb, 255:red, 0; green, 0; blue, 0 }  ][line width=0.75]    (10.93,-3.29) .. controls (6.95,-1.4) and (3.31,-0.3) .. (0,0) .. controls (3.31,0.3) and (6.95,1.4) .. (10.93,3.29)   ;
\draw   (283,66.29) -- (294.86,66.29) -- (294.86,146) -- (283,146) -- cycle ;
\draw    (283,82) -- (294.86,82) ;
\draw    (283,98) -- (294.86,98) ;
\draw    (284,114) -- (295.86,114) ;
\draw    (283,130) -- (294.86,130) ;
\draw    (249,103) -- (271.86,103) ;
\draw [shift={(273.86,103)}, rotate = 180] [color={rgb, 255:red, 0; green, 0; blue, 0 }  ][line width=0.75]    (10.93,-3.29) .. controls (6.95,-1.4) and (3.31,-0.3) .. (0,0) .. controls (3.31,0.3) and (6.95,1.4) .. (10.93,3.29)   ;
\draw   (341,65.29) -- (352.86,65.29) -- (352.86,145) -- (341,145) -- cycle ;
\draw    (307,102) -- (329.86,102) ;
\draw [shift={(331.86,102)}, rotate = 180] [color={rgb, 255:red, 0; green, 0; blue, 0 }  ][line width=0.75]    (10.93,-3.29) .. controls (6.95,-1.4) and (3.31,-0.3) .. (0,0) .. controls (3.31,0.3) and (6.95,1.4) .. (10.93,3.29)   ;

\draw (59,145) node [anchor=north west][inner sep=0.75pt]   [align=left] {Sampler};

\end{tikzpicture}

    \caption{Structure of $H$}
    \label{fig:hierarchy}
\end{figure}
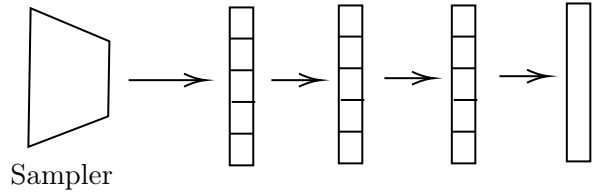

Note that our resizable sketch $H$ has two tuneable parameters: an error parameter $\epsilon > 0$, and a space parameter $s_H$ (eventually, the total space used by sketch $H$ will be $s_H \cdot \lceil\log_2 (1 / \epsilon)\rceil$). Next, we define the same set of operations for $H$ (which consists of a sampler followed by a hierarchy of elastic compactors), just as we defined earlier for each elastic compactor $C$:
\begin{itemize}
    \item[-] $\Call{Insert}{H, x}$: Add $x$ into the input stream $\pi$ of $H$. 
    \item[-] $\Call{Resize}{H, s'}$: Expand or compress the space of $H$ so that we set the space parameter to be $s_H \gets s'$.
    \item[-] $\Call{Rank}{}_H(x)$: Return an estimate for the rank of $x$. To compute the estimate $\Call{Rank}{}_H(x)$, we output the total weight of the elements in $H$ that are smaller than $x$. That is, 
    \begin{equation}\label{equ:rank-estimator-H}
        \Call{Rank}{}_H(x) \coloneqq \sum_{j=0}^{\log(1 / \epsilon) - 1} 2^j \cdot \epsilon^2 R \cdot \Call{Rank}{}_{C_j}(x) + \epsilon R \cdot \Call{Rank}{}_B(x)
    \end{equation} 
where $\Call{Rank}{}_{C_j}(x)$ and $\Call{Rank}{}_{B}(x)$ denote the rank of $x$ among all elements contained in compactor $C_j$ and buffer $B$, respectively.

Recall that for all $x$ with $\rank_{\pi}(x) \leq R$, we want 
    $$\left|\Call{Rank}{}_H(x) - \rank_\pi(x)\right| \leq O(\epsilon) \cdot R.$$

\end{itemize}

We now describe the operation of each component as the stream travels through the sketch $H$ from left to right. Additionally, for every element $x$ stored in the memory of $H$, we define the \textit{weight} of $x$ to represent the number of stream elements that are ``represented'' by $x$.

\begin{enumerate}
    \item \textbf{Sampler}: first, each element $x$ in input stream $\pi$ is passed through the sampler: with probability $\frac{1}{\epsilon^2 R}$, the sampler inserts $x$ into the first elastic compactor $C_0$, and $x$ is discarded otherwise. Note that if there are $|\pi| = n$ elements in the input stream, roughly $\ell = \frac{n}{\epsilon^2 R}$ elements will be inserted into $C_0$.  

    Additionally, when element $x$ first appears in the stream $\pi$, $x$ has weight $1$. After passing through the sampler, each sampled element carries weight $\epsilon^2 R$. 
    \item \textbf{Elastic compactors}: for each $0 \leq j < \log_2(1 / \epsilon) $, the elastic compactor $C_j$ has block-size parameter $k = 1 / \epsilon$ and space parameter $s = s_H$. After each operation to $C_j$, it might output $m$ ($m \leq s$) element $x_1, x_2, \dots, x_m$. We call $\Call{Insert}{C_{j + 1}, x_1, x_2, \dots, x_m}$ to insert them into the next compactor. Eventually, the output of the last compactor is inserted into the buffer. 

    Note that for each $0 \leq j < \log_2(1 / \epsilon)$, elements in $C_j$ all have weight $\epsilon^2 R \cdot 2^j$. 

    \item \textbf{Buffer}: the buffer $B$ stores the smallest $ 1/\epsilon $ elements inserted to it, and throws away the largest element whenever it is full. 
    
    Each element in the buffer has weight $\epsilon \cdot R$.  
\end{enumerate}

Intuitively, as the smallest $1/\epsilon$ elements in the buffer already have total weight $R$, and we only care about top-$R$ quantiles, this explains why we can afford to throw away all other elements from the buffer. 

\begin{remark} \label{remark:corner-case-structure}
    For the corner case when $R < \frac{1}{\epsilon^2}$, we omit the sampler and directly insert stream elements into the compactor $C_j$ such that $2^j \cdot \epsilon^2 R = 1$. 
\end{remark}

\paragraph{Implementation of Operations. }  
For $\Call{Insert}{H, x}$, we feed $x$ to the sampler, which inserts $x$ into $C_0$ with probability $\frac{1}{\epsilon^2 R}$. (See \Cref{alg:insertion}.) To simplify the notation, we will define $C_{\log(1/\epsilon)}$ to be the buffer $B$ at the last level of the hierarchy. 

\begin{algorithm}[h]
\caption{Insert an element.}
\label{alg:insertion}
\begin{algorithmic}[1]
\Procedure{Insert}{$H, x$}
\Withprob{$\frac{1}{\epsilon^2 R}$} 
\State \Call{Insert}{$C_0, x$}. 
\For{$i = 0, 1, \dots \log(1/\epsilon) - 1$}
\State Let $x_1, x_2, \dots, x_m$ be the elements that $C_i$ outputs after insertion. 
\State \Call{Insert}{$C_{i + 1}, x_1, x_2, \dots, x_m$}. 
\EndFor
\EndWithprob
\EndProcedure
\end{algorithmic}
\end{algorithm}

Likewise, we insert elements into the buffer $B$ (i.e. the last compactor $C_{\log(1\epsilon)})$ as follows. Here, we note that if the new insertion causes there to be more than $1/\epsilon$ elements stored in the buffer, we remove the largest elements until there are only $1/\epsilon$ remaining.

\begin{algorithm}[h]
\caption{Insert elements to buffer $B$.} \label{alg:buffer}
\begin{algorithmic}[1]
\Procedure{Insert}{$B, x_1, x_2, \dots, x_m$}
\State Add $x_1, x_2, \dots, x_m$ to the buffer $B$. 
\While{the buffer $B$ has more than $1 / \epsilon$ elements}
\State Remove the largest elements in the buffer. \label{line:throw}
\EndWhile
\EndProcedure
\end{algorithmic}
\end{algorithm}

Now, to implement $\Call{Resize}{H, s'}$ on the full sub-sketch $H$, we simply call $\Call{Resize}{C_j, s'}$ for each level $j$ of the compactor hierarchy (See \Cref{alg:resize}).

\begin{algorithm}[h]
\caption{Resize the hierarchy.} \label{alg:resize}
\begin{algorithmic}[1]
\Procedure{Resize}{$H, s'$}
\State \Call{Resize}{$C_0, s'$}. 
\For{$j = 0, 1, \dots \log(1/\epsilon) - 1$}
\State Let $x_1, x_2, \dots, x_m$ be the elements that $C_i$ outputs after insertion. 
\State \Call{Insert}{$C_{j + 1}, x_1, x_2, \dots, x_m$}. 
\State \Call{Resize}{$C_{j + 1}, s'$} unless $j = \log(1/\epsilon) - 1$. (The buffer is always of fixed size $1/\epsilon$.) 
\EndFor
\EndProcedure
\end{algorithmic}
\end{algorithm}

\paragraph{Error Guarantee.} In conclusion, we have the following lemma.

\begin{restatable}{lem}{hierarchy}  \label{lem:H}
Let $\pi$ be the input stream and $s_1, s_2, \dots, s_\ell$ be the sequence of space parameters after each resize / insertion from the sampler into $C_0$. Our randomized top-$R$ quantiles sketch $H$, conditioning on 
\begin{equation} \label{equ:condition-H}
\sum_{t = 1}^\ell 2^{- \epsilon \cdot s_t} \leq 0.25,    
\end{equation}
for any query $x$ with $\rank_\pi(x) \leq R$, achieves standard deviation
$$\E\left[|\Call{rank}{}_H(x) - \rank_\pi(x)|^2\right]^{1/2} \leq O(\epsilon) \cdot R.$$
\end{restatable}
We defer the proof to \Cref{sec:expectation}. The main idea is to apply \Cref{lem:compactor} to analyze the error introduced by each $C_j$.

\begin{remark} \label{remark:improved}
Let $n = |\pi|$. Then there are in expectation $\ell = \frac{n}{\epsilon^2 R}$ many insertions into $C_0$. Setting $s_1 = s_2 = \dots = s_\ell = \epsilon^{-1} \log \ell = O\left(\epsilon^{-1} \log n\right)$ gives an $O\left(\epsilon^{-1} \log n\right)$ space algorithm for the top-$R$ quantiles problem. We will later improve it in \Cref{sec:further}. By setting $s_1 = s_2 = \cdots = s_\ell = O(\epsilon^{-1} \sqrt{\log n})$, we obtain an optimal algorithm for Top-$R$ Quantiles problem, matching the lower bound in \Cref{lem:lb-top-quantile}. For this section, the simpler bound in \Cref{lem:H} suffices.
\end{remark}

\paragraph{Final Description of Sub-sketch $H_i$.} Finally, we augment the top-$R$ quantiles sketch $H$ described above to obtain the final version of our sketches $H_i$, which we will use to build our final algorithm for the relative error quantile estimation problem. In particular, we add three additional features, which are described in detail below. Note that we denote $R_i \coloneqq \epsilon^{-1} \cdot 2^i$ and let $\pi_{H_i}$ be the input stream of $H_i$. 

\footnotetext{For the buffer $B$, $\Call{Remove}{}_{\max}(B)$ is defined naturally as removing the largest element in the buffer and output it. }
\begin{enumerate}
    \item \textbf{Limiting the maximum total weight in $H$.} In our algorithm, we will always ensure that the total weight of elements in each $H_i$ is less than $3 R_i$. Also, we will set the buffer size to $\frac{3}{\epsilon}$. Since each element in the buffer has weight $\epsilon R_i$, the buffer will never be full (as long as the total weight is bounded by $3 R_i$). Thus, each subsketch $H_i$ no longer needs to throw away elements when the buffer is full (Remove Line~\ref{line:throw} from \Cref{alg:buffer}).

    Instead, $H_i$ operates as follows: 
    whenever the total weight exceeds $3 R_i$, the sub-sketch $H_i$ outputs the largest element $x$ stored in $H_i$ into the output stream $\pi'_{H_i}$, with multiplicity equal to the \textit{weight} of $x$. Then, we remove $x$ from $H_i$ (Note: the largest element $x$ may not be in the final buffer $B$, rather it may be in one of the elastic compactors in the hierarchy). We repeat this removal process until the total weight of $H$ becomes less than $2 R_i$ again. See \Cref{alg:Hi-weight} for the pseudocode. Observe that once the total weight (which is initially $0$) exceeds $R_i$, it will always be within $[R_i, 3R_i]$.

    \begin{algorithm}[h]
\caption{Maintaining the total weight of $H_i$.} \label{alg:Hi-weight}
\begin{algorithmic}[1]
\State $R_i \gets \epsilon^{-1} \cdot 2^i$
\If{total weight of elements in $H_i$ exceeds $3 R_i$}
\State $\Call{Reset}{H_{i + 1}}$. \label{line:reset-Hi1} \Comment \textcolor{gray}{This line is for \Cref{sec:relative-error} and should now be ignored. }
\Repeat
\State Suppose the largest element in $H_i$ is in $C_j$ ($0 \leq j \leq \log(1/\epsilon)$). 

\Comment \textcolor{gray}{Note this includes the buffer ($j = \log(1/\epsilon)$).}
\State $\Call{Remove}{}_{\max}(C_j)$ and let $x$ be the largest element that $C_j$ outputs. \footnotemark
\State Output $2^j \cdot \epsilon^2 R_i$ many copies of $x$ to an output stream $\pi'_{H_i}$.  \label{line:output-with-multiplicity}

\Comment \textcolor{gray}{Later in \Cref{sec:relative-error}, we will insert each element in $\pi'_{H_i}$ to $H_{i + 1}$. }
\Until{total weight of elements in $H_i$ is less or equal to $2 R_i$}
\EndIf
\end{algorithmic}
\end{algorithm}

    In the final algorithm, we will then insert the elements in $\pi'_{H_i}$ into the next sub-sketch $H_{i + 1}$. We observe that by repeatedly removing the largest elements from $H_i$ and adding them into $\pi'_{H_i}$, we guarantee the following ``ordering'' property:
    
    \begin{observation} \label{obs:order}
    The elements in the output stream $\pi'_{H_i}$ of $H_i$ are naturally in non-decreasing order, and are always larger than all elements in $H_i$. Furthermore, it follows that all elements in $H_{i+1}$ are always larger than all elements in $H_i$.
    \end{observation}
    
    \item \textbf{Error guarantee for all queries:} Instead of only those queries $x$ with $\rank_{\pi_{H_i}}(x) \leq R_i$, we are going to define a notion of error for \emph{arbitrary query}. We define the estimate of $\rank_{\pi_{H_i}}(x)$ by $H_i$ as 
$$\widehat{\rank_{\pi_{H_i}}(x)} \coloneqq \Call{rank}{}_{H_i}(x) + \rank_{\pi'_{H_i}}(x).$$
Here $\pi'_{H_i}$ is the ordered output we defined above. The error is therefore defined as $$\Delta_{H_i}(x) \coloneqq \widehat{\rank_{\pi_{H_i}}(x)} - \rank_{\pi_{H_i}}(x).$$

    \item \textbf{Resets:} Just as we defined a reset operation for elastic compactors, we also define resets for our sub-sketch $H_i$.
    
    \begin{itemize}
    \item $\Call{Reset}{H_i}$: This operation resets the sum in \Cref{equ:condition-H} at the cost of introducing more error. Concretely, this operation simply calls $\Call{Reset}{C_{i,j}}$ for every compactor $C_{i,j}$ ($0 \leq j < \log(1/\epsilon)$) in the sampler/compactor hierarchy defined in \Cref{sec:top-quantile}.
    \end{itemize}

    Fixing a query $x \in \U$, we say a reset is \emph{important} if at the time of reset, there exists at least one element in $H_i$ that is smaller than $x$.

\end{enumerate}

To summarize, we have the following lemma.
\begin{restatable}{lem}{subsketch}
\label{claim:sub-sketch-reset}
Consider an arbitrary query $x$. Suppose there are $t_i(x)$ important resets. Given the same condition as \Cref{lem:H}, we have 
$$\E\left[\Delta_{H_i}(x)^2\right]^{1/2} = O(\epsilon) \cdot \sqrt{R_i \cdot \rank_{\pi_{H_i}}(x) + t_i(x) \cdot R_i^2}.$$

\end{restatable} 
Notably, this lemma plays an important role in the proof of \Cref{lem:final-analysis}; we defer the proof to \Cref{sec:expectation}.

\subsection{All Quantiles Relative Error Sketch} \label{sec:relative-error}

Now, we are ready to describe the complete algorithm for the relative error quantile estimation problem in the streaming model.

\begin{figure}[H]
    \centering

\begin{center}

\tikzset{every picture/.style={line width=0.75pt}} %

\begin{tikzpicture}[x=0.75pt,y=0.75pt,yscale=-1,xscale=1]

\draw    (87.86,190.29) -- (460.86,189.29) ;
\draw [shift={(463.86,189.29)}, rotate = 179.85] [fill={rgb, 255:red, 0; green, 0; blue, 0 }  ][line width=0.08]  [draw opacity=0] (8.93,-4.29) -- (0,0) -- (8.93,4.29) -- cycle    ;
\draw  [dash pattern={on 4.5pt off 4.5pt}]  (129.86,109.29) -- (129.86,209.29) ;
\draw  [dash pattern={on 4.5pt off 4.5pt}]  (188.86,109.29) -- (188.86,210.29) ;
\draw  [dash pattern={on 4.5pt off 4.5pt}]  (280.86,109.29) -- (280.86,210.29) ;
\draw  [dash pattern={on 4.5pt off 4.5pt}]  (430.86,112.29) -- (430.86,213.29) ;

\draw (88,139) node [anchor=north west][inner sep=0.75pt]   [align=left] {$\displaystyle \cdots $};
\draw (141,129) node [anchor=north west][inner sep=0.75pt]   [align=left] {$\displaystyle H_{i-1}$};
\draw (228,130) node [anchor=north west][inner sep=0.75pt]   [align=left] {$\displaystyle H_{i} \ $};
\draw (444,137) node [anchor=north west][inner sep=0.75pt]   [align=left] {$\displaystyle \cdots $};
\draw (346,130) node [anchor=north west][inner sep=0.75pt]   [align=left] {$\displaystyle H_{i+1} \ $};
\draw (453,197) node [anchor=north west][inner sep=0.75pt]   [align=left] {rank};
\draw (95,223) node [anchor=north west][inner sep=0.75pt]  [font=\small] [align=left] {$\displaystyle \epsilon ^{-1} \cdot 2^{\ i-2\ }$};
\draw (169,222) node [anchor=north west][inner sep=0.75pt]  [font=\small] [align=left] {$\displaystyle \epsilon ^{-1} \cdot 2^{\ i-1\ }$};
\draw (260,221) node [anchor=north west][inner sep=0.75pt]  [font=\small] [align=left] {$\displaystyle \epsilon ^{-1} \cdot 2^{\ i\ }$};
\draw (407,223) node [anchor=north west][inner sep=0.75pt]  [font=\small] [align=left] {$\displaystyle \epsilon ^{-1} \cdot 2^{\ i+1\ }$};
\draw (51,85) node [anchor=north west][inner sep=0.75pt]   [align=left] {error $ $};
\draw (136,84) node [anchor=north west][inner sep=0.75pt]  [font=\small] [align=left] {$\displaystyle \leq 2^{\ i-1\ }$};
\draw (221,83) node [anchor=north west][inner sep=0.75pt]  [font=\small] [align=left] {$\displaystyle \leq 2^{\ i\ }$};
\draw (341,83) node [anchor=north west][inner sep=0.75pt]  [font=\small] [align=left] {$\displaystyle \leq 2^{\ i+1\ }$};

\end{tikzpicture}
\end{center}

    \repeatcaption{fig:basic}{Our basic strategy.}
\end{figure}
\paragraph{The Full Relative Error Quantile Sketch.} 
As explained at the beginning of the section, for each $0 \leq i \leq \log_2 (\epsilon n)$, our algorithm will maintain one sub-sketch $H_i$ for elements of rank roughly $R_i \coloneqq \epsilon^{-1} \cdot 2^i $.\footnote{For the first $\log (1 / \epsilon)$ many $H_i$'s, we have $R_i \leq \frac{1}{\epsilon^2}$, thus they have the structure described in \Cref{remark:corner-case-structure}.} We now describe the operation of the algorithm on each inserted stream element $x$.
\begin{enumerate}
    \item First, we insert $x$ into the hierarchy $H_i$ whose range contains $x$. More precisely, we feed $x$ to the sampler of the  first $H_i$ such that $H_{i + 1}$ is either empty, or contains only elements that are greater than $x$. \label{step:insret}
    \item  As we mentioned earlier, $H_i$ might have total weight exceeding $3 R_i$ after this insertion; in this case, $H_i$ will move at least $R_i$ elements into the output steam $\pi'_i$; these elements will eventually be added to the input stream of the next hierarchy $H_{i+1}$. \label{step:overflow}
    \item  Importantly, whenever \Cref{step:overflow} happens, we first \emph{reset} the sketch $H_{i + 1}$ (Line~\ref{line:reset-Hi1}, \Cref{alg:Hi-weight}) and then insert those newly added elements in $\pi'_i$ into $H_{i + 1}$. \label{step:move}
\end{enumerate}

Our estimator is a natural extension of \Cref{equ:rank-estimator-H}: to estimate the rank of any query $x$, we output 

\begin{equation}\label{equ:final-estimator}
   \widehat{\rank_{\pi}}(x) := \sum_{i=0}^{\log_2(\epsilon n)} \Call{Rank}{}_{H_i}(x)
\end{equation}

Before we proceed, we make a few observations about the structure of our overall sketch.

\begin{lem}
Our all quantiles relative error sketch maintains the following properties:
\begin{enumerate}
    \item Throughout the execution of the algorithm, all sub-sketches $H_i$ maintain the invariant that the total weight of elements stored in $H_i$ is within $[R_i, 3 R_i]$ (excluding the last non-empty sub-sketch $H_{i'}$ which may have total weight less than $R_{i'}$).
    \item  Moreover, the ranges of the $H_i$'s are disjoint, that is, all elements in $H_{i - 1}$ are always smaller than those in $H_i$. 
\end{enumerate}
\end{lem}
\begin{proof}
    We note that the first property follows directly from the definition of $H_i$. So, we proceed to check the second property, which we prove by induction. Suppose that before an operation, the $H_i$'s ranges are disjoint. Consider the following two cases:
\begin{itemize}
    \item If this operation is an insertion (\Cref{step:insret}), because we insert $x$ into the $H_i$ whose range contains $x$. The disjointness is preserved. 
    \item If this operation is moving elements from $H_i$ to $H_{i + 1}$ (\Cref{step:overflow} and \Cref{step:move}), by \Cref{obs:order}, these elements we moved to $H_{i + 1}$ are larger than all elements in $H_i$. Thus the range of $H_i$ and $H_{i + 1}$ are still disjoint. 

\end{itemize}
\end{proof}

At this point, it suffices for us to show that this sketch achieves the relative error guarantee, assuming that we can allocate an appropriate amount of space to each sub-sketch $H_i$ (we will address our space allocation strategy next, in \Cref{sec:dynamic-space}).

\begin{claim}\label{claim:important-reset}
For any sub-sketch $H_i$, the number of important reset $t_i(x) \leq 2 \cdot \frac{\rank_{\pi_{H_i}}(x)}{R_i}$.
\end{claim}
\begin{proof}
This is because whenever we have a important reset for $H_i$, there must be at least $R_{i - 1} = R_i / 2$ many elements that are smaller than $x$ being inserted into $H_i$. 
\end{proof}

\begin{lem} \label{lem:final-analysis}
For any input stream $\pi$, assuming that the space we allocate to each $H_i$ satisfies the premise of \Cref{lem:H} (\Cref{equ:condition-H}), we always have that for any query $x$,
$$\E\left[\Big|\rank_{\pi}(x) - \sum_{i=0}^{\log_2(\epsilon \cdot \rank_\pi(x))} \Call{Rank}{}_{H_i}(x)\Big|^2\right]^{1/2} \leq O(\epsilon) \cdot \rank_{\pi}(x).$$

where the expectation is taken over the randomness of the $H_i$'s. 
\end{lem}
\begin{remark} \label{reamrk:tail}
    Before we prove this lemma, we remark that it suffices to prove that our algorithm succeeds with constant probability. This is because by Chebyshev's inequality, there is an absolute constant $c > 0$ such that  $$\Pr\left[\Big|\rank_{\pi}(x) - \sum_{i=0}^{\log_2(\epsilon \cdot \rank_\pi(x))} \Call{Rank}{}_{H_i}(x)\Big| \geq c \cdot \epsilon \cdot \rank_{\pi}(x)\right] \leq \frac{1}{3}.$$
    But we know the total weight of $H_1, \dots, H_{\log_2(\epsilon \cdot \rank_{\pi}(x))}$ is at least $R_1 + R_2 + \cdots + R_{\epsilon \cdot \log_2 (\epsilon \cdot \rank_{\pi}(x))} > 1.5 \cdot \rank_{\pi}(x)$. As long as, we pick $\epsilon < 1 / (2c)$, there must be elements in these hierarchies that are larger than $x$. In this case, $\Call{Rank}{}_{H_i}(x) = 0$ for all $i > \log_2(\epsilon \cdot \rank_{\pi}(x)$. 
\end{remark}
\begin{proof}
By \Cref{claim:important-reset}, for any sub-sketch $H_i$, the number of important resets $t_i(x) =  O\left(\frac{\rank_{\pi_{H_i}}(x)}{R_i}\right).$ Together with $\E[\rank_{\pi_{H_i}}(x)] \leq \rank_{\pi}(x)$, plug in \Cref{claim:sub-sketch-reset}, we get 
$$\E\left[\Delta_{H_i}(x)^2\right]^{1/2} \leq O(\epsilon) \cdot \sqrt{R_i \cdot \rank_{\pi}(x) + \E[t_i(x)] \cdot R_i^2} = O(\epsilon) \cdot \sqrt{R_i \cdot \rank_{\pi}(x)}.$$

Finally we use \Cref{fact:stadnard-deviation} and sum over all the sub-sketches $H_1, H_2, \dots, H_{\log_2(\epsilon \cdot \rank_\pi(x))}$. 
\begin{align*}
&\E\left[\Big|\rank_{\pi}(x) - \sum_{i=0}^{\log_2(\epsilon \cdot \rank_\pi(x))} \Call{Rank}{}_{H_i}(x)\Big|^2\right]^{1/2} \\
= & \E\left[\left|\ \sum_{i=0}^{\log_2(\epsilon \cdot \rank_\pi(x))} 
\rank_{\pi_{H_i}}(x) - \rank_{\pi'_i}(x) - \Call{Rank}{}_{H_i}(x)\right|^2\right]^{1/2} \\
\leq &\sum_{i=0}^{\log_2(\epsilon \cdot \rank_\pi(x))}  \E\left[\Delta_{H_i}(x)^2\right]^{1/2} \\
\leq &\sum_{i=0}^{\log_2(\epsilon \cdot \rank_\pi(x))} O(\epsilon) \cdot \sqrt{R_i \cdot \rank_\pi(x)} \\
= &O(\epsilon) \cdot \rank_{\pi}(x).
\end{align*}
Here $\pi'_{H_i}$ is the stream of elements we move from $H_i$ to $H_{i + 1}$, and the last step follows since the values $R_i$'s are exponentially increasing.
\end{proof}

\subsection{Dynamic Space Allocation}

The only piece left from \Cref{lem:final-analysis} is to specify the allocation of space for each $H_i$. Naively, the most straightforward approach is to give each $H_i$ the same amount of space. However, since each $H_i$ solves a top quantiles problem, we need at least $O\left(\frac{\sqrt{\log (\epsilon n)}}{\epsilon}\right)$ space for each $H_i$ and can at best get an algorithm with $O\left(\frac{\log^{1.5} (\epsilon n)}{\epsilon}\right)$ space. To go beyond this lower bound, we will need to \textit{dynamically} adjust the space used by each subsketch $H_i$ \textit{on-the-fly}. Before we give a complete description of the online space allocation procedure, we describe a way to allocate space in the restricted (offline) case, when all reset times are known in advance to the algorithm.

    \paragraph{The offline space allocation problem.} Our strategy for allocating space affects the behavior of each $H_i$ and also changes the time when different levels reset. To best illustrate our idea, let us first consider the \emph{offline} version, where the compactor reset times are fixed and known in advance. 

First, we will view each operation call that changes the memory of the $H_i$'s (i.e. for instance, an insertion of a sampled element into the zero-th level compactor $C_{i,0}$ or a resize call to $H_i$) as one time step\footnote{Importantly, observe that this definition does not count those insertions that are not sampled in the ``sampler'' step, as these do not affect the state of the $H_i$'s. }. Notably, it is clear from the algorithm description that there are $O(\poly(n))$ time steps in total, where $n$ is the stream length. Also, let $t_{i,j}$ denote the time step at which $H_i$ resets for the $j$-th time. We use $W_{i,j}$ to denote the set of time points within the interval $\left[t_{i,j}, t_{i, j + 1}\right)$ at which we resize $H_i$/insert from the sampler into $C_{i,0}$, and $s_{i, t}$ is the space we allocate to $H_i$ at time $t$ (which must be positive). 

Our strategy satisfies the following two constraints:

\begin{itemize}
    \item (The total space is bounded.) At any time $t$, $$\sum_{i=0}^{\log (\epsilon n)} s_{i,t} = O(\epsilon^{-1} \log n  \log(1/\epsilon)).$$
    \item (The space sequence is feasible.) For any $H_i$ and time interval $\left[t_{i,j}, t_{i, j + 1}\right)$, we have $$\sum_{t \in W_{i,j}} 2^{-\epsilon \cdot  s_{i,t} } \leq 0.5.$$
\end{itemize}

Concretely, the offline space allocation question is defined as follows: given fixed time intervals $\left[t_{i,j}, t_{i, j + 1}\right)$ and the time of operations $W_{i,j}$'s, we need to find the proper sequence of space allocations $s_{i,t}$'s satisfying both of the constraints above.

\begin{remark} \label{remark:emptyHi-reset}
To handle the edge case where some of the $H_i$'s are empty, we reset those sub-sketches $H_i$ at every time-step until they become non-empty. Note that this does not change the state of those empty $H_i$'s at all, but ensures that when $H_i$ is empty, the corresponding time intervals between consecutive resets are all of length $1$. We will need this fact later when handling unknown stream length $n$. 
\end{remark}

\newcommand{\pa}{\mathrm{parent}}

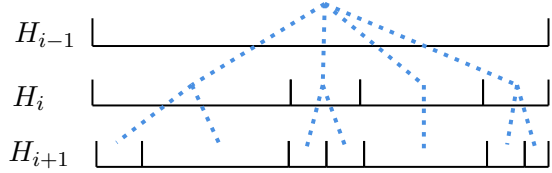
\begin{figure}[H]
    \centering

\tikzset{every picture/.style={line width=0.75pt}} %

\begin{tikzpicture}[x=0.75pt,y=0.75pt,yscale=-1,xscale=1]

\draw    (139.86,80.29) -- (369.86,80.29) ;
\draw    (139.86,66.29) -- (139.86,80.29) ;
\draw    (369.86,66.29) -- (369.86,80.29) ;
\draw    (139.86,110.29) -- (239.86,110.29) ;
\draw    (139.86,97.22) -- (139.86,110.29) ;
\draw    (239.86,97.22) -- (239.86,110.29) ;

\draw    (239.86,110.29) -- (274.86,110.29) ;
\draw    (239.86,97.22) -- (239.86,110.29) ;
\draw    (274.86,97.22) -- (274.86,110.29) ;

\draw    (274.86,110.29) -- (336.86,110.29) ;
\draw    (274.86,97.22) -- (274.86,110.29) ;
\draw    (336.86,97.22) -- (336.86,110.29) ;

\draw    (336.86,110.29) -- (369.86,110.29) ;
\draw    (336.86,97.22) -- (336.86,110.29) ;
\draw    (369.86,97.22) -- (369.86,110.29) ;

\draw    (141.86,141.29) -- (164.86,141.29) ;
\draw    (141.86,128.22) -- (141.86,141.29) ;
\draw    (164.86,128.22) -- (164.86,141.29) ;

\draw    (164.86,141.29) -- (238.86,141.29) ;
\draw    (164.86,128.22) -- (164.86,141.29) ;
\draw    (238.86,128.22) -- (238.86,141.29) ;

\draw    (238.86,141.29) -- (257.86,141.29) ;
\draw    (238.86,128.22) -- (238.86,141.29) ;
\draw    (257.86,128.22) -- (257.86,141.29) ;

\draw    (257.86,141.29) -- (276.86,141.29) ;
\draw    (257.86,128.22) -- (257.86,141.29) ;
\draw    (276.86,128.22) -- (276.86,141.29) ;

\draw    (276.86,141.29) -- (338.86,141.29) ;
\draw    (276.86,128.22) -- (276.86,141.29) ;
\draw    (338.86,128.22) -- (338.86,141.29) ;

\draw    (338.86,141.29) -- (357.86,141.29) ;
\draw    (338.86,128.22) -- (338.86,141.29) ;
\draw    (357.86,128.22) -- (357.86,141.29) ;

\draw    (357.86,141.29) -- (369.86,141.29) ;
\draw    (357.86,128.22) -- (357.86,141.29) ;
\draw    (369.86,128.22) -- (369.86,141.29) ;

\draw [color={rgb, 255:red, 74; green, 144; blue, 226 }  ,draw opacity=1 ][line width=1.5]  [dash pattern={on 1.69pt off 2.76pt}]  (257,59) -- (190,100) ;
\draw [color={rgb, 255:red, 74; green, 144; blue, 226 }  ,draw opacity=1 ][line width=1.5]  [dash pattern={on 1.69pt off 2.76pt}]  (257,59) -- (256,100) ;
\draw [color={rgb, 255:red, 74; green, 144; blue, 226 }  ,draw opacity=1 ][line width=1.5]  [dash pattern={on 1.69pt off 2.76pt}]  (257,59) -- (307,100) ;
\draw [color={rgb, 255:red, 74; green, 144; blue, 226 }  ,draw opacity=1 ][line width=1.5]  [dash pattern={on 1.69pt off 2.76pt}]  (257,59) -- (354,100) ;
\draw [color={rgb, 255:red, 74; green, 144; blue, 226 }  ,draw opacity=1 ][line width=1.5]  [dash pattern={on 1.69pt off 2.76pt}]  (190,100) -- (152,129) ;
\draw [color={rgb, 255:red, 74; green, 144; blue, 226 }  ,draw opacity=1 ][line width=1.5]  [dash pattern={on 1.69pt off 2.76pt}]  (190,100) -- (204,130) ;
\draw [color={rgb, 255:red, 74; green, 144; blue, 226 }  ,draw opacity=1 ][line width=1.5]  [dash pattern={on 1.69pt off 2.76pt}]  (256,100) -- (248,131) ;
\draw [color={rgb, 255:red, 74; green, 144; blue, 226 }  ,draw opacity=1 ][line width=1.5]  [dash pattern={on 1.69pt off 2.76pt}]  (256,100) -- (267,130) ;
\draw [color={rgb, 255:red, 74; green, 144; blue, 226 }  ,draw opacity=1 ][line width=1.5]  [dash pattern={on 1.69pt off 2.76pt}]  (307,100) -- (307,132) ;
\draw [color={rgb, 255:red, 74; green, 144; blue, 226 }  ,draw opacity=1 ][line width=1.5]  [dash pattern={on 1.69pt off 2.76pt}]  (354,100) -- (349,130) ;
\draw [color={rgb, 255:red, 74; green, 144; blue, 226 }  ,draw opacity=1 ][line width=1.5]  [dash pattern={on 1.69pt off 2.76pt}]  (354,100) -- (363,131) ;

\draw (98,99) node [anchor=north west][inner sep=0.75pt]   [align=left] {$\displaystyle H_{i} \ $};
\draw (96,127) node [anchor=north west][inner sep=0.75pt]   [align=left] {$\displaystyle H_{i+1} \ $};
\draw (99,67) node [anchor=north west][inner sep=0.75pt]   [align=left] {$\displaystyle H_{i-1} \ $};

\end{tikzpicture}

    \caption{Tree-like intervals.}
    \label{fig:tree-like}
\end{figure}

\paragraph{Space allocation for tree-like intervals.} As a warm up, we will first describe the space allocation strategy for the following (simpler) special case: suppose that for all the intervals $\left[t_{i + 1, j}, t_{i + 1, j + 1}\right)$, there always exists an interval $\left[t_{i, \pa(j)}, t_{i, \pa(j)  + 1}\right)$ that contains it. In this case, these intervals form a tree-like structure, which is shown in \Cref{fig:tree-like}. 

Our strategy for this case is easy to state: For all $t \in \left[t_{i + 1,j}, t_{i + 1, j + 1}\right)$, we set the space of $H_i$ (the parent) as 
\begin{equation} \label{equ:interval-space}
s_{i, t} = \epsilon^{-1} \cdot \left( \log \frac{t_{i, \pa(j) + 1}- t_{i, \pa(j)}}{t_{i + 1, j  + 1} - t_{i + 1, j}}  + 5\log(1/\epsilon)\right).    
\end{equation}
For the corner case when $i = \log(\epsilon n)$, we pretend that there is $H_{\log(\epsilon n) + 1}$ which resets every step. That is, there is one addition level at the bottom with all length-$1$ intervals. We know that $s_{i,t}$ is always positive because $[t_{i + 1,j}, t_{i + 1, j + 1})$ is always shorter or equal to its parent. 

We need to verify that it satisfies the two constraints:

\begin{itemize}
    \item (The total space is bounded.) This part is straightforward to verify. We first fix a time $t$. Let $j_i$ be the index of the interval $[t_{i,j_i}, t_{i, j_i + 1})$ that contains time $t$. Then, we obtain the following telescoping sum.
    \begin{align*}
\sum_{i=0}^{\log (\epsilon n)} s_{i, t} &= \sum_{i=0}^{\log (\epsilon n)} \epsilon^{-1} \cdot \left(\log \frac{t_{i, j_i + 1} - t_{i,j_i}}{t_{i + 1, j_{i + 1} + 1} - t_{i + 1, j_{i + 1}}} + 5 \log(1/\epsilon) \right) \\ &= \epsilon^{-1} \cdot \log \frac{\poly(n)}{1} + \epsilon^{-1} \cdot \log(\epsilon n) \cdot 5 \log(1/\epsilon) \\ &= O(\epsilon^{-1} \log n  \log(1/\epsilon)).        
    \end{align*}
    Thus the total space is at most $O(\epsilon^{-1} \log n  \log^2(1/\epsilon))$ because the space used by each hierarchy $H_i$ is $s_{i,t} \cdot \lceil \log_2(1/\epsilon)\rceil$. 
    \item (The space sequence is feasible.) To prove this, we need the following claim.
    \begin{restatable}{claim}{insinterval}\label{claim:ins-interval}
        Within any interval $[t_{i + 1,j}, t_{i + 1,j + 1})$, $H_i$ has at most $3 / \epsilon^2$ many insertions into $C_{i,0}$. We observe that this is true in general, i.e. not only for the case that all resets occur in tree-like intervals.
    \end{restatable} 
    \begin{proof}
        This is because each element in $C_{i,0}$ has weight $\epsilon^2 R_i$. If there are more than $3 / \epsilon^2$ such insertions, it will triger a reset of $H_{i + 1}$. As $t_{i + 1, j + 1}$ is the next reset after $t_{i + 1, j}$, we know there can be at most $3 / \epsilon^2$ insertions between them. 
    \end{proof}

    Consider any interval $[t_{i,j}, t_{i,j + 1})$. Suppose it is the union of its children  $[t_{i + 1, \ell}, t_{i + 1, \ell + 1})$, $[t_{i + 1, \ell + 1}$, $t_{i + 1, \ell + 2})$, $\dots$, $[t_{i + 1, r - 1}, t_{i + 1, r})$. We know that $H_i$ only resizes at time step $t_{i + 1,\ell}$, $t_{i + 1, \ell + 1}$, $\dots$, $t_{i + 1, r}$. Thus, together with \Cref{claim:ins-interval}, we know that for each $[t_{i + 1, m}, t_{i + 1, m + 1})$ ($l \leq m \leq r - 1$), we have $|W_{i,j} \cap [t_{i + 1, m}, t_{i + 1, m + 1})| \leq 3/\epsilon^2 + 2$. Thus, we know 
    \begin{align*}
        \sum_{t \in W_{i,j}} 2^{-\epsilon \cdot s_{i,t}} & \leq \sum_{m = \ell}^{r - 1} \sum_{\substack{t \\ t \in W_{i,j} \cap [t_{i + 1, m}, t_{i + 1, m + 1})}} 2^{- \epsilon \cdot s_{i,t}} \\
        &\leq \sum_{m =\ell}^{r - 1} (3 / \epsilon^2 + 2) \cdot \frac{t_{i + 1, m + 1} - t_{i + 1, m}}{t_{i,j + 1} - t_{i,j}} \cdot \epsilon^5 \leq 0.25 \cdot \frac{\sum_{m =\ell}^{r - 1} t_{i + 1, m + 1} - t_{i + 1, m}}{t_{i,j + 1} - t_{i,j}}  \leq 0.25
    \end{align*}

\end{itemize}

\paragraph{Space allocation for general intervals.} In the general case, a single interval for $H_{i + 1}$ may intersect with multiple ``parent'' intervals for $H_i$ above it. It might also be longer than any such ``parent'' interval, which could cause \Cref{equ:interval-space} to allocate negative space. Thus, we need to carefully generalize our strategy. 

\begin{figure}[H]
    \centering

\tikzset{every picture/.style={line width=0.75pt}} %

\begin{tikzpicture}[x=0.75pt,y=0.75pt,yscale=-1,xscale=1]

\draw    (139.86,80.29) -- (369.86,80.29) ;
\draw    (139.86,66.29) -- (139.86,80.29) ;
\draw    (369.86,66.29) -- (369.86,80.29) ;
\draw    (141,110.29) -- (313,110.29) ;
\draw    (141,97.22) -- (141,110.29) ;
\draw    (313,97.22) -- (313,110.29) ;

\draw    (313,110.29) -- (369,110.29) ;
\draw    (313,97.22) -- (313,110.29) ;
\draw    (369,97.22) -- (369,110.29) ;

\draw    (141.86,141.29) -- (179,141.29) ;
\draw    (141.86,128.22) -- (141.86,141.29) ;
\draw    (179,128.22) -- (179,141.29) ;

\draw    (179,141.29) -- (239,141.29) ;
\draw    (179,128.22) -- (179,141.29) ;
\draw    (239,128.22) -- (239,141.29) ;

\draw    (239,141.29) -- (346,141.29) ;
\draw    (239,128.22) -- (239,141.29) ;
\draw    (346,128.22) -- (346,141.29) ;

\draw    (302,141.29) -- (369.86,141.29) ;
\draw    (302,128.22) -- (302,141.29) ;
\draw    (369.86,128.22) -- (369.86,141.29) ;

\draw    (260.86,66.29) -- (260.86,80.29) ;
\draw    (209.86,66.29) -- (209.86,80.29) ;
\draw [color={rgb, 255:red, 74; green, 144; blue, 226 }  ,draw opacity=1 ][line width=1.5]  [dash pattern={on 1.69pt off 2.76pt}]  (175,68.22) -- (236,108) ;
\draw [color={rgb, 255:red, 74; green, 144; blue, 226 }  ,draw opacity=1 ][line width=1.5]  [dash pattern={on 1.69pt off 2.76pt}]  (235,69) -- (236,108) ;
\draw [color={rgb, 255:red, 74; green, 144; blue, 226 }  ,draw opacity=1 ][line width=1.5]  [dash pattern={on 1.69pt off 2.76pt}]  (308,71) -- (236,108) ;
\draw [color={rgb, 255:red, 74; green, 144; blue, 226 }  ,draw opacity=1 ][line width=1.5]  [dash pattern={on 1.69pt off 2.76pt}]  (308,71) -- (345,104) ;
\draw [color={rgb, 255:red, 74; green, 144; blue, 226 }  ,draw opacity=1 ][line width=1.5]  [dash pattern={on 1.69pt off 2.76pt}]  (236,108) -- (161,131) ;
\draw [color={rgb, 255:red, 74; green, 144; blue, 226 }  ,draw opacity=1 ][line width=1.5]  [dash pattern={on 1.69pt off 2.76pt}]  (236,108) -- (219,132) ;
\draw [color={rgb, 255:red, 74; green, 144; blue, 226 }  ,draw opacity=1 ][line width=1.5]  [dash pattern={on 1.69pt off 2.76pt}]  (236,108) -- (275,132) ;
\draw [color={rgb, 255:red, 74; green, 144; blue, 226 }  ,draw opacity=1 ][line width=1.5]  [dash pattern={on 1.69pt off 2.76pt}]  (236,108) -- (323,130) ;
\draw [color={rgb, 255:red, 74; green, 144; blue, 226 }  ,draw opacity=1 ][line width=1.5]  [dash pattern={on 1.69pt off 2.76pt}]  (345,104) -- (323,130) ;
\draw [color={rgb, 255:red, 74; green, 144; blue, 226 }  ,draw opacity=1 ][line width=1.5]  [dash pattern={on 1.69pt off 2.76pt}]  (345,104) -- (360,131) ;

\draw (98,99) node [anchor=north west][inner sep=0.75pt]   [align=left] {$\displaystyle H_{i} \ $};
\draw (96,127) node [anchor=north west][inner sep=0.75pt]   [align=left] {$\displaystyle H_{i+1} \ $};
\draw (99,67) node [anchor=north west][inner sep=0.75pt]   [align=left] {$\displaystyle H_{i-1} \ $};

\end{tikzpicture}

    \caption{In general, the incidence relation between intervals can be far more complex than a tree. }
    \label{fig:enter-label}
\end{figure}
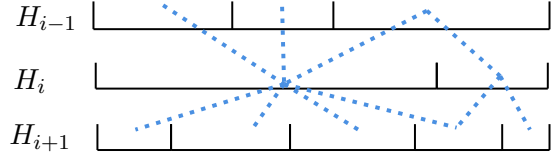

First, we need the following claim, which upper bounds the number of ``parent intervals'' for a single interval by a constant.  
\begin{claim} \label{claim:parent}
Within the time interval $\left[t^{(i)}_j, t^{(i)}_{j + 4}\right)$, $H_{i + 1}$ resets at least once. 
\end{claim}
\begin{proof}
Each time $H_i$ resets, the total weight of $H_{i}$ increases by at least $R_{i - 1} = \epsilon^{-1} \cdot 2^{i - 1}$ due to the corresponding batched insertion. Initially, the total weight of $H_i$ is at least $2^i / \epsilon$. Within at most four such resets, its weight increases to at least $3 \cdot 2^i / \epsilon$, which triggers a reset of $H_{i + 1}$. 
\end{proof}

To handle the corner case, again we imagine there is a $H_{\log(\epsilon n) + 1}$ which resets every step. So there is a imaginary level at the bottom full of length-$1$ intervals.  We define the potential $\phi_{i,j}$ for each interval $[t_{i,j}, t_{i,j + 1})$ recursively as follows:
\begin{itemize}
    \item For $i = \log(\epsilon n) + 1$ (the imaginary level at the bottom), we simply let $\phi_{i,j} = 1$ for all $j$.
    \item For $0 \leq i \leq \log(\epsilon n)$, suppose $[t_{i,j}, t_{i,j + 1})$ intersects with ``children'' intervals $[t_{i + 1, \ell}, t_{i + 1, \ell + 1}]$, $[t_{i + 1, \ell + 1}$, $t_{i + 1, \ell + 2}]$, $\dots$, $[t_{i + 1, r - 1}, t_{i + 1, r}]$. We define 
    $$\phi_{i,j} = \sum_{m = \ell}^{r - 1} \phi_{i + 1, m}.$$
\end{itemize}
Intuitively speaking, the potential $\phi_{i,j}$ is a the ``generalized length'' of each interval. We then assign the space similar to the tree-like case. For all $t \in [t_{i,j}, t_{i, j + 1}) \cap [t_{i + 1, m}, t_{i + 1, m + 1})$, set
\begin{equation} \label{equ:general-interval-space}
s_{i, t} = \epsilon^{-1} \cdot \left( \log \frac{\phi_{i,j}}{\phi_{i + 1, m}}  + 5\log(1/\epsilon)\right).    
\end{equation}

In this case, we know that $s_{i,t}$ is always positive because $\phi_{i,j}$ is always larger than $\phi_{i + 1, m}$ by definition. We defer the analysis to \Cref{sec:detailed}. In the analysis, we crucially use the fact that all $\phi_{i,j}$'s are at most $\poly(n)$ so that the telescoping sum for calculating total space will sum up to $O(\log n)$. The rest of the analysis are almost identical to the tree-like case. 

\begin{restatable}{claim}{potentialbound}\label{claim:potential-bound}
        For any $0 \leq i \leq \log(\epsilon n) + 1$, suppose there are $\ell_i$ intervals in total for $H_i$. Suppose that there are $T$ time steps (counting insertion into $C_{i,0}$'s and resizes of $H_i$'s) in total.
    
        Then,
        $$\sum_{j=1}^{\ell_i} \phi_{i,j} \leq 4^{\log(\epsilon n) + 1 - i} \cdot \ell_{\log(\epsilon n) + 1}.$$
        Specifically, the potential  $\phi_{i,j}$ of any interval is bounded by $4^{\log(\epsilon n) + 1} \cdot T = \poly(\epsilon n) \cdot T$. 
    \end{restatable}
    \begin{proof}
    We prove this by induction. When $i = \log(\epsilon n) + 1$, we know that every $\phi_{i,j} = 1$. Thus $\sum_{j=1}^{\ell_i} \phi_{i,j} = \ell_{\log(\epsilon n ) + 1}$, which equals $T$.

    Suppose this is true for $i + 1$. From \Cref{claim:parent}, we know that each interval $[t_{i + 1, j'}, t_{i + 1, j' + 1})$ intersects with at most $4$ ``parent'' intervals $[t_{i,j}, t_{i , j + 1}]$'s. 

    Thus $$\sum_{j=1}^{\ell_i} \phi_{i,j} \leq 4 \cdot \sum_{j=1}^{\ell_{i + 1}} \phi_{i,j + 1} \leq 4 \cdot 4^{\log(\epsilon n)  - i} \cdot \ell_{\log(\epsilon n) + 1},$$
    where in the last step we use the induction hypothesis. 
    \end{proof}

\paragraph{Online space allocation.} In the actual streaming model, we do not know the intervals in advance, so we need to adjust our strategy in order to allocate the space of each $H_i$ on-the-fly. 

Let $t$ be the current time. For all unfinished intervals, we pretend that the interval ends (i.e. the corresponding sub-sketch $H_i$ resets) at the \textit{current} time $t$; then, we  calculate all the potentials which we denote by $\phi^{(t)}_{i,j}$. We then round them up to the closest power of $2$, which we denote by $\cceil{\phi^{(t)}_{i,j}}$. We also calculate $\widehat{s}_{i, t} = \epsilon^{-1} \cdot \left( \log \frac{\cceil{\phi^{(t)}_{i,j}}}{\cceil{\phi^{(t)}_{i + 1, m}}}  + 5\log(1/\epsilon) + 5 \log \log n\right).$ We then allocate to $H_i$ space $\widehat{s}_{i, t}$.

\begin{figure}[H]
    \centering

\tikzset{every picture/.style={line width=0.75pt}} %

\begin{tikzpicture}[x=0.75pt,y=0.75pt,yscale=-1,xscale=1]

\draw    (139.86,80.29) -- (280,79.5) ;
\draw    (139.86,66.29) -- (139.86,80.29) ;
\draw    (141,110.29) -- (279,110) ;
\draw    (141,97.22) -- (141,110.29) ;
\draw    (141.86,141.29) -- (179,141.29) ;
\draw    (141.86,128.22) -- (141.86,141.29) ;
\draw    (179,128.22) -- (179,141.29) ;

\draw    (179,141.29) -- (239,141.29) ;
\draw    (179,128.22) -- (179,141.29) ;
\draw    (239,128.22) -- (239,141.29) ;

\draw    (239,141.29) -- (281,141.5) ;
\draw    (239,128.22) -- (239,141.29) ;
\draw    (260.86,66.29) -- (260.86,80.29) ;
\draw    (209.86,66.29) -- (209.86,80.29) ;
\draw [color={rgb, 255:red, 208; green, 2; blue, 27 }  ,draw opacity=1 ][line width=1.5]  [dash pattern={on 1.69pt off 2.76pt}]  (277.5,51) -- (280.5,169) ;

\draw (98,99) node [anchor=north west][inner sep=0.75pt]   [align=left] {$\displaystyle H_{i} \ $};
\draw (96,127) node [anchor=north west][inner sep=0.75pt]   [align=left] {$\displaystyle H_{i+1} \ $};
\draw (99,67) node [anchor=north west][inner sep=0.75pt]   [align=left] {$\displaystyle H_{i-1} \ $};
\draw (274,29) node [anchor=north west][inner sep=0.75pt]  [font=\small] [align=left] {$\displaystyle \textcolor[rgb]{0.82,0.01,0.11}{t}$};
\draw (297,95) node [anchor=north west][inner sep=0.75pt]  [color={rgb, 255:red, 208; green, 2; blue, 27 }  ,opacity=1 ] [align=left, font=\Large] {$\displaystyle \cdots  $};
\draw (334,95) node [anchor=north west][inner sep=0.75pt]  [color={rgb, 255:red, 208; green, 2; blue, 27 }  ,opacity=1 ] [align=left, font=\Large] {$\displaystyle \cdots $};

\end{tikzpicture}

    \caption{We calculate $\phi^{(t)}_{i,j}$'s using partial information. }
    \label{fig:online-allocation}
\end{figure}
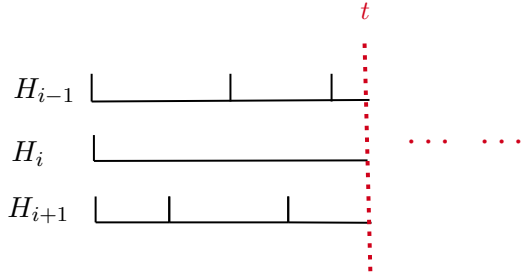

Intuitively, this works because the rounding guaranteed that the space $\widehat{s}_{i,t}$ does not change too often. Roughly speaking the number of resize operations multiplies by $O(\log^2 n)$, which offsets by the extra $\log \log n $ term in $\widehat{s}_{i,t}$. We again leave the detailed analysis to \Cref{sec:detailed}. 

\paragraph*{Handling unknown stream length.} One side benefit of our algorithm is that it handle unknown $n$ naturally. In our algorithm, none of the parameters ($k, R_i$ and buffer size) for each $H_i$ depends on $n$. Thus, we can imagine running our algorithm with infinite many subsketches, $H_0, H_1, H_2, \dots$. 

The only potential issue is that, our recursive definition of the potential $\phi^{(t)}_{i,j}$'s needs a base case. Recall that by \Cref{remark:emptyHi-reset}, all empty $H_i$'s have intervals of length $1$. We can the potential of those intervals to $1$. Because whenever $H_i$ is empty, $H_{i + 1}$ must also be empty, any such length-$1$ interval of $H_i$ can only intersect with a single length-$1$ interval of $H_{i + 1}$. So the recurrence relation still holds for all intervals. 

 \label{sec:dynamic-space}

\section{Implementation of the Elastic Compactor}
\label{sec:full-details-elastic}

In this section, we fill in the remaining implementation details for the elastic compactors $C_{i,j}$'s used in our subsketches. Before we proceed with this, we first provide a necessary overview of \emph{relative compactor}, which was introduced in \cite{cormode2023relative}; importantly, some basic operations of our elastic compactor data structure will be partially adapted from the relative compactor.\footnote{Although we presented a sketch of relative compactor in \Cref{sec:tech_overview}, the presentation here will be based on the deterministic compaction schedule of \cite{cormode2023relative} (as we mentioned in \Cref{footnote:determinstic}), which is easier to formally analyze and helps up the basic notations for our elastic compactor.}

\subsection{Overview of the Relative Compactor}\label{sec:overview_RC}

\paragraph{Blocks and Compaction.}
A relative compactor~\cite{cormode2023relative} stores $s$ elements (indexed $1$ through $s$), which are subdivided into $b = \lceil \frac{s}{k}\rceil$ blocks of size $k$. \footnote{$k$ is a even integer parameter which relates to the error incurred by compactor $C$.} Let $\{B_i\}_{i \in [b]}$ denote the blocks in $C$, and note that each block $B_i$ contains elements indexed from $(i - 1) \cdot k + 1$ to $i \cdot k$. The elements stored in $C$ are always maintained in sorted increasing order. 

As new elements arrive in the stream, they are inserted directly into the relative compactor $C$ while maintaining the increasing sorted order in $C$. When $C$ is full, the suffix of $\ell$ blocks is fed as input to the \emph{compaction} operation, which is defined below. This operation compacts the $(b - \ell + 1) \cdot k$ elements in the suffix of blocks into $(b - \ell + 1) \cdot k / 2$ outputted elements. In particular, the compaction always frees-up space for future insertions.

\begin{algorithm}[H] 
\caption{Compacting suffix $[B_{\ell}, B_{\ell + 1}, \dots, B_b]$}
\label{alg:compaction}
\begin{algorithmic}[1]
\Procedure{Compact}{$C, \ell$}
\State{Let $y \in_R \{0,1\}$ be a fair coin flip.}
\If {$y = 0$}
\State{Output all odd indexed elements in blocks $B_{\ell}, B_{\ell + 1}, \dots, B_b$.} 
\Else
\State{Output all even indexed elements in blocks $B_{\ell}, B_{\ell + 1}, \dots, B_b$.} 
\EndIf
\State{Remove all elements in suffix $[B_{\ell}, B_{\ell + 1}, \dots, B_b]$ from $C$.}
\EndProcedure
\end{algorithmic}
\end{algorithm}

\paragraph{Compaction Schedule.} However, we have not specified how to pick $\ell$ when calling $\Call{Compact}{C, \ell}$. Indeed, the work of \cite{cormode2023relative} presents an intricate method to select the number of blocks $\ell$ which will be compacted in a particular iteration; the sequence of choices of $\ell$ is called the \emph{compaction schedule}. This schedule is essential for the error analysis in ~\cite{cormode2023relative}. In the context of our work, we rephrase their strategy as follows.

Initially, each block $B_i$ is associated with a bit $z_i$ that is initialized to $0$. The compactor $C$ has a ``progress measure'', $z := \left(\overline{.z_1z_2...z_b}\right)_2$, where $0 \leq z < 1$ is a binary fraction. To ensure that the total accumulated error stays small, the compactor $C$ only has the capacity to handle a limited number of compactions. Intuitively, the bit-string $z$ indicates how much of the ``capacity'' of $C$ we have used already, and represents the ``progress'' of the compaction schedule. Eventually, if $z= 1 - 2^{-b}$, $C$ cannot perform any more compactions. The scheduling strategy of~\cite{cormode2023relative} is to do the following for each compaction\footnote{In \cite{cormode2023relative}, the authors maintain a counter $Z$ for the number of compactions that have already been performed. Then, the number of blocks to compact is determined by the \textit{number of trailing ones in the binary representation of $Z$ + 1}. We note that these two definitions are equivalent. Numerically, the sequence of these numbers are $\{1,2,1,3,1,2,1,4,...\}$.}: 

\begin{enumerate}
    \item Update $z \gets z + 2^{-(b - 1)}$. 
    \item Find the least significant bit $z_r$ such that $z_r = 1$. 
    \item Call $\Call{Compact}{C, r + 1}$.
\end{enumerate}
Note that in the furst line, we always add $2^{-(b - 1)}$ instead of $2^{-b}$ to $z$, as a result, the last bit $z_b$ is always zero. So $r \leq b - 1$, and in the last line we always compact at least one block.

For any query $x \in \mathcal{U}$, the compactor should provide an estimate of the rank of $x$ \textit{with respect to the current contents of $C$} (each element has weight 1) \textit{and with respect to the output stream $\pi'$ of $C$} (each element in $\pi'$ has weight 2). Specifically, this estimator is defined as $$\widehat \rank_{\pi}(x) = \rank_{C}(x) + 2\cdot\rank_{\pi'}(x).$$ 

Recall that to achieve the relative error guarantee, we need to ensure that our estimate $\widehat \rank_{\pi}(x)$ is very accurate when $\rank_\pi(x)$ is small (in fact, the first $1/\epsilon$ ranks should be known exactly). The compaction schedule chosen in ~\cite{cormode2023relative} facilitates this guarantee by ensuring that the  ``smaller'' elements of the stream will get compacted much less frequently than the ``larger'' elements, so they will naturally incur a smaller amount of error over all time-steps of the algorithm.

\paragraph{Error Analysis} Now, we examine the source of error for rank queries in the compaction procedure. Note that $\Call{Compact}{C, \ell}$ introduces no error for $x$ if all elements in block $B_\ell, B_{\ell + 1}, \dots, B_b$ are larger than $x$. So a compaction may introduce error, only when the involved blocks contains \textit{at least one ``important'' element} (i.e. an element that is smaller than the query $x$). This notion is defined more formally in the next definition.
\begin{definition}[Important element for query $x$]\label{def:important}
    For a fixed query $x \in \mathcal{U}$, we say a stream element $y$ is \emph{important} if and only if $y \leq x$. We also say a compaction $\Call{Compact}{C, \ell}$ is \emph{important} if the $B_\ell, B_{\ell + 1}, \dots, B_b$ contains at least one important element. 
\end{definition}
 Since each important compaction induces an independent $\pm 1$ error to the rank estimate, if there are $N$ important compactions in total, the standard deviation of our estimation will be bounded by $\pm \sqrt{N}$. Thus, it suffices for us to upper bound the total number of important compactions.
 
 In the previous work~\cite{cormode2023relative}, the authors showed that  the compaction schedule described above, the number of important compactions is upper-bounded by $\frac{\rank_\pi(x)}{k}$, and $C$ is able to handle roughly $k2^b$ insertions before the ``progress measure'' $z$ overflows. By chaining $O(\log(\epsilon n))$ such compactors together, they construct a relative-error quantile sketch using space $\tilde O(\epsilon^{-1}\log^{1.5}(\epsilon n))$.

\subsection{Implementation of Elastic Compactors}
Now, we are ready to describe the implementation of our \emph{elastic compactor}. Notably, the basic compaction operation and the compaction schedule will be generalized from those of relative compactors. Additionally, our elastic compactor is augmented with a \emph{resize} operation, which can be used to adjust the memory allocated to $C$, as well as a \emph{reset} operation which simply zero-outs the ``progress measure'' $z$ at the cost of introducing more error.

As before, we consider ${C}_{\textrm{elastic}}$ to have blocks of size $k$, but unlike the original relative compactor, we will allocate the \emph{number of blocks} in ${C}_{\textrm{elastic}}$ dynamically; as a result, we can think of ${C}_{\textrm{elastic}}$ as having possibly infinitely-many blocks $\{B_i\}_{i \geq 1}$ available, but only a specific number of blocks will actually be in-use at each time-step. As for the relative compactor, each block $B_i$ will be equipped with a bit $z_i$, which is initialized to $0$. Analogously to the progress measure for the relative compactor, the compaction procedure keeps track of its progress measure $z := \left(\overline{.z_1z_2...}\right)_2$ (observe that the number of bits is no longer fixed in advance). 

\paragraph{Basic Operations.} Recall that we have the following basic operations of the elastic compactor $\mathcal{C}_{\textrm{elastic}}$. (See \Cref{table:elas-basic-ops} on the next page.) In the description below, we let $s_t$ denote the total amount of space allocated to ${C}_{\textrm{elastic}}$ at time $t$ (i.e. if the space allocated is $s_t$, then there are $\lceil \frac{s_t}{k} \rceil$ blocks allocated to ${C}_{\textrm{elastic}}$ at time $t$).

\begin{table}[H]
    \centering
    \begin{tabular}{|p{14cm}|}
    \hline ~\\
    \multicolumn{1}{|c|}{\textbf{Basic operations for Elastic Compactor $C$}} \\
    
\begin{itemize}
    \item[-] $\Call{Compact}{C, \ell}$: This operation is the same as \Cref{alg:compaction}, while the number of blocks is $b = \lceil \frac{s_t}{k} \rceil$ for current time $t$. 
    
    \item[-] $\Call{Resize}{C, s'}$: Expand or compress the space allocated to $C$. Suppose that $C$ had space $s$ previously. Then:
    \begin{enumerate}
        \item If $s' > s$, this corresponds to \emph{expanding} the space for  $C$. Allocate $\lceil{\frac{s'}{k}}\rceil - 
        \lceil\frac{s}{k}\rceil$ additional blocks at the end of $C$. For each of these block $B_i$, we initialize $z_i$ to $0$. 
        \item If $s' < s$, this corresponds to \emph{compressing} the space for  $C$. Let $\ell = \lceil{\frac{s'}{k}}\rceil$ be the target number of blocks after the resize. We do the following:
        \begin{itemize}
            \item Increase $z$ to the next multiple of $2^{-\ell}$.
            \item Find the least significant bit $z_r$ such that $z_r = 1$. 
            \item Call $\Call{Compact}{C, r + 1}$. %
            \item Release all blocks $B_{>\ell}$.
        \end{itemize}
        See \Cref{fig:resize} for an example.
    \end{enumerate}
    \item[-] $\Call{Insert}{C, x_1,..., x_m}$: Suppose $C$ has space $s$. As described in \Cref{sec:compactor}, we implement insertion using $\Call{Resize}{C, \cdot}$ (implemented below). 
        \begin{itemize}
            \item First, we call $\Call{Resize}{C, 2s}$, and insert $x_1, x_2, \dots, x_m$ into the new open spaces in memory (note that $m \leq s$ necessarily).
            \item Then, we call $\Call{Resize}{C, s}$ to compress the space back to the original size~$s$. 
        \end{itemize}
        Note that some elements may be outputted to the output stream $\pi'$ during this last compression.

    \item[-] $\Call{Reset}{}_{}(C)$: Set $z \leftarrow 0$.

\end{itemize}

    \\
    \hline
    \end{tabular}
    \caption{Implementations of basic operations to Elastic Compactors.}
    \label{table:elas-basic-ops}
\end{table}

Recall that in \Cref{lem:compactor}, we require that ${C}_{\textrm{elastic}}$ can handle any sequence of insert and resize calls with a sequence of space constraints
$s_1, s_2..., s_m$  (after a reset) which satisfy $\sum_{t=1}^m 2^{-s_t / k} \leq 0.5$. Since we implement insertions using resizes, each ``external'' insert call with parameter $s$ generates two ``internal'' resize calls with parameters $s$ and $2s$ (See \Cref{table:elas-basic-ops}). Let sequence $s'_1, s'_2, \dots, s'_{m'}$ be the space parameters for all (internal and external) resize calls after a reset. As $2^s + 2^{-2s} \leq 2 \cdot 2^{-s}$, we know that $\sum_{t=1}^{m'} 2^{-s'_t / k} \leq 1$. 

Now, let us relate the running sum $\sum_{t=1}^{m'} 2^{-s'_t / k}$ to our ``progress measure'' $z$. 
\begin{lem} \label{lemma:bound-z}
    Let sequence $s'_1, s'_2, \dots, s'_{m'}$ be the space parameters for all (internal and external) resize calls after a reset. At any time in the operation sequence, we always have $z \leq \sum_{t=1}^{m'} 2^{-s'_t / k}$. That is, the ``progress measure'' $z$ is always upper-bounded by the running sum.
\end{lem}
\begin{proof}
We prove this by induction. Initially, after a reset, both $z$ and the running sum are zero.  

Suppose that this is true for $t - 1$. When there is a resize call $\Call{Resize}{C, s'_t}$, the running sum is increased by $2^{-s'_t / k}$. We will have $\ell = \lceil \frac{s'_t}{k}\rceil$ and $z$ is increase to the next multiple of $2^{-\ell}$. Because that multiple is at most $2^{-\ell}$ away and $2^{-\ell} \leq 2^{-s'_t / k}$, we know that this lemma is true for time~$t$.
\end{proof}

Having this lemma, we know that as long as $\sum_{t=1}^m 2^{-s_t} \leq 0.5$ (\Cref{equ:condition-H}) holds, we always have $0 \leq z < 1$. Then resizes can always be performed because the least significant bit $z_r$ such that $z_r = 1$ always exists. Moreover, we always have $r \leq \ell$ because $z$ is a multiple of $2^{-\ell}$. Then after $\Call{Compact}{C, r + 1}$, when we release the blocks $B_{> \ell}$, these blocks are guaranteed to be empty. 
Finally, $\Call{Insert}{}$ can also always be performed (again conditioning on \Cref{equ:condition-H} holds) because it is implemented by calls to \Call{resize}{}.

\begin{figure}
    \centering

\tikzset{every picture/.style={line width=0.75pt}} %

\includegraphics{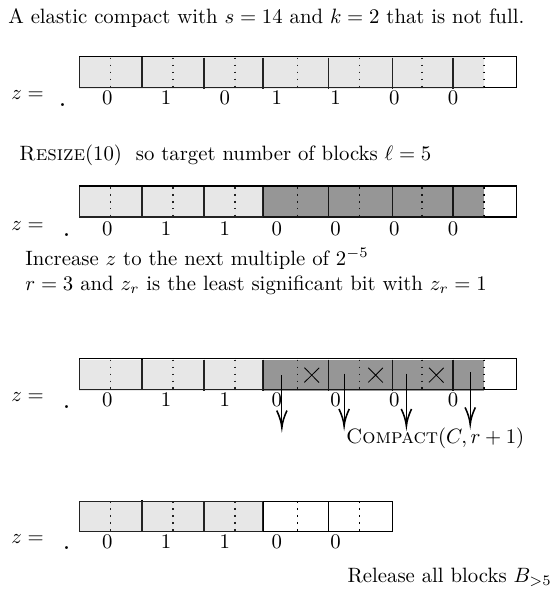}
    \caption{An example of the resize opeation.}
    \label{fig:resize}
\end{figure}

\subsection{Proof of \Cref{lem:compactor} and  \Cref{lem:compactor-reset}}

In this subsection, we will prove \Cref{lem:compactor-reset}. If we set $t = 0$ (having no reset at all) in \Cref{lem:compactor-reset}, it directly implies \Cref{lem:compactor}. 

Let $x \in \mathcal{U}$ be a fixed query. We first recall the following definition of important elements, and provide a definition for an ``important reset'':

\begin{definition} \label{def:inmportant-reset}
    Recall the following definition from \Cref{def:important}: for a fixed query $x \in \mathcal{U}$, we say a stream element $y$ is important if $y \leq x$. Also, we say a reset operation is important if, at the time of reset, at least one of the elements in the compactor is important. 
\end{definition}

\Cref{lem:compactor-reset} states the following:

\compactorreset*

Recall that the estimation error solely stems from \textit{important compactions} for query $x$ (See \Cref{def:important}). Let $N$ be the number of important compactions. Then, the error distribution $\Delta(x)$ is the sum of (at most) $N$ independent $\pm 1$ random variables. As a result, we have $\E\left[\Delta(x)^2\right] \leq N$. Therefore, it suffices show that $N \leq \frac{\rank_\pi(x) + t \cdot S}{k}$. Going forward, we will have a charging argument similar to that of~\cite{cormode2023relative}:

Initially all the blocks of the compactor are unmarked.
\begin{itemize}
    \item For each important compaction, we will ``mark'' an unmarked block. 
    \item A marked block will only be unmarked at a reset, or when k important elements are compacted from it. This allows us to upper bound the total number of marks by $\frac{\rank_\pi(x) + t \cdot S}{k}$. 
\end{itemize}
Combining the two bullets above, we obtain the desired upper bound for $N$ and conclude the proof of \Cref{lem:compactor-reset}.

\paragraph{Mark the elements.} Consider a call $\Call{Compact}{C, r + 1}$, for it to be important, there must be at least one important element (an element $y$ such that $y \leq x$) in blocks $B_{\geq r+1}$. We mark the block $B_{r}$ if $\Call{Compact}{C, r + 1}$ is important. We unmark block $B_i$ when one of the following happens:
\begin{enumerate}
\item There is a compaction call $\Call{Compact}{C, \ell}$ involving $B_i$ (i.e., $\ell \leq i$). 
\item There is a reset.
\end{enumerate}

Let us prove that in our resizes, whenever we call $\Call{Compact}{C, r + 1}$, the block $B_r$ must be unmarked before.

\begin{claim} \label{claim:mark-progress}
The block $B_i$ is marked only when $z_i = 1$.
\end{claim} 
\begin{proof}
Fisrt, whenver we call $\Call{Compact}{C, r + 1}$ and mark a block $B_r$, $z_r$ is always the least significant bit with $z_r = 1$. So when $B_i$ is marked, we must have $z_i = 1$. 

Second, when we change $z_i$ to $0$ in a resize call, suppose that the target number of blocks of the reset is $\ell$ and we call $\Call{Compact}{C, r + 1}$. If $B_i$ is not unmarked, we must have $i \leq r$. But all the bits $z_1, z_2, \dots, z_{r - 1}$ are not changed during this resize, and $z_r$ is changed to $1$. Therefore, $z_i$ must remains $1$ if $B_i$ is still marked. 
\end{proof}

\begin{cor} \label{coro:mark-once}
Whenever we call $\Call{Compact}{C, r + 1}$ (in the resize operation), the block $B_r$ must be unmarked before.
\end{cor}
\begin{proof}
In a resize, we first increase $z$ to a multiple of $2^{-\ell}$. Because $z_r$ is the least significant bit equal to $1$ after the increment, we know that before the increment, it must be the case that $z_r = 0$. Then the block $B_r$ must be unmarked by \Cref{claim:mark-progress}.
\end{proof}

\paragraph{Bound the number of marks} To bound the number of marks, we first formalize the following claim, which suggests that we can bound the numebr of marks by the number of important elements. 
\begin{claim} \label{claim:important-block}
A marked block contains only important elements.
\end{claim}
\begin{proof}
Initially, suppose that a block $B_r$ is marked because the compaction $\Call{Compact}{C, r + 1}$ is important. Because elements in $C$ are in sorted increasing order, all blocks $B_{\leq r}$ must also be smaller than or equal to $x$. Thus, $B_r$ contains only important elements. 

Afterward, if $B_r$ is not unmarked, it cannot be involved in any compaction. The only way the elements in $B_r$ can change is due to insertions. Because we always sort the elements in $C$ after insertions, the elements in $B_r$ can only monotonely become smaller. Thus $B_r$ still contains only important elements after the insertions.
\end{proof}
\begin{lem} \label{lem:mark-upperbonud}
Suppose there are $t$ important resets, the total number of blocks we mark is bounded by $\frac{\rank_\pi(x) + t\cdot S}{k}$ where $S$ is an upper bound on the number of elements in $C$. 
\end{lem}
\begin{proof}
For every block that is marked, it is either unmarked due to (1) a compaction or (2) a reset; or it stays marked at the end of the algorithm.

If it is unmarked due to compaction (or stays marked at the end of the algorithm). By \Cref{claim:important-block}, we know at the time of that compaction (or at the end of the algorithm) it must contain only important elements. These $k$ important elements are removed from $C$ during the compaction (or stays in $C$ till the end). This bounds the number of such marked blocks by $\frac{\rank_\pi(x)}{k}$.

If it is unmarked due to a reset, we note that during a reset, there are at most $\frac{S}{k}$ many blocks in $C$. So it unmarks at most $\frac{S}{k}$ blocks. There are $t$ important resets in total that umarks at most $\frac{t \cdot S}{k}$ blocks. For the unimportant resets, by \Cref{claim:important-block}, they never unmark any block.  

Adding these two together finishes the proof.
\end{proof}

Putting together \Cref{lem:mark-upperbonud} and \Cref{coro:mark-once}, we conclude that the number of important compactions can be at most $\frac{\rank_\pi(x) + t\cdot S}{k}$. So the final error is a sum of this many $\pm1$ Rademacher random variables. We conclude that the variance $\E[\Delta(x)^2] \leq \frac{\rank_\pi(x) + t\cdot S}{k}$.

\section{Analysis of the Full Sketch} \label{sec:detailed}
In this section, we fill in the gaps in our algorithm analysis: we first finish the upper bound on the expected square error in \Cref{sec:expectation}. Then, we show that our algorithm succeeds with high probability in \Cref{sec:high-prob}. Furthermore, to improve the space complexity from $\tilde{O}(\epsilon^{-1} \log n)$ to $\tilde{O}(\epsilon^{-1} \log (\epsilon n))$, we make a small modification to our algorithm, which is presented in \Cref{sec:optimize-space}. Finally, we finish the analysis of our dynamic space allocation rules and prove the final space complexity bound in \Cref{sec:space-allocation}.

\subsection{Expectation Bound} \label{sec:expectation}
In this subsection, we first provide the error analysis for a single compactor hierarchy $H$ which solves the top quantiles problem (i.e. supporting queries of rank at most $R$). Later, we generalize our analysis to our overall algorithm for the relative-error quantile estimation problem, in which sub-sketches $H_i$ also have ``reset'' operations and support arbitrary queries. 

To address the first objective above, we recall \Cref{lem:H}, which gives an upper-bound on the expected squared error of the top-$R$ quantiles sketch $H$ defined in section \Cref{sec:top-quantile}.

\hierarchy*
\begin{proof}

We first decompose the error. Again let $C_0, \dots, C_{\log(1/\epsilon) - 1}$ be the compactors in $H$, and $C_{\log (1/\epsilon)}$ denotes the buffer. We use $\pi_j$ to denote the input stream of $C_j$ (which is the output stream of $C_{j - 1}$). 
\begin{align}
    \Call{Rank}{}_H(x) - \rank_\pi(x) = &\sum_{j = 0}^{\log(1/\epsilon)} \Call{Rank}{}_{C_j}(x) \cdot {2^j} \cdot {\epsilon^2 \cdot R} - \rank_\pi(x) \notag \\
    = &\sum_{j = 0}^{\log(1/\epsilon) - 1} \left((\Call{Rank}{}_{C_j}(x)  + 2\rank_{\pi_{j + 1}}(x)) - \rank_{\pi_j}(x) \right) \cdot {2^j} \cdot {\epsilon^2 R} \notag \\
    & + \left(\rank_{\pi_0}(x) \cdot \epsilon^2 R - \rank_{\pi}(x)\right) \notag \\
    & + \left(\Call{Rank}{}_{C_{\log(1/\epsilon)}}(x) - \rank_{\pi_{\log(1/\epsilon)}}(x) \right) \cdot {\epsilon R} .\label{equ:err-decomp}
\end{align}
The first term is the error of each compactor $C_j$. The second one is the error of the sampler. Third one is the buffer. 

We first argue that the buffer (the third term) never contributes to the total error: 
\begin{itemize}
    \item If $\rank_{\pi_{\log(1/\epsilon)}}(x) \leq 1/\epsilon$, because the buffer always keep the smallest $1 / \epsilon$ elements, we always have $\rank_{C_{\log(1/\epsilon)}}(x) = \rank_{\pi_{\log(1/\epsilon)}}(x)$. Thus the third term is zero. 
    \item If $\rank_{\pi_{\log(1/\epsilon)}}(x) > 1/\epsilon$, intuitively, because $\rank_\pi(x)\leq R$, the compactors already over estimate the rank of $x$. By only keep the first $1/\epsilon$ elements, the buffer only reduces the error. 
    
    Formally, the sum of the first two terms equals  $$\left(\sum_{j = 0}^{\log(1/\epsilon) - 1}\Call{Rank}{}_{C_j}(x) \cdot {2^j}\cdot {\epsilon^2  R} + \rank_{\pi_{\log(1/\epsilon)}}(x) \cdot {\epsilon R} \right)- \rank_{\pi}(x).$$ This sum, because $\rank_{\pi}(x) \leq R$, is at least $\left(\rank_{\pi_{\log(1/\epsilon)}}(x) - 1/\epsilon\right) \cdot \epsilon R$. Thus as the buffer $C_{\log(1/\epsilon)}$ keeps the first $1/\epsilon$ elements and throw away the rest. It is negative and of a smaller magnitude than the first two terms. So it can only reduce the error. 
\end{itemize}

Second, let us analyze the first term. We apply \Cref{lem:compactor} to each compactor $C_j$. We need to verify that the condition (\Cref{equ:compactor-cond}) holds. For $\Call{Resize}{C_j, \cdot}$, we only call it when we resize $H$. For $\Call{Insert}{C_j, \cdot}$, we call it once only when either there is a resize of $H$ or an insertion into $C_0$. 
Thus we have $$\eqref{equ:compactor-cond} \leq 2 \cdot \sum_{t=1}^\ell 2^{-s_t / k} = 2 \cdot \sum_{t=1}^\ell 2^{-\epsilon \cdot s_t}  \leq 0.5.$$
Let $\pi_j$ be the input stream of $C_j$ and $$\Delta_j(x) = (\Call{Rank}{}_{C_j}(x) + 2\rank_{\pi_{j + 1}}(x)) - \rank_{\pi_j}(x)$$ be the error of $C_j$ as defined in \Cref{equ:Error-Compactor}. As $\E[\rank_{\pi_j}(x)] = \frac{\rank_{\pi}(x)}{\epsilon^2 R \cdot 2^j} \leq \frac{1}{\epsilon^2 \cdot 2^j}$ (because $\rank_{\pi}(x) \leq R$), we know (from \Cref{lem:compactor}) that, 
$$\E\left[\Delta_j(x)^2\right]^{1/2} \leq \E\left[\frac{\rank_{\pi_j}(x)}{k}\right]^{1/2} \leq  \sqrt{\frac{ 1}{k \epsilon^2 \cdot 2^j}} = \sqrt{\frac{ 1}{ \epsilon \cdot 2^j}}.$$
The contribution from the first term is therefore (by \Cref{fact:stadnard-deviation}),    $$\sum_{j=0}^{\log_2(1/\epsilon) - 1} \epsilon^2 R \cdot 2^j \cdot \E\left[\Delta_j(x)^2\right]^{1/2} \leq \sum_{j=0}^{\log_2(1/\epsilon) - 1} \epsilon^2 R \cdot 2^j \cdot \sqrt{\frac{ 1}{ \epsilon \cdot 2^j}} = O(\epsilon) \cdot R.$$

Thirdly, we will also verify the error of the sampler (the second term) is $O(\epsilon) \cdot R$. This is because we sample with probability $p = \frac{1}{\epsilon^2 R}$. We have that the rank in the sampled stream $\rank_{\pi_0}(x) = \sum_{t = 1}^{\rank_\pi(x)} \mathrm{Bernoulli}(p)$. Thus 
    $$\E\left[\left(\rank_{\pi_0}(x) \cdot \epsilon^2 R - \rank_{\pi}(x)\right)^2\right]^{1/2} \leq \epsilon^2 R \cdot \sqrt{R \cdot p (1 - p)} \leq \epsilon \cdot R.$$

Finally, we use \Cref{fact:stadnard-deviation} again to add these parts together. This concludes our proof.
\end{proof}

We can extend this proof to our subsketch $H_i$'s which have error guarantee for arbitrary queries and can handle resets. Recall \Cref{claim:sub-sketch-reset} which is for our subsketch $H_i$'s.

\subsketch*

\begin{proof}
Recall that we use $\pi_{H_i}$ to denote the input stream to $H_i$ and $\pi'_{H_i}$ to denote the elements that we move from $H_i$ to $H_{i + 1}$. Further more, we use $\pi_{i,j}$ to denote the input stream of compactor $C_{i,j}$ and $\pi'_{i,j}$ to denote the output stream. Specially, $\pi^{\mathrm{remove}}_{i,j}$ denotes the stream of maximum elements that we removed from $C_{i,j}$. Note that each element in $\pi^{\mathrm{remove}}_{i,j}$ is added $2^j \cdot \epsilon^2 R$ times to $\pi'_{H_i}$ (Line~\ref{line:output-with-multiplicity}, \Cref{alg:Hi-weight}). 

We perform a similar decomposition of error that is slightly more complicated than \eqref{equ:err-decomp} and get
\begin{align*}
\Delta_{H_i}(x) =& \Call{rank}{}_{H_i}(x) + \rank_{\pi'_{H_i}}(x) - \rank_{\pi_{H_i}}(x) \\
= &  \sum_{j=0}^{\log(1/\epsilon)} \left(\Call{Rank}{}_{C_{i,j}}(x) + \rank_{\pi^\mathrm{remove}_{i,j}}(x)\right) \cdot 2^j \cdot \epsilon^2 R_i - \rank_{\pi_{H_i}}(x) \\
= & \sum_{j=0}^{\log(1/\epsilon) - 1} \left( (\Call{Rank}{}_{C_{i,j}}(x) + 2 \rank_{\pi_{i,j+1}}(x) + \rank_{\pi^\mathrm{remove}_{i,j}}(x)) - \rank_{\pi_{i,j}}(x) \right) \cdot 2^j \cdot \epsilon^2 R_i\\
&+ \left(\rank_{\pi_{i,0}}(x) \cdot \epsilon^2 R - \rank_{\pi_{H_i}}(x)\right)\\
&+ \left((\Call{Rank}{}_{C_{i, \log(1/\epsilon)}}(x) + \rank_{\pi^\mathrm{remove}_{i,\log(1/\epsilon)}}(x)) - \rank_{\pi_{i,\log(1/\epsilon)}}(x)\right) \cdot \epsilon R 
\end{align*}

In this case, the error from the buffer (the third term) is identically zero. This is because the buffer can store $3/\epsilon$ many elements each of weight $\epsilon R_i$, and we keep the total weight in $H_i$ below $3 R_i$. As a result, the buffer is never full. The elements in the buffer ($C_{i,\log(1/\epsilon)}$) plus the elements remove from the buffer ($\pi^\mathrm{remove}_{i,\log(1/\epsilon)}$) equal exactly the elements that are ever inserted into the buffer~($\pi_{i,\log(1/\epsilon)}$). 

Then for the first term, we use \Cref{lem:compactor-reset} instead of \Cref{lem:compactor}. We plug the following definition from \Cref{equ:compactor-with-removal-error} into \Cref{lem:compactor-reset}.
$$\Delta_{i,j}(x) \coloneqq (\Call{Rank}{}_{C_{i,j}}(x) + 2 \rank_{\pi_{i,j+1}}(x) + \rank_{\pi^\mathrm{remove}_{i,j}}(x)) - \rank_{\pi_{i,j}}(x).$$
Also we note because the total weight is at most $3R_i$, each compactor $C_{i,j}$ has at most $\frac{3R_i}{2^j \cdot \epsilon^2 \cdot R_i} = \frac{3}{2^j \cdot \epsilon^2}$ elements in it. Same as the previous proof, we have $\E[\rank_{\pi_{i,j}}(x)] \leq \frac{\rank_{\pi_{H_i}}(x)}{2^j \cdot \epsilon^2 R_i}$.

We get the contribution from the first term satisfies (using \Cref{fact:stadnard-deviation}),
\begin{align*}
\sum_{j=0}^{\log(1/\epsilon) - 1} \epsilon^2 R_i \cdot 2^j \cdot \E\left[\Delta_{i,j}(x)^2\right]^{1/2} &\leq \sum_{j=0}^{\log(1/\epsilon) - 1} 
 \epsilon^2 R_i \cdot 2^j \cdot \E\left[\frac{\rank_{\pi_{i,j}}(x) + t_i(x) \cdot \frac{3}{2^j \cdot \epsilon^2}}{k} \right]^{1/2} \\
&\leq \sum_{j=0}^{\log(1/\epsilon) - 1} 
 \epsilon^2 R_i \cdot 2^j \cdot \left(\frac{\rank_{\pi_{H_i}}(x)}{2^j \cdot \epsilon^2 R_i \cdot k } + \frac{3 t_i(x)}{2^j \cdot \epsilon^2 \cdot k}\right)^{1/2} \\
&= O(\epsilon)\cdot \sqrt{R_i \cdot \rank_{\pi_{H_i}}(x) + t_i(x) \cdot R_i^2}.
\end{align*}
Here in the last step, we plug in $k = 1/\epsilon$ and sum over all $j$. 

Finally, for the second term, we have that the rank in the sampled stream $\rank_{\pi_{i,0}}(x) = \sum_{t = 1}^{\rank_{\pi_{H_i}}(x)} \mathrm{Bernoulli}(p_i)$ where $p_i = \frac{1}{\epsilon^2 R_i}$. Thus, 
$$\E\left[\left(\rank_{\pi_{i,0}}(x) \cdot \epsilon^2 R_i - \rank_{\pi_{H_i}}(x)\right)^2\right]^{1/2} \leq \epsilon^2 R_i \cdot \sqrt{\rank_{\pi_{H_i}}(x) \cdot p (1 - p)} \leq \epsilon \cdot \sqrt{R_i \cdot \rank_{\pi_{H_i}}(x)}.$$

We conclude the proof by applying \Cref{fact:stadnard-deviation} to sum up the contribution from different parts. 
\end{proof}

Now that we have error bounds for the $H_i$'s, \Cref{lem:final-analysis} tells us, for any query $x$, the total error of the algorithm is at most
$$\E\left[\Big|\rank_{\pi}(x) - \sum_{i=0}^{\log_2(\epsilon n)} \Call{Rank}{}_{H_i}(x)\Big|^2\right]^{1/2} \leq O(\epsilon) \cdot \rank_{\pi}(x).$$
This finishes the error analysis.

\subsection{Extremely High Probability Bound} \label{sec:high-prob}

Now let us prove that our algorithm answers a single query $x$ approximately correct with high probability. Note from the expectation bound and Chebyshev's inequality, we know that our algorithm outputs an estimate that is within $(1 \pm \epsilon)\rank_{\pi}(x)$  with constant probability. The  na\"ive amplification of maintaining $\log(1/\delta)$ copies of our algorithm in parallel and outputting their median already gives an algorithm that succeeds with $1 - \delta$ probability. 

In the following, we will change a few parameters of our algorithm and focus on getting an extremely high probability bound (double logarithmic dependency on $\delta$). If combined with a small optimization which we will explain in \Cref{sec:optimize-space}, our final space complexity will be $$\tilde{O}(\epsilon^{-1} \cdot \log (\epsilon n) \cdot (\log \log 1/\delta)^3).$$
The analysis uses the idea of analyzing top compactors deterministically, which first appears in~\cite{karnin2016optimal}. 

\begin{remark}[Connection to deterministic algorithms]
We note that the previous work~\cite{cormode2023relative}, also gives an extremely high probability bound with space $$O(\epsilon^{-1} \cdot \log^2(\epsilon n) \cdot \log \log (1/\delta)).$$ 
Our algorithm improves the previous bound for not-too-small $\delta$. 

In comparison model, there are at most $n!$ many possible input streams $\pi$. So, for a single query, if the success probability $\delta < 1 / n!$ (that is, $\log \log (1/\delta) = O(\log n)$), there must exist a fixing of the randomness that works for all possible input streams $\pi$. Then, by selecting $\delta < 1/n!$, we can obtain  a deterministic algorithm. Using this connection, the previous algorithm implies the same $\tilde{O}(\epsilon^{-1} \cdot \log^3 (\epsilon n))$ space deterministic upper bound, matching the state-of-the-art result of Zhang and Wang~\cite{zhang2007efficient}. 

However, since our improvement is conditioned on $\delta$ being not-too-small, we are not able to directly improve upon the known upper bound for deterministic algorithms; in fact, our bound is actually worse in this extreme setting. Thus, improving the upper bound for deterministic quantile estimation in the relative-error regime remains an interesting open problem. We will discuss this more in the Open Problem section (\Cref{sec:open-prob}). 
\end{remark}

Now let us specify the parameter changes for each $H_i$. 
\begin{itemize}
    \item We will set the block length of compactors $k = c \cdot \frac{ (\log \log (1/ \delta))^2}{\epsilon}$ with a large enough constant $c$ (say $c = 128$) throughout the algorithm. 
    \item For the sampler, we adjust the sampling probability of each sampler to $\frac{c \cdot \log (1/\delta)}{\epsilon^2 \cdot R_i}$, as a result, now each element outputted by the sampler has weight $\frac{\epsilon^2 \cdot R_i}{c \cdot \log (1/\delta)}$. 
    \item We set $h_i \coloneqq \log \log (1/\delta) + \log(c / (k\epsilon^2))$. For each $0 \leq j < h_i$, the compactor $C_{i,j}$ has elements of weight $w_{i,j} \coloneqq \frac{\epsilon^2 \cdot R_i \cdot 2^j}{c \cdot \log(1/\delta)}$. 
    \item The buffer now has size $3 k$, each element in the buffer (which we denote by $C_{i,h_i}$) has weight $w_{i, h_i} = \frac{\epsilon^2 \cdot R_i \cdot 2^{h_i}}{c \cdot \log(1/\delta)} = \frac{R_i}{k}$. 
\end{itemize}
The rest of the algorithm stays the same.

In our analysis we define $h'_i = h_i - \log \log (1/\delta) = \log(1/(k\epsilon^2))$. For levels $0 \leq j \leq h'_i$, we are going to analyze compactor $C_{i,j}$'s using Chernoff bounds. But, for the top level compactors with $h'_i < j \leq h_i$, we analyze the error deterministically as done in~\cite{karnin2016optimal} and \cite{cormode2023relative}. We defer the full analysis of the high probability bound to \Cref{appendix:high-probability}.

\subsection{Optimizing Space with Batch Insertions} \label{sec:optimize-space}

The algorithm we described in \Cref{sec:algo-description} gives a relative-error solution to the quantile estimation problem using space $\tilde{O}(\epsilon^{-1} \log n)$ with constant success probability. In this subsection, we add a simple optimization to our algorithm which reduces the space to $\tilde{O}(\epsilon^{-1} \log (\epsilon n))$. In particular, consider the following adjustment to our original sub-sketches $H_i$, which were described in \Cref{sec:algo-description}: 

\begin{itemize}
    \item For each $H_i$, we create an additional ``temporary'' buffer $B'_{i}$ which will store incoming stream elements \textit{before} they are passed as a ``batch'' to the sampler. We set the batch-size $|B_i'| = 1/\epsilon$. 

    \item To insert an element $x$ into $H_i$, we first place $x$ into the temporary buffer $B_i'$, and once $B_i'$ becomes full, we pass all of the elements stored in $B_i'$ to the sampler (as a batch). The sampler operates in the same way as defined previously, i.e. it samples the elements in $B'_i$ with probability $\frac{1}{\epsilon^2 R_i}$ (or alternatively, with probability $\frac{c \cdot \log(1/\delta)}{\epsilon^2 R_i}$ to obtain an extremely high-probability bound), and then inserts the sampled elements into the first level compactor $C_{i,0}$ together. 
\end{itemize}

Now, recall that in \Cref{sec:dynamic-space}, we defined a single time step to correspond to an insertion operation to $C_{i,0}$ or a resize of $H_i$. Also, recall that operation $\Call{Insert}{C_{i, 0}, x_1, x_2, \dots, x_m}$ can handle $m \leq s$ elements in one step where $s$ is the current space for $C_{i,0}$. Since we never resize $C_{i,0}$ to a have space smaller than $1 / \epsilon$ (see e.g. \Cref{equ:general-interval-space}), the elements from $B'_i$ can be inserted together in one time step even when they are all sampled. 

As a result of the observation above, the total number of time steps goes from $\poly(n)$ to $\poly(\epsilon n)$. We will see in \Cref{sec:space-allocation} that this reduces the space to $\tilde{O}(\epsilon^{-1} \log (\epsilon n))$ as claimed.

\subsection{Space Allocation Analysis} \label{sec:space-allocation}

In this section, we will assume the optimization in \Cref{sec:optimize-space} is applied so that the total number of time steps is $T = \poly(\epsilon n)$. When proving the total space is bounded, we now aim for a space bound of $\tilde{O}(\epsilon^{-1} \log (\epsilon n))$. 

\paragraph*{Analysis for offline general intervals.} Recall that $t_{i,j}$ denotes the time step at which $H_i$ resets for the $j$-th time. We defined the potential function  $\phi_{i,j} = 1$ for $i = \log(\epsilon n) + 1$ and $\phi_{i,j} = \sum_{m = \ell}^{r - 1} \phi_{i + 1, m}$ when 
$[t_{i,j}, t_{i,j + 1})$ intersects with ``children'' intervals $[t_{i + 1, \ell}, t_{i + 1, \ell + 1}]$, $[t_{i + 1, \ell + 1}$, $t_{i + 1, \ell + 2}]$. Also the choice of space parameters $s_{i,t} = \epsilon^{-1} \cdot \left( \log \frac{\phi_{i,j}}{\phi_{i + 1, m}} +5\log(1/\epsilon)\right).$
    
We will need \Cref{claim:ins-interval} and \Cref{claim:potential-bound}, which we recall below:

\insinterval*
\potentialbound*

We need to verify the two constraints are satisfied:

\begin{itemize}
    \item (The total space is bounded.) Fix a time $t$. We also let $j_i$ be the index of the interval $[t_{i,j_i}, t_{i,j_i + 1})$ that contains $t$. 
    
    From a similar telescoping sum as the tree-like case, we have 
    \begin{align*}
        \sum_{i=0}^{\log (\epsilon n)} s_{i,t} &= \sum_{i=0}^{\log(\epsilon n)}   \epsilon^{-1} \cdot \left( \log \frac{\phi_{i,j_i}}{\phi_{i + 1, j_{i + 1}}}  + 5\log(1/\epsilon)\right) \\
        &= \epsilon^{-1} \cdot \log_2 \frac{\phi_{0,j_0}}{1} + \epsilon^{-1} \cdot \log(\epsilon n) \cdot 5 \log(1/\epsilon)
    \end{align*}
    To bound the total space,$\lceil \log_2(1/\epsilon) \rceil \cdot \sum_{i=0}^{\log(\epsilon n)} s_{i,t}$\footnote{Note each hierarchy $H_i$ has $\lceil \log_2(1/\epsilon) \rceil$ compactors in it and uses space $\lceil \log_2(1/\epsilon) \rceil \cdot s_{i,t}$}, by $O(\epsilon^{-1} \log (\epsilon n)  \log^2(1/\epsilon))$, we only need that $\phi_{0,j_0} = \poly(\epsilon n)$, which is exactly \Cref{claim:potential-bound} with $T = \poly(\epsilon n)$.  
    \item (The space sequence is feasible.) Similar as the tree-like case, we consider any interval $[t_{i,j}, t_{i,j + 1})$. Suppose it now intersects with children  $[t_{i + 1, \ell}, t_{i + 1, \ell + 1})$, $[t_{i + 1, \ell + 1}$, $t_{i + 1, \ell + 2})$, $\dots$, $[t_{i + 1, r - 1}, t_{i + 1, r})$. Again by \Cref{claim:ins-interval}, we know that $|W_{i,j} \cap [t_{i + 1, m}, t_{i + 1, m + 1})| \leq 3/\epsilon^2 + 2$. Then,
        \begin{align*}
        \sum_{t \in W_{i,j}} 2^{-\epsilon \cdot s_{i,t}} & \leq \sum_{m = \ell}^{r - 1} \sum_{\substack{t \\ t \in W_{i,j} \cap [t_{i + 1, m}, t_{i + 1, m + 1})}} 2^{- \epsilon \cdot s_{i,t}} \\
        &\leq \sum_{m =\ell}^{r - 1} (3 / \epsilon^2 + 2) \cdot \frac{\phi_{i + 1,m}}{\phi_{i,j}} \cdot \epsilon^5 \leq 0.25 \cdot \frac{\sum_{m =\ell}^{r - 1} \phi_{i + 1,m}}{\phi_{i,j}}  \leq 0.25.
    \end{align*}
    Thus, the space we allocate satisfies the premise of \Cref{lem:H} (\Cref{equ:condition-H}). 
\end{itemize}

\paragraph*{Analysis for online space allocation.} Recall that for all unfinished intervals, we pretend that the interval ends (i.e. the corresponding sub-sketch $H_i$ resets) at the \textit{current} time $t$; then, we  calculate all the potentials which we denote by $\phi^{(t)}_{i,j}$. We then round them up to the closest power of $2$, which we denote by $\cceil{\phi^{(t)}_{i,j}}$, and let $\widehat{s}_{i, t} = \epsilon^{-1} \cdot \left( \log \frac{\cceil{\phi^{(t)}_{i,j}}}{\cceil{\phi^{(t)}_{i + 1, m}}}  + 5\log(1/\epsilon) + 5 \log \log n\right).$

To see why it works, we again need to verify the two constraints:

\begin{itemize}
    \item (The total space is bounded.) This follows from exactly the same telescoping sum as before with $\phi_{i,j}$'s replaced by $\phi^{(t)}_{i,j}$'s, and the total space is now $O(\epsilon^{-1} \cdot \log(1/\epsilon) \cdot \log (\epsilon n) (\log(1/\epsilon) + \log \log n))$. 
    \item (The space sequence is feasible.) This is the more subtle part. Again suppose interval $[t_{i,j}, t_{i,j + 1})$ intersects with children  $[t_{i + 1, \ell}, t_{i + 1, \ell + 1})$, $[t_{i + 1, \ell + 1}$, $t_{i + 1, \ell + 2})$, $\dots$, $[t_{i + 1, r - 1}, t_{i + 1, r})$. First, we need to prove that we do not have too many resizes within each children interval. That is, 
    \begin{equation} \label{equ:online-allocation-intersection}
    |W_{i,j} \cap [t_{i + 1, m}, t_{i + 1, m + 1})| \leq 3/\epsilon^2 + 2 + O(\log^2 (\epsilon n))    
    \end{equation}
    
    This is because the only changing term in $\widehat{s}_{i,t}$ is $\log \frac{\cceil{\phi^{(t)}_{i,j}}}{\cceil{\phi^{(t)}_{i + 1, m}}}$. Within each intersection $[t_{i,j}, t_{i,j + 1}) \cap [t_{i + 1, m}, t_{i + 1, m + 1})$, both numerator and denominator are increasing, and by \Cref{claim:potential-bound} have at most $O(\log (\epsilon n))$ different values. This means we have at most $O(\log^2 (\epsilon n))$ extra resizes due to online allocation. This will be offset by the extra $5 \log \log n$ term in $\widehat{s}_{i,t}$. 

    Second, we need to upper bound 
    \begin{align*}
        \sum_{t \in W_{i,j}} 2^{-\epsilon \cdot s_{i,t}} & \leq \sum_{m = \ell}^{r - 1} \sum_{\substack{t \\ t \in W_{i,j} \cap [t_{i + 1, m}, t_{i + 1, m + 1})}} 2^{- \epsilon \cdot s_{i,t}} \\
        &= \sum_{m = \ell}^{r - 1} \sum_{\substack{t \\ t \in W_{i,j} \cap [t_{i + 1, m}, t_{i + 1, m + 1})}} \frac{\cceil{\phi^{(t)}_{i + 1, m}}}{\cceil{\phi^{(t)}_{i,j}}} \cdot \epsilon^{-5} \cdot (\log n)^{-5}
    \end{align*}
    Note $\cceil{\phi^{(t)}_{i,j}}$ is monotone in $t$ and has $O(\log (\epsilon n))$ many different values. For each possible value $2^x$, we let $[a(x), b(x)] \subseteq [t_i, t_{j + 1})$ be the time interval such that $\cceil{\phi^{(t)}_{i,j}} = 2^x$, and suppose it intersects with children $[t_{i + 1, \ell(x)}, t_{i + 1, \ell(x) + 1})$, $[t_{i + 1, \ell(x) + 1}, t_{i + 1, \ell(x) + 2})$, $\dots$, $[t_{i + 1, r(x) - 1}, t_{i + 1, r(x)})$.
    
    Then for every $0 \leq x \leq O(\log(\epsilon n))$, we have
    \begin{align*}
    &\sum_{m=\ell(x)}^{r(x)- 1} \sum_{\substack{t  \\ t \in W_{i,j} \cap [t_{i + 1, m}, t_{i + 1, m + 1})\\ \cap [a(x), b(x)]}}\frac{\cceil{\phi^{(t)}_{i + 1, m}}}{\cceil{\phi^{(t)}_{i,j}}} \cdot \epsilon^{-5} \cdot (\log n)^{-5}  \\
    \leq &\sum_{m=\ell(x)}^{r(x) - 1} \left(3/\epsilon^2 + 2 + O(\log^2 (\epsilon n))\right) \cdot \frac{2 \cdot \phi^{(b(x))}_{i + 1, m}}{2^x} \cdot \epsilon^{-5} \cdot (\log n)^{-5} \tag{Here we use $|W_{i,j} \cap [t_{i + 1, m}, t_{i + 1, m + 1})| \leq 3/\epsilon^2 + 2 + O(\log^2 (\epsilon n))$.} \\
    \leq &0.25 \cdot \frac{\sum_{m = \ell(x)}^{r(x) - 1} \phi^{b(x)}_{i + 1, m}}{2^x} \leq 0.25
    \end{align*}
    Here the last step is because, by definition of $\cceil{\cdot}$, we know that $\phi^{b(x)}_{i,j} \leq 2^x$. On the other hand, we know $\phi^{b(x)}_{i,j}  \geq \sum_{m=\ell(x)}^{r(x) - 1} \phi^{b(x)}_{i + 1, m}$. Thus, the space sequence is always feasible. 
\end{itemize}

This finishes our proof. For the space allocation of the algorithm with extremely-high success probability, see \Cref{appendix:high-probability}.

\section{Lower and Upper bounds for the Top-$R$ Quantiles Problem} \label{sec:lb} \label{sec:further}

\subsection{Lower bound}

This subsection is devoted to the proof of \Cref{lem:lb-top-quantile}. We first recall its statement:

\lbtopquantile*
\newcommand{\mem}{\mathcal{M}}
\newcommand{\alg}{\mathcal{A}}
The idea is to construct the hard input stream $\pi = x_1 x_2 \cdots x_n$ recursively, randomly, and adaptively. For any (possibly randomized) algorithm $\mathcal{A}$, we use $\mem_t$ to denote the set of elements in its memory after reading $x_t$. We say that $\mathcal{A}$ remembers an element $e$ at time $t$ if and only if $\Pr[e \in \mem_t] \geq 1/2$ where the probability is taken both over the randomized algorithm and the potentially randomized input stream. 

\paragraph{Construction of the hard input stream.} For any randomized algorithm $\mathcal{A}$, we can construct a length $2^k - 1$ (partial) input stream $\pi_k^\alg$ as follows. Suppose here $A$ takes $n$ elements as input and we allow $n$ to be larger or equal to $2^k - 1$.
\begin{itemize}
    \item We first insert one element $e$ as the first element in $\pi^\alg_k$. 
    \item Let $\alg'$ be the algorithm $\alg$ with the first input hard-coded to $e$. It takes $n - 1$ inputs. We insert the stream $\pi^{\alg'}_{k - 1}$ and let all these elements be strictly larger than $e$. 
    \item Now we have two cases:
    \begin{itemize}
        \item Case 1: The algorithm $\alg$ \emph{remembers} $e$ (by the definition above) at this time ($1+|\pi^{\alg'}_{k-1}| = 2^{k - 1}$). In this case, we simply insert arbitrary $2^{k-1} - 1$ elements that are larger than element $e$ and those elements in $\pi^{\alg'}_{k-1}$.
        \item Case 2: Otherwise, we let $\alg''$ be the algorithm with input prefix hard-coded to the length-$2^{k -1}$ partial stream we have constructed so far. 
        We then insert $\pi^{\alg''}_{k - 1}$. With probability $1/2$, we let elements in it be greater than $e$ but smaller than elements in $\pi^{\alg'}_{k-1}$. With probability $1/2$, we let elements in $\pi^{\alg'}_{k-1}$ be smaller than $e$. 
    \end{itemize}
\end{itemize}

\begin{figure}
    \centering
\tikzset{every picture/.style={line width=0.75pt}} %
\begin{subfigure}{0.5\textwidth}
\begin{tikzpicture}[x=0.75pt,y=0.75pt,yscale=-1,xscale=1]

\draw    (51,245) -- (297,245) ;
\draw [shift={(299,245)}, rotate = 180] [color={rgb, 255:red, 0; green, 0; blue, 0 }  ][line width=0.75]    (10.93,-3.29) .. controls (6.95,-1.4) and (3.31,-0.3) .. (0,0) .. controls (3.31,0.3) and (6.95,1.4) .. (10.93,3.29)   ;
\draw    (74,265) -- (74,127) ;
\draw [shift={(74,125)}, rotate = 90] [color={rgb, 255:red, 0; green, 0; blue, 0 }  ][line width=0.75]    (10.93,-3.29) .. controls (6.95,-1.4) and (3.31,-0.3) .. (0,0) .. controls (3.31,0.3) and (6.95,1.4) .. (10.93,3.29)   ;
\draw   (116,204) -- (91,204) -- (91,224) -- (116,224) -- cycle ;
\draw   (116,185) -- (204,185) -- (204,204) -- (116,204) -- cycle ;
\draw   (204,166) -- (292,166) -- (292,185) -- (204,185) -- cycle ;

\draw (96,210) node [anchor=north west][inner sep=0.75pt]   [align=left] {$\displaystyle e$};
\draw (142,185) node [anchor=north west][inner sep=0.75pt]   [align=left] {$\displaystyle \pi^{\alg'}_{k-1}$};
\draw (215,170) node [anchor=north west][inner sep=0.75pt]   [align=left] {{\fontfamily{pcr}\selectfont {\footnotesize arbitrary}}};
\draw (246,245) node [anchor=north west][inner sep=0.75pt]   [align=left] {{\fontfamily{pcr}\selectfont {\large time}}};
\draw (5,118) node [anchor=north west][inner sep=0.75pt]   [align=left] {{\fontfamily{pcr}\selectfont larger }\\{\fontfamily{pcr}\selectfont element}};
\draw (150,103) node [anchor=north west][inner sep=0.75pt]   [align=left] {Case 1};
\end{tikzpicture}
\end{subfigure}
\\
\begin{subfigure}{0.5\textwidth}
\tikzset{every picture/.style={line width=0.75pt}} %
\begin{tikzpicture}[x=0.75pt,y=0.75pt,yscale=-1,xscale=1]

\draw    (347,244) -- (606,244) ;
\draw [shift={(608,244)}, rotate = 180] [color={rgb, 255:red, 0; green, 0; blue, 0 }  ][line width=0.75]    (10.93,-3.29) .. controls (6.95,-1.4) and (3.31,-0.3) .. (0,0) .. controls (3.31,0.3) and (6.95,1.4) .. (10.93,3.29)   ;
\draw    (370,264) -- (370,126) ;
\draw [shift={(370,124)}, rotate = 90] [color={rgb, 255:red, 0; green, 0; blue, 0 }  ][line width=0.75]    (10.93,-3.29) .. controls (6.95,-1.4) and (3.31,-0.3) .. (0,0) .. controls (3.31,0.3) and (6.95,1.4) .. (10.93,3.29)   ;
\draw   (412,204) -- (387,204) -- (387,224) -- (412,224) -- cycle ;
\draw   (413,165) -- (501,165) -- (501,184) -- (413,184) -- cycle ;
\draw   (500,184) -- (588,184) -- (588,203) -- (500,203) -- cycle ;
\draw   (499,225) -- (587,225) -- (587,244) -- (499,244) -- cycle ;
\draw    (610,208) -- (592.57,194.24) ;
\draw [shift={(591,193)}, rotate = 38.29] [color={rgb, 255:red, 0; green, 0; blue, 0 }  ][line width=0.75]    (10.93,-3.29) .. controls (6.95,-1.4) and (3.31,-0.3) .. (0,0) .. controls (3.31,0.3) and (6.95,1.4) .. (10.93,3.29)   ;
\draw    (609,223) -- (594.7,231.94) ;
\draw [shift={(593,233)}, rotate = 327.99] [color={rgb, 255:red, 0; green, 0; blue, 0 }  ][line width=0.75]    (10.93,-3.29) .. controls (6.95,-1.4) and (3.31,-0.3) .. (0,0) .. controls (3.31,0.3) and (6.95,1.4) .. (10.93,3.29)   ;

\draw (392,210) node [anchor=north west][inner sep=0.75pt]   [align=left] {$\displaystyle e$};
\draw (437,164) node [anchor=north west][inner sep=0.75pt]   [align=left] {$\displaystyle \pi^{\alg'}_{k-1}$};
\draw (542,244) node [anchor=north west][inner sep=0.75pt]   [align=left] {{\fontfamily{pcr}\selectfont {\large time}}};
\draw (302,117) node [anchor=north west][inner sep=0.75pt]   [align=left] {{\fontfamily{pcr}\selectfont larger }\\{\fontfamily{pcr}\selectfont element}};
\draw (528,224) node [anchor=north west][inner sep=0.75pt]   [align=left] {$\displaystyle \pi^{\alg''}_{k-1}$};
\draw (527,183) node [anchor=north west][inner sep=0.75pt]   [align=left] {$\displaystyle \pi^{\alg''}_{k-1}$};
\draw (451,103) node [anchor=north west][inner sep=0.75pt]   [align=left] {Case 2};
\draw (612,195) node [anchor=north west][inner sep=0.75pt]   [align=left] {{\fontfamily{pcr}\selectfont with equal }\\{\fontfamily{pcr}\selectfont probability}};
\end{tikzpicture}
\end{subfigure}

    \caption{Construction of the hard stream $\pi^\alg_k$. }
    \label{fig:hard-stream}
\end{figure}
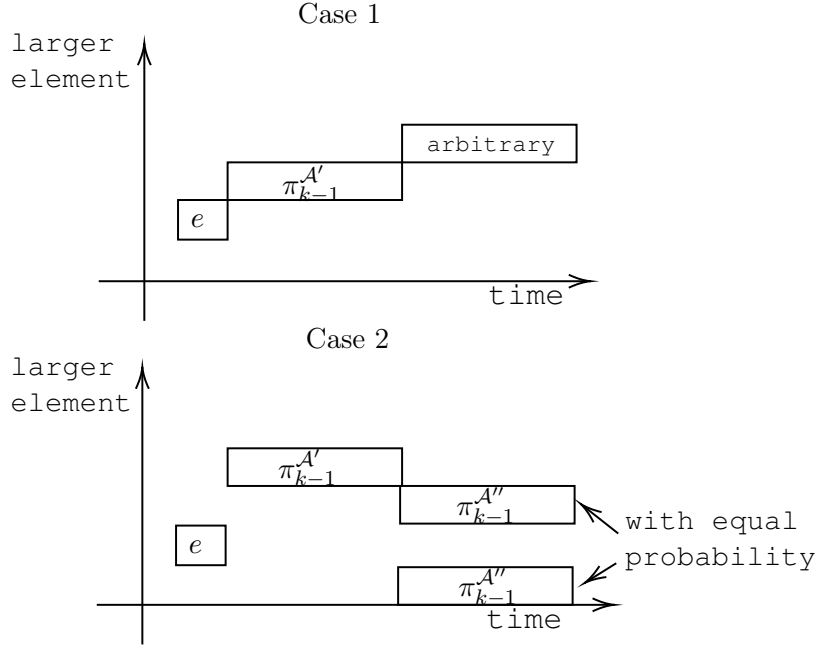

In the base case, when $k = 0$, the stream is just empty. It is easy to see that $|\pi_k| = 1 + 2|\pi_{k - 1}| = 1 + 2 \cdot (2^{k - 1} - 1) = 2^k - 1$. See \Cref{fig:hard-stream} for an illustration. 

The idea behind this construction is as follows. In Case 1, the algorithm $\alg$ remembers $e$ throughout the first half of the stream. Intuitively, this means that the algorithm $\alg'$ has one less space in memory to store other elements in $\pi^{\alg'}_{k - 1}$. Alternatively, in Case 2, the algorithm $A$ does not remember $e$ through the first half.  By letting $e$ be either greater than $\pi^{\alg''}_{k - 1}$ or less than $\pi^{\alg''}_{k - 1}$ with equal probability, we create more uncertainty for rank queries. Thus, the algorithm $\alg''$ must estimate the ranks in $\pi^{\alg''}_{k - 1}$ with a smaller error. In conclusion, from level $k$ to $k - 1$, either the space or error constraint of the algorithm are more strained. If we further expand out the recursion like this, finally in the base case, neither the space nor the error can be negative. Thus, the original algorithm $\alg$ must either have a large enough space or large error in the beginning. 

\paragraph*{Space \& Error Analysis.} Now, we carry out the idea described above and prove the lower bound via induction. 

\begin{lem}
    For any algorithm $\alg$, let $S_t$ be the space it use at time $t$. After reading the (partial) input stream $\pi^{\alg}_k = x_1 x_2 \dots x_{2^k - 1}$, we will choose an index $i$ ($1 \leq i \leq |\pi_k^{\alg}|$) depending on $\alg$ but independent of the randomness of $\pi^{\alg}_k$ and $\alg$. 
    
    Let $\widehat{\rank}_{\pi^{\alg}_k}(x_i)$ be any estimator for the rank of $x_i$ based on the memory content of $\alg$ (after reading this partial stream). We can choose $i$ so that we always have $\rank_{\pi^{\alg}_k}(x_i) \leq k$ and 
    $$\max_{1 \leq t \leq |\pi^{\alg}_k|} \E[S_t] + \E\left[\left(\widehat{\rank}_{\pi^{\alg}_k}(x_i) - \rank_{\pi^{\alg}_k}(x_i)\right)^2\right] \geq 0.1k.$$
\end{lem}
\begin{proof}
    We prove this lemma by induction. In the base case where $k = 0$, this is true because none of the terms can be negative. 

    Now we verify it is true for both cases:
    \begin{itemize}
        \item Case 1: In this case, the first element $e$ satisfies $\Pr[e \in \mem_{2^{k - 1}}] \geq 1/2$. Let $\alg'$ be the algorithm $\alg$ with the first element fixed to $e$ as before, and $S'_1, S'_2, \dots, S'_{2^{k - 1} - 1}$ be its space after reading each element in $\pi^{\alg'}_{k - 1}$ (not counting the element $e$). 

        Let $\pi_{k - 1}^{\alg'} = x'_1 x'_2 \dots x'_{2^{k - 1} - 1}$. By our induction hypothesis, we know that we can choose an $i$ with $\rank_{\pi^{\alg'}_k}(x'_i) \leq k - 1$ and 
    \begin{equation} \label{equ:case1}
    \max_{1 \leq t \leq 2^{k - 1} - 1} \E[S'_t] + \E\left[\left(\widehat{\rank}_{\pi^{\alg'}_k}(x'_i) - \rank_{\pi^{\alg'}_k}(x'_i)\right)^2\right] \geq 0.1 (k - 1)
    \end{equation}

    for any estimator $\widehat{\rank}_{\pi^{\alg}_k}(x'_i)$ that only depends on the memory of $\alg'$ after reading $\pi^{\alg'}_{k - 1}$.

   For all $1 \leq t \leq 2^{k - 1} - 1$, we have $\E[S_{t + 1}] = \E[S'_t] + \Pr[e \in \mathcal{M}_t]$, and $\Pr[e \in \mathcal{M}_t] \geq \Pr[e \in \mathcal{M}_{2^{k - 1}}] \geq 1/2$ (because once $\alg$ removes one element from the memory it cannot get it back). Also, the second half of the stream consists of only elements with rank larger than $e$ and $\pi^{\alg'}_{k - 1}$, so they are irrelavant for the rank of $x'_i$ and the second term of \eqref{equ:case1} does not change. Thus, we conclude that, for $\pi^A_k = x_1 x_2 \dots x_k$ (with $x_1 = e$), we have
   $$\max_{1 \leq t \leq |\pi^{\alg}_k|} \E[S_t] + \E\left[\left(\widehat{\rank}_{\pi^{\alg}_k}(x_{i + 1}) - \rank_{\pi^{\alg}_k}(x_{i + 1})\right)^2\right] \geq 0.1(k - 1) + 1/2 \geq 0.1k.$$

   Finally, we verify that $\rank_{\pi^{\alg}_k}(x_{i + 1}) = \rank_{\pi^{\alg'}_k}(x'_i) + 1 \leq k$. 
    \item Case 2: In this case, the first element $e$ satisfies $\Pr[e \in \mem_{2^{k - 1}}] \leq 1/2$. Let $\alg''$ be the algorihtm $\alg''$ with prefix fixed to the concatenation of $e$ and $\pi^{\alg'}_{k - 1}$, and $S''_1, S''_2, \dots, S''_{2^{k - 1} - 1}$ be the space after reading each element in $\pi^{\alg''}_{k - 1}$ (not counting the element $e$ nor the elements in $\pi^{\alg'}_{k - 1}$). 

    Let $\pi_{k - 1}^{\alg''} = x''_1 x''_2 \dots x''_{2^{k - 1} - 1}$. Again, by our induction hypothesis, we can select an index $i$ (independent of all the randomness) such that 
    $$\max_{1 \leq t \leq 2^{k - 1} - 1} \E[S''_t] + \E\left[\left(\widehat{\rank}_{\pi^{\alg''}_k}(x''_i) - \rank_{\pi^{\alg''}_k}(x''_i)\right)^2\right] \geq 0.1 (k - 1).
    $$
    For all $1 \leq t \leq 2^{k - 1} - 1$, we have $\E[S_{t + 2^{k - 1}}] \geq \E[S''_t]$. For the second term, let $E_k$ be the event that $e \in \mathcal{M}_{2^k}$. We know that 
    \begin{align*}
    &\E\left[\left(\widehat{\rank}_{\pi^{\alg}_k}(x_{i + 2^{k - 1}}) - \rank_{\pi^{\alg}_k}(x_{i + 2^{k - 1}})\right)^2 \ \ \middle\vert \ \  E_k\right] \\
    \geq &\E\left[\left(\widehat{\rank}_{\pi^{\alg''}_k}(x''_i) - \rank_{\pi^{\alg''}_k}(x''_i)\right)^2\ \ \middle\vert \ \  E_k\right]
    \end{align*}
    because $x_{i + 2^{k - 1}}$ and $x''_i$ are the same element and $e$ is hard-coded in $A''$. For any estimator $\widehat{\rank}_{\pi^{\alg}_k}(x_{i + 2^{k - 1}})$, we can always compare $e$ with $x_{i + 2^{k - 1}}$ for free and get the estimator $\widehat{\rank}_{\pi^{\alg''}_k}(x''_i)$. 

    On the other hand, we have 
    \begin{align*}
    &\E\left[\left(\widehat{\rank}_{\pi^{\alg}_k}(x_{i + 2^{k - 1}}) - \rank_{\pi^{\alg}_k}(x_{i + 2^{k - 1}})\right)^2 \ \ \middle\vert \ \  \lnot E_k\right] \\
    \geq &\E\left[\left(\widehat{\rank}_{\pi^{\alg''}_k}(x''_i) - \rank_{\pi^{\alg''}_k}(x''_i)\right)^2\ \ \middle\vert \ \  \lnot E_k\right] + 0.25
    \end{align*}
    The reasoning is as follows. First, the event $E_k$ only depends on $e$ and $\pi^{\alg'}_{k - 1}$ and is independent of $\pi^{\alg''}_{k - 1}$. Second, conditioning on $\lnot E_k$, any estimator $\widehat{\rank}_{\pi^{\alg}_k}(x_{i + 2^{k - 1}})$ can never distinguish the two possibilities of the relative order between $e$ and $\pi^{\alg''}_{k - 1}$. (Because the element $e$ is never able to be compared elements in $\pi^{\alg''}_{k - 1}$, including $x_{i + 2^{k - 1}}$.). 
    
    So, the optimal strategy for the estimator is to simply estimate the rank of $x_{i + 2^{k - 1}}$ in $\pi_{k}^{\alg''}$ and plus $1/2$. The squared error of this is lower bounded by $\E\left[\left(\widehat{\rank}_{\pi^{\alg''}_k}(x''_i) - \rank_{\pi^{\alg''}_k}(x''_i)\right)^2\ \ \middle\vert \ \  \lnot E_k\right]$ plus $1/4$ squared error because the ground truth $\rank_{\pi^{\alg}_k}(x_{i + 2^{k - 1}})$ differs by $1$ in these two indistinguishable possibilities. 

    Combining these and $\Pr[E_k] \leq 1/2$, we conclude that 
    $$\E\left[\left(\widehat{\rank}_{\pi^{\alg}_k}(x_{i + 2^{k - 1}}) - \rank_{\pi^{\alg}_k}(x_{i + 2^{k - 1}})\right)^2\right] \geq \E\left[\left(\widehat{\rank}_{\pi^{\alg''}_k}(x''_i) - \rank_{\pi^{\alg''}_k}(x''_i)\right)^2\right] + 0.125.$$

    Thus we have, 
   $$\max_{1 \leq t \leq |\pi^{\alg}_k|} \E[S_t] + \E\left[\left(\widehat{\rank}_{\pi^{\alg}_k}(x_{i + 1}) - \rank_{\pi^{\alg}_k}(x_{i + 1})\right)^2\right] \geq 0.1(k - 1) + 0.125 \geq 0.1k.$$
    \end{itemize}
\end{proof}

Having this lemma and letting $k = \Theta(\log n)$, we see that for any algorithm $\alg$, it either uses $\Omega(\log n)$ space, or for an element of rank at most $\log n$, introduce an error of at least $\Omega(\sqrt{\log n})$ with $0.99$ probability (by Chebyshev's inequality). This proves \Cref{lem:lb-top-quantile}.

\subsection{Upper bound}
In this subsection, we provide a matching upper bound for the top-$R$ quantiles problem, as defined in \Cref{table:top-quantiles}. For this problem, recall that the algorithm should provide the following guarantee: given any query $x \in \mathcal{U}$ such that $\rank(x) \leq R$, the algorithm should return an estimate $\widehat \rank(x)$ such that $|\widehat \rank(x) - \rank(x)| \leq \epsilon \cdot R$. Our algorithm for this problem will closely mimic the construction given earlier in \Cref{sec:algo-description}, but we will replace the resizable elastic compactors with ordinary relative compactors (as originally defined in \cite{cormode2023relative}, see the description in \Cref{sec:detailed}). 

\begin{figure}[H]
    \centering
\tikzset{every picture/.style={line width=0.75pt}} %

\begin{tikzpicture}[x=0.75pt,y=0.75pt,yscale=-1,xscale=1]

\draw   (70.58,67.68) -- (110.31,84.39) -- (109.69,122.26) -- (69.43,137.67) -- cycle ;
\draw    (120,104) -- (156.86,104) ;
\draw [shift={(158.86,104)}, rotate = 180] [color={rgb, 255:red, 0; green, 0; blue, 0 }  ][line width=0.75]    (10.93,-3.29) .. controls (6.95,-1.4) and (3.31,-0.3) .. (0,0) .. controls (3.31,0.3) and (6.95,1.4) .. (10.93,3.29)   ;
\draw   (171,67.29) -- (182.86,67.29) -- (182.86,147) -- (171,147) -- cycle ;
\draw    (171,83) -- (182.86,83) ;
\draw    (171,99) -- (182.86,99) ;
\draw    (172,115) -- (183.86,115) ;
\draw    (171,131) -- (182.86,131) ;
\draw   (226,66.29) -- (237.86,66.29) -- (237.86,146) -- (226,146) -- cycle ;
\draw    (226,82) -- (237.86,82) ;
\draw    (226,98) -- (237.86,98) ;
\draw    (227,114) -- (238.86,114) ;
\draw    (226,130) -- (237.86,130) ;
\draw    (192,104) -- (214.86,104) ;
\draw [shift={(216.86,104)}, rotate = 180] [color={rgb, 255:red, 0; green, 0; blue, 0 }  ][line width=0.75]    (10.93,-3.29) .. controls (6.95,-1.4) and (3.31,-0.3) .. (0,0) .. controls (3.31,0.3) and (6.95,1.4) .. (10.93,3.29)   ;
\draw   (283,66.29) -- (294.86,66.29) -- (294.86,146) -- (283,146) -- cycle ;
\draw    (283,82) -- (294.86,82) ;
\draw    (283,98) -- (294.86,98) ;
\draw    (284,114) -- (295.86,114) ;
\draw    (283,130) -- (294.86,130) ;
\draw    (249,103) -- (271.86,103) ;
\draw [shift={(273.86,103)}, rotate = 180] [color={rgb, 255:red, 0; green, 0; blue, 0 }  ][line width=0.75]    (10.93,-3.29) .. controls (6.95,-1.4) and (3.31,-0.3) .. (0,0) .. controls (3.31,0.3) and (6.95,1.4) .. (10.93,3.29)   ;
\draw   (341,65.29) -- (352.86,65.29) -- (352.86,145) -- (341,145) -- cycle ;
\draw    (307,102) -- (329.86,102) ;
\draw [shift={(331.86,102)}, rotate = 180] [color={rgb, 255:red, 0; green, 0; blue, 0 }  ][line width=0.75]    (10.93,-3.29) .. controls (6.95,-1.4) and (3.31,-0.3) .. (0,0) .. controls (3.31,0.3) and (6.95,1.4) .. (10.93,3.29)   ;

\draw (59,145) node [anchor=north west][inner sep=0.75pt]   [align=left] {Sampler};

\end{tikzpicture}

    \caption{Structure of sketch $\mathcal{M}$ for Top Quantiles}
    \label{fig:hierarchy-top-quantiles}
\end{figure}
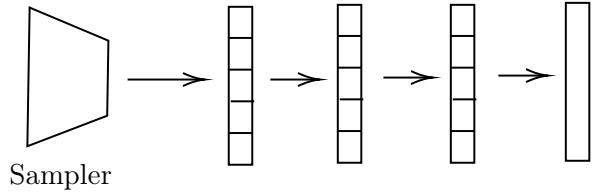

As shown above, the sketch $\mathcal{M}$ will consist of the following components:

\begin{enumerate}
    \item \textbf{Sampler:} For each new stream element $x$ in $\pi$, the sampler inserts $x$ into the first relative compactor $C_0$ with probability $\frac{1}{\epsilon^2 R}$, and discards $x$ otherwise. 
    
    As before, we attach a weight to each new element $x$: when $x$ first appears in the stream $\pi$, $x$ initially has weight $1$. After passing through the sampler, each sampled element has weight $\epsilon^2 R$. 
    \item \textbf{Relative compactors:} for each $0 \leq j < O(\log_2(1/\epsilon))$, we maintain a relative compactor $C_j$ with block-size $k = O\left(\frac{1}{\epsilon\sqrt{\log(n)}}\right)$, such that each $C_j$ can store at most $s = O\left(\frac{\sqrt{\log(n)}}{\epsilon}\right)$ elements.\footnote{Throughout this subsection, we assume the range of parameter $\epsilon \leq \frac{d}{\sqrt{\log n}}$ for some sufficiently small universal constant $d > 0$. This bound enables us to pick $k$ to be an integer.} After each compaction of $C_j$, $C_j$ may output some elements $x_1, x_2, \dots, x_m$, which we insert into the next compactor $C_{j+1}$ in the hierarchy. Eventually, the output of the last compactor is inserted into the buffer. 

    For each $0 \leq j < \log_2(1 / \epsilon)$, elements in $C_j$ all have weight $\epsilon^2 R \cdot 2^j$.

    \item \textbf{Buffer:} At the end of the compactor hierarchy (i.e. after compactor $C_{O(\log(1/\epsilon))-1}$), we have a  buffer $B$ which  stores $O(1/\epsilon)$ elements. 
    
    Note that each element in the buffer has weight $\epsilon R$.
\end{enumerate}

Additionally, observe that since the smallest $1/\epsilon$ elements in the buffer already have total weight $R$ and we only need to estimate the first $R$ ranks, we can afford to throw away other elements from the buffer (just as we did previously, in \Cref{sec:top-quantile}).

To estimate the rank for any query $x \in \mathcal{U}$, we use the same estimator as we previously in \Cref{sec:top-quantile}: let 

\begin{equation}\label{equ:rank-estimator-UB}
        \Call{Rank}{}_{\mathcal{M}}(x) \coloneqq \sum_{j=0}^{O(\log(1 / \epsilon)) - 1} \frac{2^j}{\epsilon^2 R} \cdot \Call{Rank}{}_{C_j}(x) + \frac{1}{\epsilon R} \cdot \Call{Rank}{}_B(x)
    \end{equation} 
    
Now, we show that this construction will solve the Top-$R$ Quantiles problem using space $\tilde O(\epsilon^{-1} \sqrt{\log n})$ with constant probability. 

\begin{lem}
    Let $\pi$ be the input stream and let $s = O(\epsilon^{-1} \sqrt{\log n})$ be the space allocated to each compactor $C_j$ in the hierarchy, for $0 \leq j \leq O(\log(1/\epsilon)) -1$. Then, we have that for any query $x \in \mathcal{U}$ such that $\rank_{\pi}(x) \leq R$, our sketch $\mathcal{M}$ achieves standard deviation

    $$\mathbb{E}\left[|\Call{Rank}{}_{\mathcal{M}}(x) - \rank_{\pi}(x)|^2\right] \leq O(\epsilon) \cdot R$$

    with constant probability. \label{lem:UB-M}
\end{lem}
\begin{proof}
    Just as we did in the proof of \Cref{lem:H}, we begin by decomposing the error incurred by the rank estimate $\Call{Rank}{}_{\mathcal{M}}(x)$. Let $C_0, \dots, C_{O(\log(1/\epsilon)) - 1}$ be the compactors in $\mathcal{M}$, and $C_{O(\log (1/\epsilon))}$ denotes the buffer. As before, we use $\pi_j$ to denote the input stream of $C_j$ (which is the output stream of $C_{j - 1}$). 
\begin{align}
    \Call{Rank}{}_{\mathcal{M}}(x) - \rank_\pi(x) = &\sum_{j = 0}^{O(\log(1/\epsilon))} \Call{Rank}{}_{C_j}(x) \cdot {2^j} \cdot {\epsilon^2 \cdot R} - \rank_\pi(x) \notag \\
    = &\sum_{j = 0}^{O(\log(1/\epsilon)) - 1} \left((\Call{Rank}{}_{C_j}(x)  + 2\rank_{\pi_{j + 1}}(x)) - \rank_{\pi_j}(x) \right) \cdot {2^j} \cdot {\epsilon^2 R} \notag \\
    & + \left(\rank_{\pi_0}(x) \cdot \epsilon^2 R - \rank_{\pi}(x)\right) \notag \\
    & + \left(\Call{Rank}{}_{C_{\log(1/\epsilon)}}(x) - \rank_{\pi_{\log(1/\epsilon)}}(x) \right) \cdot {\epsilon R} .\label{equ:err-decomp-M}
\end{align}

We examine each term in \Cref{equ:err-decomp-M} above: the first term is the error contribution from the compactor hierarchy, and the second and third terms represent the error from the sampler and buffer, respectively. By the same argument as given in the proof of \Cref{lem:H}, we note that the third term (corresponding to the buffer) will never contribute to the error of the sketch. Next, we take a closer look at the first term. Let us denote 

$$\Delta_{j}(x) = (\Call{Rank}{}_{C_j}(x) + 2\rank_{\pi_{j+1}}(x)) - \rank_{\pi_j}(x)$$

Since each $C_j$ is a relative compactor, we can apply Lemma 5 from \cite{cormode2023relative} to bound the standard deviation. Moreover, recall that since we are guaranteed that $\rank_{\pi}(x) \leq R$ and stream elements are sampled with probability $\frac{1}{\epsilon^2 R}$, we have $\mathbb{E}[\rank_{\pi_j}(x)] = \frac{\rank_{\pi_j}(x)}{\epsilon^2 R 2^j} \leq \frac{1}{\epsilon^2 2^j}$. Thus, we get

$$\mathbb{E}\left[\Delta_{j}(x)^2\right]^{1/2} \leq \mathbb{E}\left[\ \frac{\rank_{\pi_j}(x)}{k} \right]^{1/2} \leq \sqrt{\frac{1}{\epsilon^2 \cdot 2^j \cdot k}}$$

Then, we see that the first term incurs standard deviation

$$\sum_{j = 0}^{O(\log(1/\epsilon))-1} \mathbb{E}[\Delta_j(x)^2]^{1/2} \cdot 2^j \cdot \epsilon^2 R \leq \sum_{j = 0}^{O(\log(1/\epsilon))-1} \sqrt{\frac{1}{\epsilon^2 \cdot 2^j \cdot k}} \cdot 2^j \cdot \epsilon^2 R \leq O(\epsilon) R \cdot  \sqrt{\textrm{poly}(1/\epsilon) \cdot \sqrt{\log n}}$$ $$\leq O(\epsilon) \cdot R$$

Finally, we upper bound the error resulting from the sampler, i.e. the second term. Since we sample each stream element with probability $p = \frac{1}{\epsilon^2 R}$, we have that $\rank_{\pi_0}(x) = \sum_{t = 1}^{\rank_{\pi}(x)} \textrm{Bernoulli}(p)$, i.e. we get 

$$\mathbb{E}\left[(\rank_{\pi_0}(x) \cdot \epsilon^2 R - \rank_{\pi}(x))^2 \right]^{1/2} \leq \epsilon^2 R \cdot \sqrt{R\cdot p(1-p)} \leq \epsilon R$$

Therefore, by applying \Cref{fact:stadnard-deviation}, we obtain the desired bound on the standard deviation. 
\end{proof}

\begin{cor}
    Suppose that $\epsilon \leq \frac{d}{\sqrt{\log n}}$ for some sufficiently small universal constant $d > 0$. Then the sketch $\mathcal{M}$ solves the Top-$R$ Quantiles problem using space $O(\epsilon^{-1} \sqrt{\log(n)} \cdot \log(1/\epsilon))$ with constant probability.
\end{cor}
\begin{proof}
    This follows directly from \Cref{lem:UB-M} together with Chebyshev's inequality.
\end{proof}
 
\begin{remark}
    Using the analysis in \Cref{sec:high-prob}, we can similarly obtain the same result with a high success probability by analyzing the top compactors deterministically. %
\end{remark}

\section{Open Problems} \label{sec:open-prob}

Several variants of the quantile estimation problem in the streaming model still remain open. In the next few paragraphs, we describe some open problems which relate to our work. 

\paragraph{Obtaining optimal bounds for relative-error quantile estimation.} Our algorithm achieves a $(1\pm\epsilon)$ relative-error guarantee using total space $O(\epsilon^{-1}\log(\epsilon n) \cdot (\log\log n + \log(1/\epsilon)) \cdot (\log\log 1/\delta)^3)$. The main open question in the randomized, relative-error regime is to close the gap  between our upper bound and the best known lower bound $\Omega(\epsilon^{-1} \log(\epsilon n))$. 

\paragraph{Full-mergeability.} Before we state the next open question, we recall the definition of a \textit{fully-mergeable} sketch. Suppose we arbitrarily partition the input stream  $\pi = \bigsqcup_{i = 1}^{\ell} \pi_i$ and summarize each piece $\pi_i$ of the stream using a separate sketch $\mathcal{M}_i$. Then, a sketch is fully-mergeable if any sequence of pairwise merging operations that merges these $\ell$ sketches $\mathcal{M}_1,..., \mathcal{M}_{\ell}$ together results in a new sketch with essentially the same error and space guarantees as if the stream had been summarized using one sketch directly. Notably, this full-mergeability property is very important in practice: since massive datasets are often partitioned and stored on multiple servers, it is often useful to sketch the data stored locally on each device, and re-combine the sketches later to obtain a summary of the entire dataset. Since our sketch relies on an intricate dynamic space allocation scheme for the compactor hierarchy $H_i$ stored for each scale $[\epsilon^{-1} 2^i, \epsilon^{-1} 2^{i+1}]$, it is not immediately clear how to ``merge'' the separate flushing patterns of the two sketches in order to obtain a valid resizable sketch $H_i$ for each scale, while maintaining total space $\tilde O(\epsilon^{-1} \log(\epsilon n))$. With this in mind, we ask whether our sketch is fully mergeable, or if it is not, we ask if there is a way to adjust our construction to prove a mergeability guarantee.

\paragraph{Comparison-based, deterministic relative-error quantiles.}
Additionally, we ask whether it is possible to close the gap for comparison-based, deterministic relative-error quantile estimation. Currently, the only known lower bound is $\Omega\left(\frac{\log^2(\epsilon n)}{\epsilon}\right)$~\cite{cormode2020tight}, and the best known upper bound is $O\left(\frac{\log^3 (\epsilon n)}{\epsilon}\right)$, via the merge-and-prune algorithm of \cite{zhang2007efficient}. To this end, we ask whether it is possible to apply our dynamic space allocation and sketch resizing techniques to improve the space used in the deterministic setting. Perhaps, as an intermediate step, it is an interesting open problem to resolve the deterministic space complexity for the Top Quantiles problem that we defined in \Cref{table:top-quantiles}. Again, the only known lower bound is $\Omega\left(\frac{\log (\epsilon n)}{\epsilon}\right)$ from the additive error setting~\cite{cormode2020tight}, and the best upper bound is $O\left(\frac{\log^2 (\epsilon n)}{\epsilon}\right)$ from the same merge-and-prune algorithm.

\paragraph{Non-comparison-based relative-error quantiles.} Although comparison-based algorithms are more powerful and can handle quantile queries for an unknown (totally-ordered) universe, there are many applications where the universe in fixed and known in advance. Then, the goal shifts from minimizing the number of elements stored to a more fine-grained measure: the number of bits used in total. 

Previously, ~\cite{cormode2006space} gave a non-comparison-based bq-summary algorithm using $O\left(\frac{\log (\epsilon n) \log U}{\epsilon}\right)$ words of memory (where each word is $\log U + \log (\epsilon n)$ bits), while the offline lower bound is $\Omega \left(\frac{\log n}{\epsilon}\right)$ words. Although our algorithm almost closes the gap for randomized relative-error quantile estimation, we ask whether there is a non-comparison-based approach that could result in simpler deterministic algorithm. 

As a remark, the bq-summary algorithm is based on the q-digest, which is a deterministic non-comparison-based algorithm for the additive-error quantiles problem~\cite{shrivastava2004medians}. Recently, the upper bound for q-digest  was improved from  $O\left(\frac{\log U}{\epsilon}\right)$ to only $O\left(\frac{1}{\epsilon}\right)$ words ~\cite{gupta2024optimal}, which is optimal. We ask whether it is possible to use similar ideas as in \cite{gupta2024optimal} to improve the upper bound in the deterministic relative-error setting as well.

\paragraph{Tight characterization for resizable sketches.} Suppose a resizable sketch is given space $s_t$ after reading the $t$-th element in the input. Our resizable sketch for the Top Quantiles problem (\Cref{lem:H}) works for space sequences $s_1, s_2, \dots, s_n$ satisfying $\sum_{t=1}^n 2^{-s_i} \leq 0.5$. By following a more careful analysis and adjusting the block size $k$, one could actually improve this condition to $\sum_{t=1}^n  2^{-\epsilon^2 s_t^2}\leq 0.5$. On the other hand, it is conceivable that one might be able to prove a lower bound saying that under the opposite condition for the space sequence (i.e. $\sum_{t= 1}^n 2^{-\epsilon^2 s_t^2} > 0.5$), there is no comparison-based algorithm for the Top Quantiles Problem.

Motivated by this, we ask whether one can give such a tight characterization for resizable sketches for the Top Quantile Problem, or analogously, if we can prove similar space-sequence bounds for other natural streaming problems.  
\bibliography{main}

\begin{thebibliography}{10}

\bibitem{agrawal1995}
Rakesh Agarwal and Arun Swami.
\newblock A one-pass space-efficient algorithm for finding quantiles.
\newblock In {\em Proceedings of 7th International Conference on Management of
  Data (COMAD-95)}, 1995.

\bibitem{manku2004}
Arvind Arasu and Gurmeet~Singh Manku.
\newblock Approximate counts and quantiles over sliding windows.
\newblock In {\em Proceedsing of the 23rd AC SIGMOD-SIGACT-SIGART symposium on
  Principles of Database Systems (PODS 2004)}, pages 286--296, 2004.

\bibitem{cormode2023relative}
Graham Cormode, Zohar Karnin, Edo Liberty, Justin Thaler, and Pavel Vesel{\`y}.
\newblock Relative error streaming quantiles.
\newblock {\em Journal of the ACM}, 70(5):1--48, 2023.

\bibitem{cormode2005effective}
Graham Cormode, Flip Korn, S~Muthukrishnan, and Divesh Srivastava.
\newblock Effective computation of biased quantiles over data streams.
\newblock In {\em 21st International Conference on Data Engineering (ICDE'05)},
  pages 20--31. IEEE, 2005.

\bibitem{cormode2006space}
Graham Cormode, Flip Korn, S~Muthukrishnan, and Divesh Srivastava.
\newblock Space-and time-efficient deterministic algorithms for biased
  quantiles over data streams.
\newblock In {\em Proceedings of the twenty-fifth ACM SIGMOD-SIGACT-SIGART
  symposium on Principles of database systems}, pages 263--272, 2006.

\bibitem{cormode2020tight}
Graham Cormode and Pavel Vesel{\`y}.
\newblock A tight lower bound for comparison-based quantile summaries.
\newblock In {\em Proceedings of the 39th ACM SIGMOD-SIGACT-SIGAI Symposium on
  Principles of Database Systems}, pages 81--93, 2020.

\bibitem{infinifilter}
Niv Dayan, Ioana Bercea, Pedro Reviriego, and Rasmus Pagh.
\newblock Infinifilter: Expanding filters to infinity and beyond.
\newblock In {\em Proc. ACM Manag. Data, Vol. 1, No. 2, Article 140}, 2023.

\bibitem{ostrovsky2015}
David Felber and Rafail Ostrovsky.
\newblock A randomized online quantile summary in o(1/epsilon * log(1/epsilon))
  words.
\newblock In {\em Approximation, Randomization, and Combinatorial Optimization.
  Algorithms and Techniques (APPROX/RANDOM 2015), volume 40 of Leibniz
  International Proceedings in Informatics}, pages 775--785, 2015.

\bibitem{greenwald2001space}
Michael Greenwald and Sanjeev Khanna.
\newblock Space-efficient online computation of quantile summaries.
\newblock {\em ACM SIGMOD Record}, 30(2):58--66, 2001.

\bibitem{simpleGK}
Elena Gribelyuk, Pachara Sawettamalya, Hongxun Wu, and Huacheng Yu.
\newblock Simple \& optimal quantile sketch: Combining greenwald-khanna with
  khanna-greenwald.
\newblock In {\em Proceedings of the ACM on Management of Data, Volume 2, Issue
  2 (PODS 2024)}, 2024.

\bibitem{gupta2003counting}
Anupam Gupta and Francis Zane.
\newblock Counting inversions in lists.
\newblock In {\em SODA}, volume~3, pages 253--254, 2003.

\bibitem{gupta2024optimal}
Meghal Gupta, Mihir Singhal, and Hongxun Wu.
\newblock Optimal quantile estimation: beyond the comparison model.
\newblock {\em arXiv preprint arXiv:2404.03847}, 2024.

\bibitem{karnin2016optimal}
Zohar Karnin, Kevin Lang, and Edo Liberty.
\newblock Optimal quantile approximation in streams.
\newblock In {\em 2016 ieee 57th annual symposium on foundations of computer
  science (focs)}, pages 71--78. IEEE, 2016.

\bibitem{manku1998}
Gurmeet~Singh Manku, Sridhar Rajagopalan, and Bruce~G. Lindsay.
\newblock Approximate medians and other quantiles in one pass with limited
  memory.
\newblock In {\em ACM SIGMOD record, volume 27}, pages 426--435, 1998.

\bibitem{pagh2013}
Rasmus Pagh, Gil Segev, and Udi Wieder.
\newblock How to approximate a set without knowing its size in advance.
\newblock In {\em IEEE 54th Annual Symposium on Foundations of Computer Science
  (FOCS)}, 2013.

\bibitem{ioannidis1999}
Viswanath Poosala, Venkatesh Ganti, and Yannis~E. Ioannidis.
\newblock Approximate query answering using histograms.
\newblock In {\em IEEE Data Eng. Bull., 22(4):5-15}, 1999.

\bibitem{shrivastava2004medians}
Nisheeth Shrivastava, Chiranjeeb Buragohain, Divyakant Agrawal, and Subhash
  Suri.
\newblock Medians and beyond: new aggregation techniques for sensor networks.
\newblock In {\em Proceedings of the 2nd international conference on Embedded
  networked sensor systems}, pages 239--249, 2004.

\bibitem{simonsworkshop}
Simons Institute Workshop:~Data Structures and Optimization for
  Fast~Algorithms.
\newblock Sketching and algorithm design, October 2023.
\newblock
  \url{https://simons.berkeley.edu/talks/rasmus-pagh-university-copenhagen-2023-10-13}.

\bibitem{zhang2007efficient}
Qi~Zhang and Wei Wang.
\newblock An efficient algorithm for approximate biased quantile computation in
  data streams.
\newblock In {\em Proceedings of the sixteenth ACM conference on Conference on
  information and knowledge management}, pages 1023--1026, 2007.

\bibitem{zhang2006space}
Ying Zhang, Xuemin Lin, Jian Xu, Flip Korn, and Wei Wang.
\newblock Space-efficient relative error order sketch over data streams.
\newblock In {\em 22nd International Conference on Data Engineering (ICDE'06)},
  pages 51--51. IEEE, 2006.

\end{thebibliography}
\bibliographystyle{plain}

\appendix

\section{Proof of the Extremely High Probability Bound}
\label{appendix:high-probability}
We first recall the parameter changes for extremely high probability bound:
\begin{itemize}
    \item We will set the block length of compactors $k = c \cdot \frac{ (\log \log (1/ \delta))^2}{\epsilon}$ with a large enough constant $c$ (say $c = 128$) throughout the algorithm. 
    \item For the sampler, we adjust the sampling probability of each sampler to $\frac{c \cdot \log (1/\delta)}{\epsilon^2 \cdot R_i}$, as a result, now each element outputted by the sampler has weight $\frac{\epsilon^2 \cdot R_i}{c \cdot \log (1/\delta)}$. 
    \item We set $h_i \coloneqq \log \log (1/\delta) + \log(c / (k\epsilon^2))$. For each $0 \leq j < h_i$, the compactor $C_{i,j}$ has elements of weight $w_{i,j} \coloneqq \frac{\epsilon^2 \cdot R_i \cdot 2^j}{c \cdot \log(1/\delta)}$. 
    \item The buffer now has size $3 k$, each element in the buffer (which we denote by $C_{i,h_i}$) has weight $w_{i, h_i} = \frac{\epsilon^2 \cdot R_i \cdot 2^{h_i}}{c \cdot \log(1/\delta)} = \frac{R_i}{k}$. 
\end{itemize}

\paragraph*{Error Analysis} Recall that $\pi_{i,j}$ is the input stream of compactor $C_{i,j}$ and $\pi_{H_i}$ is the input stream of sub-sketch $H_i$ while $\pi'_{H_i}$ is the output of the sub-sketch $H_i$. $w_{i,j}$ is the weight of elements in $C_{i,j}$. 

In addition, we let $\pi^I_{H_i}$ be the elements in the original input stream $\pi$ that we directly inserts into $H_i$. The difference with $\pi_{H_i}$ is that, $\pi^I_{H_i}$ do not contains those elements in $\pi'_{H_{i - 1}}$. 

We first show that with probability $1 - \delta$,  the rank of query $x$ in every $\pi_{i,j}$ cannot be too large. This will later be used to upper bound the number of relevant compactions.

\begin{lem} \label{lem:rank-rough-bound}

With probability $1 - O(\delta)$, for all $0 \leq i \leq \log(\epsilon \cdot \rank_{\pi}(x))$ and $0 \leq j \leq h_i$, we have 
\begin{equation}
\rank_{\pi_{i,j}}(x) \leq 2 \cdot \frac{\rank_{\pi}(x)}{w_{i,j}} \text{ \ \  and \ \ } \rank_{\pi_{H_i}}(x) \leq 2 \cdot \rank_{\pi}(x). \label{equ:rank-bound}
\end{equation}
where $w_{i,j} \coloneqq \frac{\epsilon^2 \cdot R_i \cdot 2^j}{\log(1/\delta)}$ is the weight of elements in $C_{i,j}$.
\end{lem}
\begin{proof}
We prove by induction for  $0 \leq i <  \log(\epsilon \cdot \rank_{\pi}(x))$ that, conditioning on \Cref{equ:rank-bound} holds for all $i'$ with $0 \leq i' < i$, it holds for $i$ with probability $1 - \delta \cdot \frac{R_i}{\rank_\pi(x)}$ for some absolute constant $c$. Then a union bound over all $i$ proves this lemma.  

Now suppose this holds for all $i'$ with $0 \leq i' < i$. We first completely unroll the expression for $\rank_{\pi_{i,j}}(x)$ and $\rank_{\pi_{H_i}}(x)$:

First for $\pi_{H_i}$, we decompose the rank of $x$ similar to \Cref{claim:sub-sketch-reset}: 
\begin{align*}
\rank_{\pi_{H_i}}(x) &= \rank_{\pi^I_{H_i}}(x) + \rank_{\pi'_{H_{i - 1}}}(x) \\
&= \rank_{\pi^I_{H_i}}(x) + \sum_{j = 0}^{h_{i - 1}} w_{i - 1, j} \cdot \rank_{\pi_{i,j}}^{\mathrm{remove}}(x) \\
&\leq \rank_{\pi^I_{H_i}}(x) + w_{i - 1,0} \cdot \rank_{\pi_{i-1,0}}(x) \\ 
&\phantom{= \rank_{\pi^I_{H_i}}(x)}+ \sum_{j = 0}^{h_{i - 1} - 1} w_{i - 1, j + 1} \cdot (2\rank_{\pi_{i-1,j + 1}}(x) +  \rank_{\pi_{i-1,j}}^{\mathrm{remove}}(x) - \rank_{\pi_{i-1,j}}(x)) \tag{Note $w_{i-1,j}=2w_{i-1,j-1}$ and $\rank^{\mathrm{remove}}_{\pi_{i,h_i}}(x) \leq 2 \rank_{\pi_{i,h_i - 1}}(x)$.} \\
&\leq \rank_{\pi^I_{H_i}}(x) + w_{i - 1,0} \cdot \rank_{\pi_{i-1,0}}(x) + \sum_{j = 0}^{h_{i - 1}} w_{i - 1, j} \cdot \Delta_{i - 1, j}(x)
\end{align*}
where $\pi^I_{H_i}$ and $\pi'_{H_i}$ are defined at the beginning of this section. In the last line we use the fact that 
$$\Delta_{i,j}(x) \coloneqq \Big(\Call{Rank}{}_{C_{i,j}}(x) + 2 \cdot \rank_{\pi_{i,j + 1}}(x) + \rank_{\pi^{\mathrm{remove}}_{i,j}}(x) \Big) - \rank_{\pi_{i,j}}(x).$$

For the third term $\sum_{j = 0}^{h_{i - 1}} w_{i - 1, j} \cdot \Delta_{i - 1, j}(x)$, note there are at most $\frac{\rank_{{\pi_{i - 1,j}}}(x) + t_i(x) \cdot S}{{k}}$ important flushes for $C_{i,j}$. Here $S \leq \frac{3 R_i}{w_{i,j}}$ is a upper bound on the number of elements in $C_{i,j}$.  By \Cref{claim:important-reset}, we know $t_i(x) \leq 2 \cdot \frac{\rank_{H_i}(x)}{R_i}$. Together with our inductive hypothesis, we get 
$$\frac{\rank_{{\pi_{i - 1,j}}}(x) + t_i(x) \cdot S}{{k}} \leq \frac{2 \cdot \rank_{\pi}(x) + 6 \cdot  \rank_{H_i(x)} }{w_{i,j} \cdot k} \leq \frac{14 \cdot \rank_{\pi}(x)}{w_{i,j} \cdot k}.$$
Thus conditioning on any fixed realization of $\pi_{i - 1, j}$, $\Delta_{i - 1, j}(x)$ is still statistically dominated by 
\begin{equation}
\sum_{t = 1}^{{14 \cdot \rank_{\pi}(x)}/(w_{i,j} \cdot k)} \left(\mathrm{Bernoulli}(1/2) - 1/2)\right). \label{equ:bound-random}    
\end{equation}

On the other hand, $\Delta_{i - 1, j}(x)$ is always deterministically at most 
\begin{equation} 
\frac{14 \cdot \rank_\pi(x)}{w_{i,j} \cdot k} \cdot w_{i,j} = \frac{14 \cdot \rank_\pi(x)}{k}. \label{equ:bound-deterministic}
\end{equation}

We will choose when to use these two bounds. Then, we keep unrolling the second term. 
\begin{align*}
w_{i - 1,0} \cdot \rank_{\pi_{i-1,0}}(x) = w_{i - 1, 0} \cdot \sum_{t=1}^{\rank_{\pi_{H_{i - 1}}}(x)} \mathrm{Bernoulli}(1 / w_{i,0}),
\end{align*}
which by our inductive hypothesis, is stochastic dominated by $$\rank_{\pi_{H_{i - 1}}}(x) + w_{i,0} \cdot \left(\sum_{t=1}^{2 \cdot \rank_{\pi}(x)} \mathrm{Bernoulli}(1 / w_{i,0}) - 1/w_{i,0}\right)$$
even when conditioned on any realization on $\pi_{H_{i - 1}}$. 

We keep unrolling $\rank_{\pi_{H_{i - 1}}}(x) $ in the same way as we unroll $\rank_{\pi_{H_i}}(x)$. Eventually, we get that $\rank_{\pi_{H_i}}(x)$ is stochastically dominated by the following sum (which we will explain how we get it):
\begin{align*}
Z_i = &\sum_{i'=1}^i \rank^I_{H_{i'}}(x) \\
&+ \sum_{i'=1}^{i - 1} \left( \sum_{j=0}^{h_{i'}- 1} w_{i',j} \cdot \left(\sum_{t=1}^{{14 \cdot \rank_{\pi}(x)}/(w_{i',j} \cdot k)} \mathrm{Bernoulli}(1/2) - 1/2\right) \cdot \mathbbm{1}[w_{i',j} \leq w_{i,h'_i}]\right. \\
&+ \sum_{j=0}^{h_{i'} - 1} \frac{14 \rank_{\pi(x)}}{k} \cdot \mathbbm{1}[w_{i',j} > w_{i,h'_i}]\\
&+ \left. w_{i',0} \cdot \left(\sum_{t=1}^{2 \cdot \rank_{\pi}(x)} \mathrm{Bernoulli}(1 / w_{i',0}) - 1/w_{i',0}\right)\right) 
\end{align*}
Here we are using bound~\eqref{equ:bound-random} for those $\Delta_{i',j}(x)$ with $w_{i',j} \leq w_{i,h'_i}$ and bound~\eqref{equ:bound-deterministic} for those with $w_{i',j} > w_{i,h'_i}$. This is the idea of analyzing top compactors deterministically in~\cite{karnin2016optimal}. 

We know that $h'_i = h_i - \log \log 1 /\delta$. Thus there are $\log \log 1/\delta - (i - i')$ many $j$'s pairs with $w_{i',j} > w_{i,h'_i}$ for each $i$. In total, there are $(\log \log 1 / \delta)^2$ many such $(i', j)$ pairs. As a result,
\begin{equation} \label{equ:deterministic-part}
\sum_{j=0}^{h_{i'} - 1} \frac{14 \rank_\pi(x)}{k} \cdot \mathbbm{1}[w_{i',j} > w_{i,h'_i}] \leq \frac{14 \rank_\pi(x)}{k} \cdot (\log \log 1/\delta)^2 \leq \epsilon \cdot \rank_\pi(x). 
\end{equation} 
We can without loss of generality assume that $\epsilon < 0.1$, so that this is less than $0.1 \cdot \rank_{\pi}(x)$. 

Note we also have $\sum_{i'=1}^i \rank^I_{H_{i'}}(x) \leq \rank_{\pi}(x)$. The rest is a sum of the deviation of Bernoulli random variables. We upper bound it using the Chernoff bound for subguassian random variables. Because Bernoulli random variable $X$ is $\mathrm{Var}[X]$ subgaussian, by \Cref{fact:sum-subgaussian}, we know $Z$ is also $\mathrm{Var}[Z]$-subgaussian. 

The following calculation bounds its variance:
\begin{align*} 
\mathrm{Var}[Z_i] &= \sum_{i'=1}^{i - 1} \left( \sum_{j=0}^{h_i - 1} w_{i',j}^2 \cdot \frac{14 \cdot \rank_{\pi}(x)}{w_{i',j} \cdot k} \cdot \mathbbm{1}[w_{i',j} \leq w_{i,h'_i}]\right)  + w_{i',0}^2 \cdot 2 \rank_{\pi}(x) \cdot \frac{1}{w_{i',0}}. \\
&= O(1) \cdot \sum_{i'=1}^{i - \log \log (1/\delta)}  w_{i',h_{i'}} \cdot \frac{\rank_{\pi}(x)}{k} + \log\log(1/\delta) \cdot w_{i,h'_i} \cdot \frac{\rank_{\pi}(x)}{k} + w_{i',0} \cdot \rank_{\pi}(x) \tag{Because $w_{i',j}$'s are exponential in $j$.}\\
&= O(1) \cdot \left(\log \log (1/\delta) \cdot w_{i, h'_{i}} \cdot \frac{\rank_\pi(x)}{k} + w_{i - 1,0} \cdot \rank_{\pi}(x)\right). \tag{Because $w_{i',j}$'s are exponential in $i'$ as well.}
\end{align*}
Since we know $\frac{w_{i, h'_i}}{k} \leq \frac{R_i}{k^2 \log (1/\delta)} \leq \frac{\epsilon^2}{c \cdot \log(1/\delta) \cdot \log \log (1/\delta)} \cdot R_i$ and $w_{i - 1, 0} \leq \frac{\epsilon^2}{c \cdot \log (1 /\delta)} \cdot R_i$, plug them in we get $\mathrm{Var}[Z_i] \leq O(1) \cdot \frac{\epsilon^2}{c \cdot \log(1/\delta)} \cdot R_i \cdot \rank_{\pi}(x).$ (Here $c$ is the constant factor we picked in our algorithm.)
Then, we upper bound the error we have using the Chernoff bound for subgaussian random variables. We conclude that there exists an absolute constant $d > 0$ such that 
\begin{align*}
\Pr[Z_i \geq 0.1 \cdot \rank_{\pi_i}(x)] &\leq  \exp\left(- \frac{\left(0.1 \cdot \rank_{\pi}(x)\right)^2}{2 \mathrm{Var[Z_i]}}\right) \\ 
&\leq \exp\left(- d \cdot \frac{c \cdot \log(1/\delta)}{\epsilon^2} \cdot \frac{\rank_{\pi}(x)}{R_i}\right) \\
&\leq \delta \cdot \frac{R_i}{\rank_\pi(x)} \cdot \frac{1}{\log(1/\epsilon) + \log \log(1/\delta)} \tag{As long as we pick $c > 2 / d$.}
\end{align*}

This finishes the proof for $\rank_{\pi_{H_i}}(x)$. 
For those $\rank_{\pi_{i,j}}(x)$, we note that the same decomposition shows that $\rank_{\pi_{i,j}}(x)$ is stochastically dominated by $Z_{i + 1}$. Since there are at most $O(\log(1/\delta) + \log \log(1/\delta))$ many $j$'s for each $i$, we can simply union bound over all those $j$'s. This finishes the proof of the lemma. 
\end{proof}

Then let us prove that, conditioning on $\rank_{H_i}(x)$'s and $\rank_{\pi_{i,j}}(x)$'s being not too large, our estimate $\widehat{\rank_\pi(x)}$ is approximately correct with probability $1 - O(\delta)$. 

\begin{lem}
For any input stream $\pi$, assuming that the space we allocate to each $H_i$ satisfies the premise of \Cref{lem:H} (\Cref{equ:condition-H}),  For any query $x$, with probability $1 - O(\delta)$, we have 
$$\left|\rank_{\pi}(x) - \sum_{i=0}^{\log_2(\epsilon n)} \Call{Rank}{}_{H_i}(x)\right| \leq O(\epsilon) \cdot \rank_{\pi}(x).$$

\end{lem}
\begin{proof}
First of all, let $\ell = \log(\epsilon \cdot \rank_{\pi}(x))$. From the same logic as \Cref{reamrk:tail}, we know  that it suffices to prove that with probability $1 - O(\delta)$, 
$$\left|\rank_{\pi}(x) - \sum_{i=0}^{\ell} \Call{Rank}{}_{H_i}(x)\right| \leq O(\epsilon) \cdot \rank_{\pi}(x).$$

We first unroll the error similar to \Cref{lem:final-analysis} and \Cref{lem:rank-rough-bound}.
\begin{align*}
&\rank_\pi(x) - \sum_{i=0}^{\ell} \Call{Rank}{}_{H_i}(x)  \\
= &\sum_{i=0}^{\ell} \rank_{\pi_{H_i}}(x) - \rank_{\pi'_i}(x) - \Call{Rank}{}_{H_i}(x) \\
= &\sum_{i=0}^{\ell} \left( \left(\rank_{\pi_{H_i}}(x) - w_{i,0} \cdot \rank_{\pi_{i,0}}(x)\right) + \sum_{j=0}^{h_i} w_{i,j} \cdot \left( \rank_{\pi_{i,j}}(x) - 2 \rank_{\pi_{i,j+1}}(x) - \rank_{\pi^{\mathrm{remove}}_{i,j}}(x) \right)\right)
\end{align*}
From \Cref{lem:rank-rough-bound}, we know that with probability $1 - O(\delta)$, \Cref{equ:rank-bound} holds. 
Conditioning on it holds and follow the same argument as \Cref{lem:rank-rough-bound}, we know that the absolute value 
 of error is stochastically dominated by $\left|Z'_{\ell}\right| + \sum_{i=0}^{\ell}\sum_{j=0}^{h_i - 1} \frac{14 \rank_{\pi(x)}}{k} \cdot \mathbbm{1}\left[w_{i,j} > w_{\ell, h'_{\ell}}\right]$, where $Z'_{\ell}$ is defined as following:
\begin{align*}
Z'_{\ell} = &\sum_{i=0}^{\ell} \left( \sum_{j=0}^{h_i - 1} w_{i,j} \cdot \left(\sum_{t=1}^{14 \cdot \rank_\pi (x) / (w_{i,j} \cdot k)} \mathrm{Bernoulli}(1/2) - 1/2\right) \cdot \mathbbm{1}\left[w_{i,j} \leq w_{\ell, h'_{\ell}}\right] \right.\\
&+ \left. w_{i,0} \cdot \left(\sum_{t=1}^{2\rank_\pi(x)} \mathrm{Bernoulli}(1/w_{i,0}) - 1/w_{i,0}\right)\right).
\end{align*}
Since $Z'_{\ell}$ is only $Z_{\ell}$ shifted, they have the same variance. We get that 
\begin{align*}
\mathrm{Var}\left[Z'_{\ell}\right] = \mathrm{Var}\left[Z_{\ell}\right] &\leq  O(1) \cdot \frac{\epsilon^2}{c \cdot \log(1/\delta)} \cdot R_\ell \cdot \rank_{\pi}(x). \\
&= O(1) \cdot \frac{\epsilon^2}{c \cdot \log(1/\delta)} \cdot R_\ell^2\tag{Because $R_{\ell} = \rank_{\pi}(x)$ by defintion.}
\end{align*}
As $Z'_{\ell}$ is the sum of Bernoulli random variables, it is $\mathrm{Var}\left[Z'_{\ell}\right]$-subgaussian. We then apply Chernoff bound for subgaussian random variables and conclude that when constant $c$ is picked to be large enough, we have
$$\Pr\left[Z'_{\ell} \geq \epsilon \cdot \rank_{\pi}(x) \right] \leq \delta.$$

For the $\sum_{i=0}^{\ell}\sum_{j=0}^{h_i - 1} \frac{14 \rank_{\pi(x)}}{k} \cdot \mathbbm{1}\left[w_{i,j} > w_{\ell, h'_{\ell}}\right]$ part, we use \Cref{equ:deterministic-part} to conclude that it is at most $\epsilon \cdot \rank_{\pi}(x)$. This finishes the proof. 
\end{proof}

\paragraph*{Space Allocation} Since now we set $k = c \cdot \frac{ (\log \log (1/ \delta))^2}{\epsilon}$, the condition for each compactor $C_{i,j}$ becomes as follows: Before it resets, let $s_1, s_2, \dots, s_\ell$ be the space parameters after each resize/insert operation. Then we need $\sum_{t = 1}^\ell 2^{-s_t / k} = \sum_{t=1}^\ell 2^{- c \cdot \frac{\epsilon}{(\log \log(1/\delta))^2} \cdot s_t} \leq 1$. (Note this is the same condition as \Cref{lem:compactor}.)

For each subset $H_i$, suppose $s_1, s_2, \dots, s_\ell$ is the sequence of space parameters after each reize / insertion from the sampler into $C_{i,0}$, instead of $\sum_{t=1}^\ell 2^{-\epsilon \cdot s_t} \leq 0.5$ (\Cref{equ:condition-H}), we now require that $\sum_{t=1}^\ell 2^{- c \cdot \frac{\epsilon}{(\log \log(1/\delta))^2} \cdot s_t} \leq 0.5$. Same as the proof of \Cref{lem:H}, as long as this condition is satisfied for each $H_i$, the condition for each compactor $C_{i,j}$ will be then satisfied. 

When allocating space, we now define $$\widehat{s}_{i, t} = \epsilon^{-1} \cdot (\log \log (1/\delta))^3 \cdot \left( \log \frac{\cceil{\phi^{(t)}_{i,j}}}{\cceil{\phi^{(t)}_{i + 1, m}}}  + 5\log(1/\epsilon) + 5 \log \log n\right).$$

Here $(\log \log (1/\delta))^2$ comes from the larger value for $k$, the one extra $\log \log (1/\delta)$ factor is because the weight of elements in $C_{i,0}$ now reduces by a factor of $\log (1/\delta)$, so the bound in \Cref{equ:online-allocation-intersection} now becomes $|W_{i,j} \cap [t_{i+1,m}, t_{i+1,m+1})| \leq 3\cdot \frac{\log (1/\delta)}{\epsilon^2} + 2 + O(\log^2(\epsilon n))$. Then we need the extra $\log \log (1/\delta)$ factor to account for this.

With these changes, the rest of the proofs follow from the exact same calculations as that of \Cref{sec:space-allocation}. For completeness, we repeat the calculations here:
\begin{itemize}
    \item (The total space is bounded.) Fix a time $t$. Let $j_i$ be the index of the interval $[t_{i,j_i}, t_{i,j_i + 1})$ that contains $t$. 
    \begin{align*}\sum_{i=0}^{\log (\epsilon n)} \widehat{s}_{i, t} &= \sum_{i=0}^{\log(\epsilon n)}  \epsilon^{-1} \cdot (\log \log (1/\delta))^3 \cdot \left( \log \frac{\cceil{\phi^{(t)}_{i,j}}}{\cceil{\phi^{(t)}_{i + 1, m}}}  + 5\log(1/\epsilon) + 5 \log \log n\right) \\
        &= O\left(\epsilon^{-1} \cdot(\log \log (1/\delta))^3 \cdot \log_2 \frac{\phi_{0,j_0}}{1} + \epsilon^{-1} \cdot (\log \log (1/\delta))^3 \cdot \log(\epsilon n) \cdot (\log(1/\epsilon) + \log \log n)\right) \\
        &= O\left(\epsilon^{-1} \cdot \log(\epsilon n) \cdot (\log \log n + \log(1/\epsilon)) \cdot (\log \log 1 / \delta)^3 \right)
    \end{align*}
    The last step follow from $\phi_{0,j_0} = \poly(\epsilon n)$ (invoking \Cref{claim:potential-bound} with $T = \poly(\epsilon n)$). 
    The total space is then $O\left(\epsilon^{-1} \cdot \log(1/\epsilon) \cdot \log(\epsilon n) \cdot (\log \log n + \log(1/\epsilon)) \cdot (\log \log 1 / \delta)^3 \right)$ because each $H_i$ uses space $\lceil \log_2(1/\epsilon) \rceil \cdot \widehat{s_{i,t}}$. 
    \item (The space sequence is feasible.) Again suppose interval $[t_{i,j}, t_{i,j + 1})$ intersects with children  $[t_{i + 1, \ell}, t_{i + 1, \ell + 1})$, $[t_{i + 1, \ell + 1}$, $t_{i + 1, \ell + 2})$, $\dots$, $[t_{i + 1, r - 1}, t_{i + 1, r})$. We use the new bound $$|W_{i,j} \cap [t_{i+1,m}, t_{i+1,m+1})| \leq 3\cdot \frac{\log (1/\delta)}{\epsilon^2} + 2 + O(\log^2(\epsilon n))$$ Then, 
    \begin{align*}
        \sum_{t \in W_{i,j}}2^{- c \cdot \frac{\epsilon}{(\log \log(1/\delta))^2} \cdot s_t} & \leq \sum_{m = \ell}^{r - 1} \sum_{\substack{t \\ t \in W_{i,j} \cap [t_{i + 1, m}, t_{i + 1, m + 1})}} 2^{- c \cdot \frac{\epsilon}{(\log \log(1/\delta))^2} \cdot s_t}\\
        &= \sum_{m = \ell}^{r - 1} \sum_{\substack{t \\ t \in W_{i,j} \cap [t_{i + 1, m}, t_{i + 1, m + 1})}} \frac{\cceil{\phi^{(t)}_{i + 1, m}}}{\cceil{\phi^{(t)}_{i,j}}} \cdot \epsilon^{-5} \cdot (\log n)^{-5} \cdot (\log(1/\delta))^{-1} \tag{$c \geq 1$.}
    \end{align*}
    Exactly the same as \Cref{sec:space-allocation}, note $\cceil{\phi^{(t)}_{i,j}}$ is monotone in $t$ and has $O(\log (\epsilon n))$ many different values. For each possible value $2^x$, we let $[a(x), b(x)] \subseteq [t_i, t_{j + 1})$ be the time interval such that $\cceil{\phi^{(t)}_{i,j}} = 2^x$, and suppose it intersects with children $[t_{i + 1, \ell(x)}, t_{i + 1, \ell(x) + 1})$, $[t_{i + 1, \ell(x) + 1}, t_{i + 1, \ell(x) + 2})$, $\dots$, $[t_{i + 1, r(x) - 1}, t_{i + 1, r(x)})$.
    
    Then for every $0 \leq x \leq O(\log(\epsilon n))$, we have
    \begin{align*}
    &\sum_{m=\ell(x)}^{r(x)- 1} \sum_{\substack{t  \\ t \in W_{i,j} \cap [t_{i + 1, m}, t_{i + 1, m + 1})\\ \cap [a(x), b(x)]}}\frac{\cceil{\phi^{(t)}_{i + 1, m}}}{\cceil{\phi^{(t)}_{i,j}}} \cdot \epsilon^{-5} \cdot (\log n)^{-5} \cdot (\log n)^{-5} \cdot (\log(1/\delta))^{-1}   \\
    \leq &\sum_{m=\ell(x)}^{r(x) - 1} \left(3 \cdot \log(1/\delta)/\epsilon^2 + 2 + O(\log^2 (\epsilon n))\right) \cdot \frac{2 \cdot \phi^{(b(x))}_{i + 1, m}}{2^x} \cdot \epsilon^{-5} \cdot (\log n)^{-5}  \cdot (\log n)^{-5} \cdot (\log(1/\delta))^{-1} \tag{Here we use $|W_{i,j} \cap [t_{i+1,m}, t_{i+1,m+1})| \leq 3\cdot \frac{\log (1/\delta)}{\epsilon^2} + 2 + O(\log^2(\epsilon n))$.} \\
    \leq &0.25 \cdot \frac{\sum_{m = \ell(x)}^{r(x) - 1} \phi^{b(x)}_{i + 1, m}}{2^x} \leq 0.25
    \end{align*}
    Thus, the space sequence is always feasible. 
\end{itemize}

 \end{document}